\documentclass[10pt,aps,prb,twocolumn,preprintnumbers,amsmath,amssymb,floatfix,showpacs,citeautoscript,superscriptaddress]{revtex4-2}
\usepackage[latin1]{inputenc}
\usepackage[T1]{fontenc}
\usepackage[english]{babel}
\usepackage[pdftex]{graphicx}
\usepackage{amsmath}
\usepackage{times}
\usepackage[usenames,dvipsnames,svgnames,table]{xcolor}
\usepackage[colorlinks,plainpages=false,linkcolor=blue,urlcolor=blue,citecolor=blue,pdfpagemode=UseNone]{hyperref}
\usepackage{hyperref}

\renewcommand{\AA}{\text{\r{A}}}

\definecolor{forestgreen(web)}{rgb}{0.13, 0.55, 0.13}

\definecolor{magenta}{rgb}{0.9,0.0,0.9}

\begin{document}

\title
{
\boldmath
Structural transitions, octahedral rotations, and electronic properties of $A_3$Ni$_2$O$_7$ rare-earth nickelates under high pressure
}

\author{Benjamin Geisler}
\email{benjamin.geisler@ufl.edu}
\affiliation{Department of Physics, University of Florida, Gainesville, Florida 32611, USA}
\affiliation{Department of Materials Science and Engineering, University of Florida, Gainesville, Florida 32611, USA}
\author{James J. Hamlin}
\affiliation{Department of Physics, University of Florida, Gainesville, Florida 32611, USA}
\author{Gregory R. Stewart}
\affiliation{Department of Physics, University of Florida, Gainesville, Florida 32611, USA}
\author{Richard G. Hennig}
\affiliation{Department of Materials Science and Engineering, University of Florida, Gainesville, Florida 32611, USA}
\affiliation{Quantum Theory Project, University of Florida, Gainesville, Florida 32611, USA}
\author{P.J. Hirschfeld}
\affiliation{Department of Physics, University of Florida, Gainesville, Florida 32611, USA}

\date{\today}

\begin{abstract}
Motivated by the recent observation of superconductivity with $T_c \sim 80$~K in pressurized La$_3$Ni$_2$O$_7$~\cite{Sun-327-Nickelate-SC:23},
we explore the structural and electronic properties in %
$A_3$Ni$_2$O$_7$ bilayer nickelates ($A=$~La-Lu, Y, Sc)
as a function of hydrostatic pressure (0-150~GPa)
from first principles including a Coulomb repulsion term.
At $\sim 20$~GPa, we observe an orthorhombic-to-tetragonal transition in La$_3$Ni$_2$O$_7$
at variance with recent x-ray diffraction data,
which points to so-far unresolved complexities at the onset of superconductivity, e.g., charge doping by variations in the oxygen stoichiometry. %
We compile a structural phase diagram
with particular emphasis on the $b/a$ ratio, octahedral anisotropy, and octahedral rotations. %
Intriguingly, chemical and external pressure emerge as two distinct and counteracting control parameters.
We find unexpected
correlations between $T_c$ and the \textit{in-plane} Ni-O-Ni bond angles for La$_3$Ni$_2$O$_7$. %
Moreover, two novel structural phases with significant $c^+$ octahedral rotations and in-plane bond disproportionations are uncovered for $A=$~Nd-Lu, Y, Sc %
that exhibit a surprising pressure-driven electronic reconstruction in the Ni~$e_g$ manifold.
By disentangling the involvement of basal versus apical oxygen  states at
the Fermi surface,
we identify Tb$_3$Ni$_2$O$_7$ as an interesting candidate for superconductivity at ambient pressure.
These results suggest a profound tunability of the structural and electronic phases in this novel materials class %
and are key for a fundamental understanding of the superconductivity mechanism.
\end{abstract}

\maketitle

\section{Introduction}

The discovery of superconductivity in Sr-doped infinite-layer NdNiO$_2$ films grown on SrTiO$_3$(001)~\cite{Li-Supercond-Inf-NNO-STO:19, Li-Supercond-Dome-Inf-NNO-STO:20, Zeng-Inf-NNO:20}
introduced a new class of formally Ni$^{1+}$ ($3d^9$) cuprate-like superconductors~\cite{Nomura-Inf-NNO:19, JiangZhong-InfNickelates:19, Sakakibara:20, JiangBerciuSawatzky:19, Botana-Inf-Nickelates:19, GeislerPentcheva-InfNNO:20, Lechermann-Inf:20, NNO-SC-Thomale:20, Gu-NNO2:20, Lu-MagExNdNiO2:21, Ortiz-NNO:21, Lechermann-Inf:21, SahinovicGeisler:21, Wang-IL-Pauli:21, GeislerPentcheva-NNOCCOSTO:21, Zeng-Inf-NNO:22, GoodgeGeisler-NNO-IF:22, KreiselLechermann-IL:22, Rossi-IL-CO:22, Fowlie-IL-IntrinsicMag:22, SahinovicGeisler:22, SahinovicGeislerPentcheva:23, Geisler-Rashba-NNOSTOKTO:23}.
Subsequent work rapidly increased the family of superconducting nickelate compounds, for instance, 
to PrNiO$_2$ and LaNiO$_2$ films~\cite{Osada-PrNiO2-SC:20, Osada-LaNiO2-SC:21}
as well as to quintuple Ruddlesden-Popper-derived films~\cite{Pan-ILSC:22}.
While epitaxial strain constitutes a successful and well-established route to realize novel quantum phases in rare-earth nickelates~\cite{OxideRoadmap:16, RENickelateReview:16, Viewpoint:19, TE-Oxides-Review-Geisler:21},
the impact of extreme pressure conditions is far less explored.
Intriguingly, a pressure-driven increase of the superconducting transition temperature $T_c$ in Sr-doped infinite-layer PrNiO$_2$ has been measured~\cite{Wang-Pressure-PNO:22}.

Very recently, signatures of superconductivity with a high $T_c \sim 80$~K were reported in the bilayer nickelate
La$_3$Ni$_2$O$_7$ subject to $\sim 14$-$43.5$~GPa external pressure in a diamond anvil cell~\cite{Sun-327-Nickelate-SC:23}.
This structure is a member of the Ruddlesden-Popper family, %
$A_{n+1}$Ni$_n$O$_{3n+1}$ ($n=2$),
and contributes another exciting perspective to this field,
specifically due to the formal valence of Ni$^{2.5+}$. %
Independent experimental works have already claimed confirmation of high-temperature superconductivity in this system in a similar pressure range~\cite{Hou-LNO327-ExpConfirm:23, Zhang-LNO327-ZeroResistance:23}.

This observation instantly spurred a sizable amount of initial theoretical literature on La$_3$Ni$_2$O$_7$~\cite{Luo-LNO327:23, Gu-LNO327:23, Yang-LNO327:23, Lechermann-LNO327:23, Sakakibara-LNO327:23, Shen-LNO327:23, Christiansson-LNO327:23, Shilenko-LNO327:23, Liu-LNO327-Optics:23, Wu-LNO327:23, Cao-LNO327:23, Chen-LNO327:23, Liu-LNO327-OxVacDestructive:23, Lu-LNO327-InterlayerAFM:23, ZhangDagotto-LNO327:23, OhZhang-LNO327:23, Liao-LNO327:23, Qu-LNO327:23}.
The general superconductivity mechanism that has been put forward involves a pressure-driven transition of the Fermi surface topology,
featuring an emerging hole pocket of Ni~$3d_{z^2}$ character~\cite{Sun-327-Nickelate-SC:23, Luo-LNO327:23, Gu-LNO327:23, Yang-LNO327:23, Lechermann-LNO327:23, Liu-LNO327-OxVacDestructive:23, ZhangDagotto-LNO327:23}.
This  typically results in a superconducting pairing with leading $s^{\pm}$ eigenvalue~\cite{Gu-LNO327:23, Yang-LNO327:23, Liu-LNO327-OxVacDestructive:23, Lu-LNO327-InterlayerAFM:23, ZhangDagotto-LNO327:23},
rendering the bilayer nickelates distinct from cuprates~\cite{Keimer:15} and Fe-based superconductors~\cite{ChubukovHirschfeld-FeSC:15}.
This electronic transition is believed to be accompanied %
by a structural transition from a $Cmcm$ (or $Amam$; space group No.~63) to a $Fmmm$ (No.~69) phase,
\textit{both} orthorhombic~\cite{Sun-327-Nickelate-SC:23},
with a concomitant suppression of the NiO$_6$ octahedral tilts that are characteristic of the $Cmcm$ phase.

This raises the fundamental questions
(i)~how the structural transition and the accompanying lack of octahedral rotations are related to the emergence of superconductivity, and
(ii)~if we can reduce the critical pressure by chemical precompression (a key concept in superconducting hydrides~\cite{Drozdov:15, Drozdov:19}) via isoelectronic $A$-site variation. %
In oxides, octahedral rotations are known to be sensitively related to the electronic structure and the degree of electronic correlation:
In the isomorphic bilayer compound Sr$_3$Ru$_2$O$_7$,
for example, similar octahedral rotations of around $\sim$6$^\circ$~\cite{Hu2011,Puetter2010}
are suppressed by Mn doping,
which drives the emergence
of antiferromagnetic order~\cite{Hu2011,Mesa2012,Mukherjee2016} and a low-temperature Mott insulating phase~\cite{Nakayam2018,Mathieu2005}.
Thus, the avoidance of octahedral relaxations may be coupled to %
unconventional superconductivity, although the exact mechanism remains elusive.

Here we provide a comprehensive and consistent exploration of the structural and electronic properties in %
$A_3$Ni$_2$O$_7$ bilayer nickelates ($A=$~La-Lu, Y, Sc)
as a function of hydrostatic pressure (0-150~GPa)
from first principles including a Coulomb repulsion term.
We compile a structural phase diagram
with particular emphasis on the orthorhombic distortion ($b/a$ ratio), octahedral anisotropy, and octahedral rotations. %
Surprisingly, the results establish chemical and external pressure as two distinct and counteracting control parameters,
which limits the perspectives of chemical precompression in this system.
We trace this phenomenon back to the enhancement of octahedral rotations with reducing $A$-site ionic radius.
The response of the lattice parameters to external pressure is found to be highly anisotropic. %
In La$_3$Ni$_2$O$_7$ at $\sim 20$~GPa, %
we observe an orthorhombic-to-tetragonal transition to an $I4/mmm$ phase at variance with recent x-ray diffraction data,
which points to yet unresolved complexities near the onset of superconductivity, e.g., electron- or hole-doped samples due to variations in the oxygen stoichiometry,
and suggests a careful reassessment of the so-far proposed superconductivity mechanisms.
The critical pressure associated with this transition, %
which coincides universally with vanishing octahedral tilts,
is found to increase quadratically over the rare-earth series. %
For $A=$~Nd-Lu, Y, Sc, two novel structural phases are uncovered at ambient conditions that are characterized by the emergence of significant in-plane $c^+$ octahedral rotations as well as in-plane bond disproportionations
and exhibit a surprising pressure-driven electronic reconstruction %
involving the rotation of the Ni~$d_{z^2}$ orbital.
The successive quenching of the distinct rotational degrees of freedom by pressure %
demonstrates that the potential energy landscape of the in-plane octahedral rotations is significantly shallower than for the octahedral tilts.
Moreover, we highlight unexpected correlations
between $T_c$ and the \textit{in-plane} Ni-O-Ni bond angles for La$_3$Ni$_2$O$_7$
and discuss their possible relation to superconductivity. %
Finally, by disentangling the involvement of basal versus apical oxygen ions in the Fermi surface,
we identify Tb$_3$Ni$_2$O$_7$ as interesting candidate for superconductivity at ambient pressure.

\begin{figure*}
\begin{center}
\includegraphics[width=\linewidth]{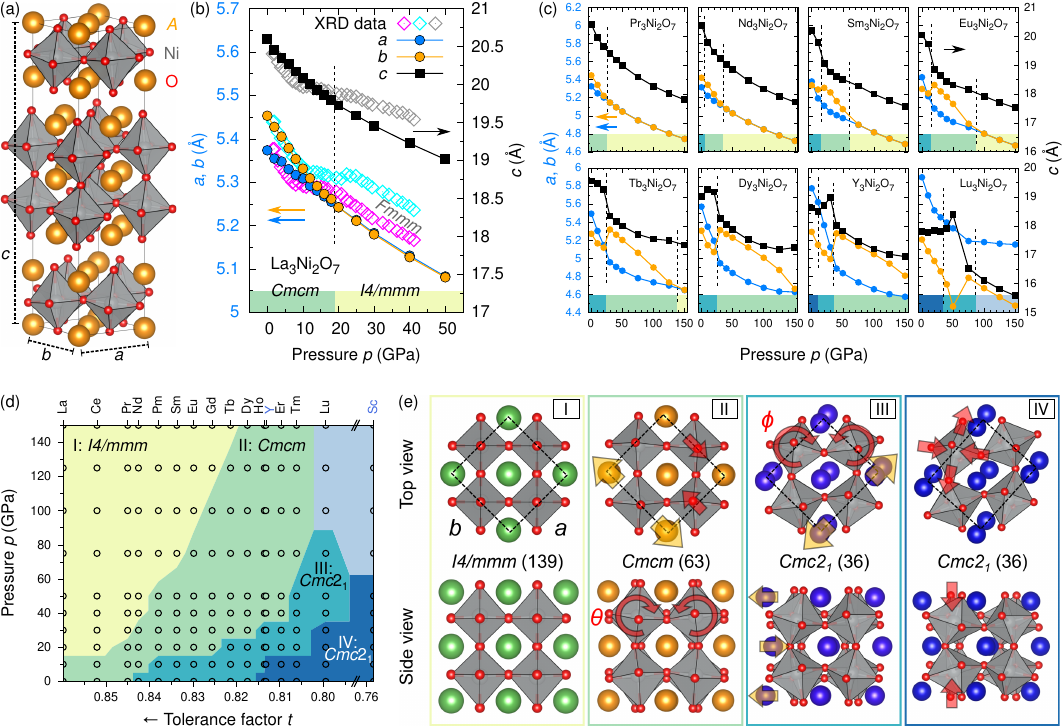}
\caption{\label{fig:XRDabcPD}(a)~Structure of the bilayer Ruddlesden-Popper nickelate $A_3$Ni$_2$O$_7$.
(b)~Lattice parameter analysis for La$_3$Ni$_2$O$_7$ as a function of the external pressure~$p$.
While our DFT$+U$ predictions agree closely with experimental observations~\cite{Sun-327-Nickelate-SC:23} for low pressure,
we observe an unexpected transition to a tetragonal $I4/mmm$ ($a = b$) instead of an orthorhombic $Fmmm$ phase ($a \ne b$) around $p \sim 20$~GPa.
(c)~Lattice parameters for further selected $A_3$Ni$_2$O$_7$ nickelates.
The vertical dashed lines mark structural phase boundaries [see panel (d)].
(d)~Structural phase diagram, contrasting the impact of chemical versus external pressure in $A_3$Ni$_2$O$_7$.
It has been compiled from the DFT$+U$ data discussed in Figs.~\ref{fig:StructTrends} and~\ref{fig:OctRot} (marked by small circles).
(e)~Corresponding top and side views of representative bilayer slabs.
The arrows mark characteristic differences between the distinct structural phases, involving octahedral rotations~$\phi$ and tilts~$\theta$, distinct orthorhombic distortions~$b/a$, typical displacements of the $A$-site ions, and different bond disproportionations of the NiO$_6$ octahedra.
}
\end{center}
\end{figure*}

\section{Methodology}

We performed first-principles simulations in the framework of  density functional theory (DFT~\cite{KoSh65}) %
as implemented in the \textit{Vienna Ab initio Simulation Package} (VASP)~\cite{USPP-PAW:99, PAW:94}, 
employing a wave-function cutoff of $520$~eV.
Exchange and correlations were described by using the generalized gradient approximation as parameterized by Perdew, Burke, and Ernzerhof~\cite{PeBu96}.
The rare-earth $4f$ electrons were frozen in the core.
We focus on the nonmagnetic phase here, but discuss additional results on the magnetic interactions in the Appendix.
Static correlation effects were considered within the DFT$+U$ formalism~\cite{LiechtensteinAnisimov:95, Dudarev:98},
employing an effective $U=3$~eV at the Ni sites,
in line with previous nickelate work~\cite{Liu-NNO:13, Botana-Inf-Nickelates:19, Geisler-LNOSTO:17, WrobelGeisler:18, GeislerPentcheva-LNOLAO:18, GeislerPentcheva-LNOLAO-Resonances:19, GeislerPentcheva-InfNNO:20, GeislerPentcheva-NNOCCOSTO:21, Geisler-VO-LNOLAO:22, Geisler-Rashba-NNOSTOKTO:23}
and a recent analysis of the optical spectrum of La$_3$Ni$_2$O$_7$~\cite{Geisler-LNO327-Optical:24}.

To account for octahedral rotations and in-plane bond disproportionations,
the $A_3$Ni$_2$O$_7$ bilayer nickelates ($A =$~La-Lu, Y, Sc) were modeled by using 24-atom unit cells.
In this geometry, the Brillouin zone was sampled employing
\mbox{$8\times8\times8$} Monkhorst-Pack $\Vec{k}$-point grids~\cite{MoPa76}
in conjunction with a Gaussian smearing of $5$~mRy.
We confirmed that these parameters provide converged energies and lattice parameters.
Accurate densities of states were obtained on \mbox{$12\times12\times12$} $\Vec{k}$-point grids.
The compounds can equivalently be described by orthorhombic 48-atom unit cells [Fig.~\ref{fig:XRDabcPD}(a)];
for these, we used \mbox{$8\times8\times2$} $\Vec{k}$-point grids.

The lattice parameters $a$, $b$ and $c$ (given with respect to the more convenient orthorhombic representation in the following)
and the internal ionic positions were accurately optimized in each case in DFT$+U$
under zero and finite external pressure, %
reducing ionic forces below $1$~mRy/a.u.

\section{\boldmath Structural properties of $A_3$N\lowercase{i}$_2$O$_7$ rare-earth nickelates under high pressure}

Over the rare-earth series and the considered pressure range ($p = 0$-$150$~GPa), 
we observe an unexpected richness of the structural properties of $A_3$Ni$_2$O$_7$ (Fig.~\ref{fig:XRDabcPD}).
Our comprehensive study puts us in position to compile an \textit{ab initio} structural phase diagram [Fig.~\ref{fig:XRDabcPD}(d)],
which underlines the possible potential of high-pressure %
experiments to realize exotic new quantum states
in the bilayer nickelates in particular and
in correlated transition metal oxides in general.
We will explore the phase diagram and the underlying data step by step in the following and discuss their implications for superconductivity.

\subsection{\boldmath Pressure-induced orthorhombic-to-tetragonal transition in La$_3$Ni$_2$O$_7$}

We begin by analyzing the pressure dependence of the lattice parameters of La$_3$Ni$_2$O$_7$ [Fig.~\ref{fig:XRDabcPD}(a,b)].
The fully \textit{ab initio} relaxed results for $a,b$ and $c$ agree nicely with
the experimental lattice parameters refined from x-ray diffraction (XRD)~\cite{Sun-327-Nickelate-SC:23} %
between 0-10 GPa.
In particular, the difference between $a$ and $b$, i.e., the degree of orthorhombic distortion, is accurately captured.
This further corroborates that the Hubbard~$U=3$~eV employed here is appropriate. %
Consistent with previous experimental~\cite{LNO-327-Diffraction-Zhang:94, LNO-327-Diffraction-Ling:00}
and theoretical assessments~\cite{Sun-327-Nickelate-SC:23, ZhangDagotto-LNO327:23},
we identify a structure  of $Cmcm$ space group (or $Amam$, No.~63) at low pressure,
referred to as phase~II in the following [Fig.~\ref{fig:XRDabcPD}(d)]. %
It is characterized by finite octahedral tilts~$\theta$ and $b/a > 1$,
where we define $b$ as the axis along which the tilts are expressed [Fig.~\ref{fig:XRDabcPD}(e)].
Simultaneously, octahedral rotations in the basal plane are absent ($\phi = 0$, $c^0$).
Notably, without octahedral rotations of any kind ($Fmmm$, $I4/mmm$), the assignment of $a$ and $b$ is arbitrary. %

Between 10-20~GPa, the experimental lattice parameters $a$ and $b$
display the unambiguous trend to converge, accompanied by a plateau in the $c$ curve.
However, a sudden increase can be observed at $\sim 20$~GPa,
after which they continue to decrease monotonically with a finite separation of around $0.075~\AA$.
It has been suggested earlier that the octahedral tilts are quenched in this pressure range ($\theta = 0$),
corresponding to inter-layer Ni-O-Ni bond angles of $180^\circ$ 
and resulting in a geometry with orthorhombic $Fmmm$ space group (No.~69)~\cite{Sun-327-Nickelate-SC:23}.
This geometry was explored in detail in subsequent work~\cite{Luo-LNO327:23, Yang-LNO327:23, Lechermann-LNO327:23, Sakakibara-LNO327:23, Christiansson-LNO327:23, Shilenko-LNO327:23, Wu-LNO327:23, Cao-LNO327:23, Liu-LNO327-OxVacDestructive:23, ZhangDagotto-LNO327:23}.

Surprisingly, we observe a clear orthorhombic-to-tetragonal transition at $p \sim 20$~GPa,
resulting in a structure of $I4/mmm$ symmetry (phase~I, space group No.~139) rather than the so-far reported $Fmmm$ geometry.
The enthalpy difference
between the $Fmmm$ (lattice parameters from XRD~\cite{Sun-327-Nickelate-SC:23})
and the $I4/mmm$ geometry obtained here, exemplarily for La$_3$Ni$_2$O$_7$ at 30~GPa,
amounts to $\Delta H = 90$~meV$/$Ni, %
which additionally corroborates our findings.
Moreover, we confirmed the orthorhombic-to-tetragonal transition
by variation of $U$ at the Ni site,
additional application of $U$ to the La~$4f$ and~$5d$ states,
and different patterns of the octahedral rotations ($c^0$, $c^+$, $c^-$).
We recently became aware of higher-resolution x-ray experiments that report the observation of the $I4/mmm$ phase above $p \sim 19$~GPa~\cite{Wang-LNO327-I4mmm:23}.

We demonstrate below for La$_3$Ni$_2$O$_7$ that the key features of the electronic structure are preserved between the $I4/mmm$ and $Fmmm$ phases.
Nevertheless, these observations suggest so-far unresolved complexities at the onset of superconductivity        %
which may be key in identifying the underlying mechanism:
On the one hand, the application of pressure  may not be uniform, %
as indicated by the role of different pressure-transmitting media employed in the diamond anvil cells~\cite{Sun-327-Nickelate-SC:23, Hou-LNO327-ExpConfirm:23, Zhang-LNO327-ZeroResistance:23}.
On the other hand, these results may indicate that the experimental samples are electron or hole doped,
most probably due to variations in the oxygen stoichiometry.
There is considerable evidence that the electronic properties are extremely sensitive to small changes in the oxygen vacancy concentration~\cite{LNO-327-Diffraction-Zhang:94, Taniguchi-LNO327:95}.
Due to the strong coupling of the charge and lattice degrees of freedom in correlated transition metal oxides~\cite{Varignon:17, Catalano:18, GeislerPentcheva-LCO:20, Radhakrishnan-Geisler-YVOLAO:21, Radhakrishnan-Geisler-YVOLAO:22},
this would alter their structural response to pressure and potentially lead to the experimentally observed finite orthorhombic distortion at higher pressure.
In that case, the so-far proposed $s^{\pm}$ superconductivity mechanism needs a careful reassessment.

\subsection{Can we substitute external by chemical pressure?}

Next, we explore the role of chemical pressure by isoelectronic variation of the $A$-site element.
The ionic radii of the considered elements range from
$1.172$ (La$^{3+}$) to $1.001$ (Lu$^{3+}$), as well as
$1.04$ (Y$^{3+}$) and
$0.885$ (Sc$^{3+}$).
As a measure of this chemical pressure, we employ the Goldschmidt tolerance factor, a traditional descriptor for perovskite oxides:
$$
t(A) = \frac{r_A + r_\text{O}}{\sqrt{2} \, (r_\text{Ni} + r_\text{O}) } \ \text{,}
$$
where $t=1$ corresponds to size-balanced $A$ and Ni sites.
It is reasonable to order the $A$-site elements with decreasing $t$ [Fig.~\ref{fig:XRDabcPD}(d)].
Here we find consistently $t<1$. %
Can we exploit this strategy to lower the critical external pressure driving the structural phase transition
and thus facilitate the emergence of superconductivity in the bilayer nickelates?

We can directly conclude
from the phase diagram in Fig.~\ref{fig:XRDabcPD}(d), which is not symmetric,
that this hypothesis does not hold.
The transition from the $Cmcm$ to the $I4/mmm$ phase occurs at critical pressures that increase monotonically
from $p \sim 20$ ($A=$~La, Ce), $30$ (Pr), and $40$~GPa (Nd)
to $p \sim 125$-$150$~GPa (Tb), as shown in Fig.~\ref{fig:XRDabcPD}(d),
with a parabola-shaped phase boundary as a function of $t$
[see also the $b/a$ ratios in Fig.~\ref{fig:StructTrends}(a) and the octahedral tilts~$\theta$ in Fig.~\ref{fig:OctRot}(a)].
While finalizing this manuscript, we became aware of another very recent work that suggests a similar shape of this phase boundary,
albeit classifying the high-pressure phase as $Fmmm$~\cite{ZhangDagotto-RE-LNO327:23}.
If superconductivity is linked to this structural phase transition,
our results imply that continuously increasing external pressures are required
as $A$ progresses through the rare-earth series.
In the following, we will identify enhanced octahedral rotations as the impeding physical mechanism. %

These results unambiguously show that chemical and external pressure generally constitute two distinct control parameters
that independently and non-interchangeably allow the designing of the quantum state in bilayer rare-earth nickelates.

\subsection{Anisotropic pressure response of the lattice parameters}

Figure~\ref{fig:XRDabcPD}(c) shows that
the cell height $c$ varies overall from $\sim 15$ ($A=$~Sc at 150~GPa) to $20.6~\AA$ (La at 0~GPa)
and generally reduces with decreasing ionic radius of the $A$-site element.
Some pronounced discontinuities can be observed at the phase boundaries, e.g., between phases II and III [Fig.~\ref{fig:XRDabcPD}(c)].
Surprisingly, a counter-intuitive \textit{increase} of $c$ can be identified with increasing pressure %
for $\sim 0$-$50$~GPa in Dy$_3$Ni$_2$O$_7$, Y$_3$Ni$_2$O$_7$, and Lu$_3$Ni$_2$O$_7$ (phase~III).

For the in-plane lattice parameters $a$ and $b$, which range from $4.41$ ($A=$~Sc at 150~GPa) to $5.95~\AA$ (Tm at 0~GPa),
a simple decreasing $A$-site dependence can only be identified
for the early rare-earth metals up to Nd [Fig.~\ref{fig:XRDabcPD}(c)].
Between Pm and Tm, even a sharp upturn in the average of $a$ and $b$ can be observed,
which also translates to the basal Ni-Ni distances (not shown).

\begin{figure*}
\begin{center}
\includegraphics[width=\linewidth]{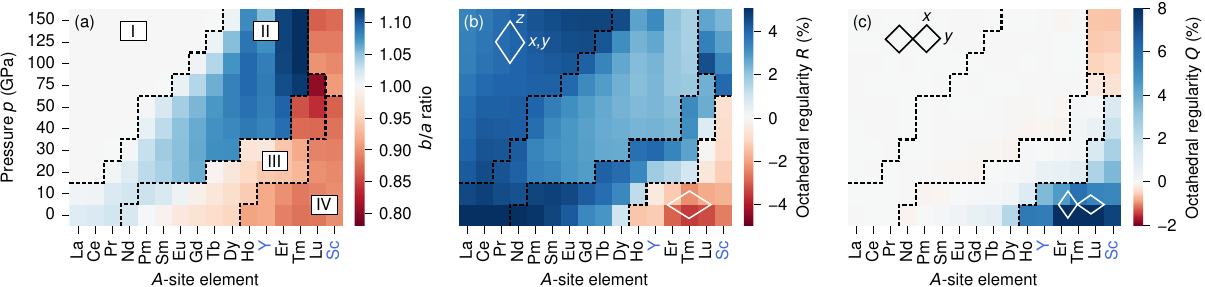}
\caption{\label{fig:StructTrends}Structural trends of $A_3$Ni$_2$O$_7$ across the rare-earth series (including Y and Sc) for varying external pressure~$p$ from 0-150~GPa.
The $A$-site elements are ordered according to their ionic radius. %
The dashed lines separate the distinct structural phases [Fig.~\ref{fig:XRDabcPD}(d)].
(a)~The $b/a$ ratio clearly identifies the tetragonal $I4/mmm$ phase~I
and separates it from the $Cmcm$ phase~II with $b/a > 1$. The remaining phases show largely $b/a < 1$.
(b)~Apical versus basal extension of the octahedra ($R$).
In phases I-III, the octahedra are elongated in vertical $z \sim c$ direction, particularly for La$_3$Ni$_2$O$_7$ at ambient conditions.
In contrast, phase~IV exhibits vertically compressed octahedra.
(c)~The in-plane disproportionation of the O-Ni-O distances ($Q$)
is very pronounced in phase~IV, which is characterized by cigar-shaped octahedra alternating in the plane, reminiscent of LaMnO$_3$,
and a considerable electronic reconstruction of the Ni~$3d_{z^2}$ orbital.
}
\end{center}
\end{figure*}

This coincides with the stabilization of the novel structural phases~III and~IV [Fig.~\ref{fig:XRDabcPD}(d)].
Both have $Cmc2_1$ symmetry (space group No.~36)
and are characterized by finite ferrodistortive ($c^+$) octahedral rotations arising as an additional degree of freedom ($\phi > 0$; Fig.~\ref{fig:OctRot}).
These are accompanied by substantial $A$-site displacements in the basal plane [Fig.~\ref{fig:XRDabcPD}(e)],
which are typical for this octahedral rotation pattern~\cite{Radhakrishnan-Geisler-YVOLAO:21, Radhakrishnan-Geisler-YVOLAO:22}
and a useful fingerprint in transmission electron microscopy. %
The transition to the $Cmcm$ phase~II is associated with pronounced jumps of $a$ and $b$ [Fig.~\ref{fig:XRDabcPD}(c)],
resulting generally in a reversal of the $b/a$ ratio [Fig.~\ref{fig:StructTrends}(a)].

We see that the response of the lattice parameters to uniform external pressure is highly anisotropic.
The cell height $c$ is generally compressed more strongly than $a$ and $b$.
This can be traced back to a partial accommodation of the pressure by the more elastic bilayer separation. %
For example, we find that
the latter reduces from $6.4$ to $5.34~\AA$ for 0-150~GPa in La$_3$Ni$_2$O$_7$,
which clearly exceeds the concomitant compression of the bilayer height from $3.91$ to $3.54~\AA$
(each measured between the Ni planes).

Inspection of the phase diagram in Fig.~\ref{fig:XRDabcPD}(d) reveals
that the boundaries between phases II-III and III-IV are roughly parabola-shaped as a function of $t$,
similar to the boundary between phases I-II.
This further corroborates the opposing impact of external pressure versus $A$-site variation.
Beyond the phases I-IV [top-right corner of Fig.~\ref{fig:XRDabcPD}(d)],
we found that these counteracting forces lead to bizarre deformations of the NiO$_6$ octahedra (not shown),
which indicates that the compounds become unstable under such extreme conditions.

The richness of the structural phase diagram and the unexpected transitions identified in the later rare-earth nickelates
provide additional opportunities for e.g.\ superconducting phases.
We will shed more light on this aspect below.

\subsection{Octahedral anisotropy and bond disproportionation}

The anisotropic pressure response of the lattice parameters is directly related to shape modifications of the NiO$_6$ octahedra,
which in turn determine the spatial orientation of the two Ni~$e_g$ orbitals and their relative occupation, i.e., the orbital polarization~\cite{WuBenckiser:13, GeislerPentcheva-LNOLAO-Resonances:19, Geisler-VO-LNOLAO:22}.
We measure the deviations from an ideal, regular octahedron  by defining %
$$
R = \frac{d_{z} - d_{x,y}}{d_{z} + d_{x,y}}  \quad \text{and} \quad  Q = \frac{d_x-d_y}{d_x+d_y}
$$
from the in-plane O-Ni-O distances $d_x$ and $d_y$ (which may alternate among the Ni sites in a checkerboard pattern, resulting in $Q \ne 0$),
their average $d_{x,y}$, and the out-of-plane O-Ni-O distances $d_{z}$.

The octahedral anisotropy $R$  is a measure of the apical versus basal extension of the octahedra [Fig.~\ref{fig:StructTrends}(b)].
It ranges overall from $5.3\%$ ($A=$~La at 0~GPa) to $-3.5\%$ (Tm at 0~GPa).
We find $R > 0$ for most rare-earth nickelates, implying an elongation of the Ni octahedra in the $z \sim c$ direction
and thus a Ni~$d_{z^2}$ orbital that points in the apical direction, whereas the Ni~$d_{x^2-y^2}$ orbital is oriented in the basal plane.

Notably, the broken octahedral connectivity between the bilayer slabs
results in an internal asymmetry of each NiO$_6$ octahedron, visible in the structural side views in Fig.~\ref{fig:XRDabcPD}(e).
Specifically, for La$_3$Ni$_2$O$_7$,
the Ni-O bond lengths involving the 'outer' apical oxygen ions
(pointing into the structural gap) are generally enhanced~\cite{Sun-327-Nickelate-SC:23, Lechermann-LNO327:23}.
Simultaneously, we find that these Ni-O bonds are reduced more strongly due to pressure than their 'inner' analogs.
For instance, we observe
$2.31$, $1.97~\AA$ (0~GPa),
$2.08$, $1.90~\AA$ (30~GPa), and %
$1.91$, $1.81~\AA$ (100~GPa)
for the outer and inner apical Ni-O bonds, respectively.
This mechanism facilitates the vertical elongation of the octahedra %
and thus the occupation of the Ni~$d_{z^2}$ orbital.

The early rare-earth elements exhibit a monotonically decreasing trend of $R$ with increasing pressure, i.e., the anisotropy is reduced [Fig.~\ref{fig:StructTrends}(b)].
This is in line with the more rapid decrease of $c$ with respect to $a$ and $b$. %
Specifically, for La$_3$Ni$_2$O$_7$,
we find that $R$ decreases from $5.3\%$ at 0~GPa to $3.3\%$ at 150~GPa. %
The central rare-earth elements exhibit a more complex pressure dependence [Fig.~\ref{fig:StructTrends}(b)].
Sharp discontinuities at the phase transition II-III can be observed,
and correlations with the orthorhombic distortion ($b/a$ ratio) are clearly visible, particularly in phase~II [Fig.~\ref{fig:StructTrends}(a)].

At zero pressure, $R$ decreases rapidly over the rare-earth series [Fig.~\ref{fig:StructTrends}(b)].
Intriguingly, it becomes negative for $A=$~Ho-Lu, Y, Sc,
which corresponds to vertically compressed octahedra (phase~IV).
The respective compounds show a pressure-induced transition from $R<0$ to $R>0$ %
at critical pressures ranging from $\sim 10$-$75$~GPa.
This behavior is highly distinct from all other bilayer rare-earth nickelates.

\begin{figure*}
\begin{center}
\includegraphics[width=\linewidth]{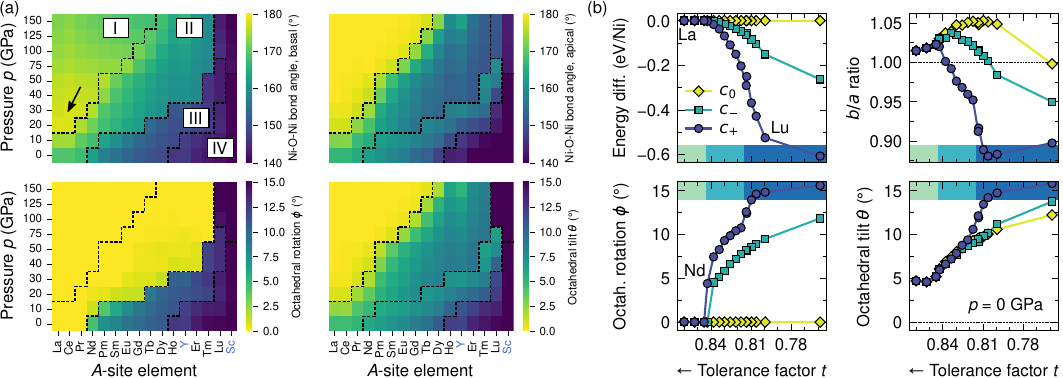}
\caption{\label{fig:OctRot}(a)~Analysis of the Ni-O-Ni bond angles and octahedral rotations in $A_3$Ni$_2$O$_7$.
The dashed lines separate the distinct structural phases [Fig.~\ref{fig:XRDabcPD}(d)].
Surprisingly, we observe an overall maximum in the \textit{in-plane} Ni-O-Ni bond angles for $A=$~La at 20~GPa, i.e., near the experimental onset of superconductivity (marked by the arrow).
For higher pressure, the in-plane bond angles are enhanced again, which correlates with the experimentally observed reduction of $T_c$.
(b)~Structural impact of the in-plane octahedral rotations~$\phi$ in $A_3$Ni$_2$O$_7$ at $p = 0$~GPa.
While absent for $A=$~La-Pr, ferrodistortive $c^+$ rotations stabilize for Nd-Lu, Y, Sc
and induce a reversal from $b/a>1$ to $b/a<1$.
Moreover, they couple strongly to the octahedral tilts~$\theta$ and enhance them considerably for the later rare-earth compounds ($t<0.82$).
}
\end{center}
\end{figure*}

Fig.~\ref{fig:StructTrends}(c) quantifies the in-plane disproportionation of the O-Ni-O distances ($Q$). %
We observe that it vanishes in phases I and II.
In phase III, where $R>0$, typical values are finite but small, e.g., $-0.06\%$ ($A=$~Nd) and $0.79\%$ (Tb) at zero pressure.
Surprisingly, $Q$ becomes very pronounced in phase~IV, which coincides with $R<0$, and reaches $9\%$ at $A=$~Tm  at zero pressure.
This corresponds to cigar-shaped octahedra that alternate in the plane in conjunction with $c^+$ octahedral rotations (Fig.~\ref{fig:OctRot}), a pattern that closely resembles LaMnO$_3$~\cite{Raman-Manganates-Iliev:98, Baldini-LMO:15}.
Most importantly, this coincides with a considerable electronic reconstruction involving the rotation of the Ni~$d_{z^2}$ orbital into the basal plane. %
Intriguingly, we see that this phase can be lifted by pressure of around 10-40~GPa [Fig.~\ref{fig:StructTrends}(c)].
Such extreme tunability of the geometry and the electronic structure is highly promising
and deserves further exploration.

\subsection{The pivotal role of octahedral rotations}

Figure~\ref{fig:OctRot}(a) analyzes the basal and apical Ni-O-Ni bond angles
as well as the octahedral rotations $\phi$ and tilts $\theta$ [defined in Fig.~\ref{fig:XRDabcPD}(e)] in $A_3$Ni$_2$O$_7$,
which provide insightful and distinct perspectives on the rotational degrees of freedom.

In the $Cmcm$ phase~II, where $\phi=0$, we find that $\theta$ ranges from $\sim 2$-$8^{\circ}$.
In the $Cmc2_1$ phases~III and~IV, $\phi$ and $\theta$ are both enhanced to $\sim 4$-$15^{\circ}$.
With increasing pressure, the in-plane rotations $\phi$, if present, %
are quenched first (for $A=$~Nd-Tm at 10-75~GPa),
which corresponds to the boundary between phases II and III. %
For Lu and Sc, the octahedral rotations are so pronounced that we observe finite values up to 150~GPa.
Subsequently, the octahedral tilts $\theta$
are quenched at yet higher pressure values (for $A=$~La-Tb at 20-150~GPa).
Intriguingly, we find that these vanishing tilts coincide universally with the orthorhomic-to-tetragonal transition from phase~II to phase~I.

This demonstrates that external pressure generally reduces or even quenches the octahedral rotations, 
whereas chemical pressure rather enhances them.
The subsequent suppression of the distinct rotational degrees of freedom indicates that different energy scales are involved,
with a significantly shallower potential energy landscape related to $\phi$ than to $\theta$.

The superconducting transition has been so far associated with a straightening
of the inter-layer (apical) Ni-O-Ni bond angles towards $180^{\circ}$ \cite{Sun-327-Nickelate-SC:23}.
The respective top-right panel in Fig.~\ref{fig:OctRot}(a)
closely resembles the bottom-right panel displaying the octahedral tilts~$\theta$
and shows that pressures up to 150~GPa successfully engineer this state for $A=$~La-Tb.
In particular, $\sim 20$~GPa for La$_3$Ni$_2$O$_7$ are sufficient to obtain inter-layer Ni-O-Ni bond angles close to $180^{\circ}$.
For Ce$_3$Ni$_2$O$_7$, already slightly lower pressures result in this state,
whereas Pr$_3$Ni$_2$O$_7$ and Nd$_3$Ni$_2$O$_7$ require rather $\sim 30$-$40$~GPa.
At zero pressure, we observe a value of $168^{\circ}$ for La$_3$Ni$_2$O$_7$, in perfect agreement with previous work~\cite{Sun-327-Nickelate-SC:23}.
The late rare-earth nickelates exhibit considerably smaller values (i.e., more pronounced bond angles),
reaching even below $140^{\circ}$ for Lu and Sc.

\begin{figure*}
\begin{center}
\includegraphics[width=\linewidth]{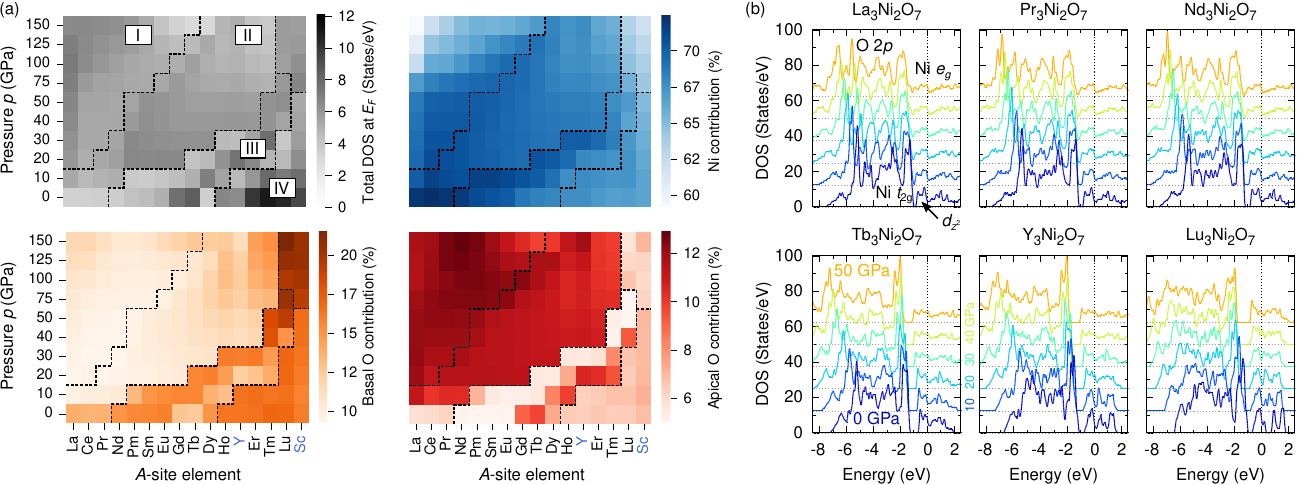}
\caption{\label{fig:ElStr}Electronic structure of $A_3$Ni$_2$O$_7$ rare-earth nickelates.
(a)~Total DOS at the Fermi energy
and relative contributions from the Ni, basal oxygen, and apical oxygen ions.
The dashed lines separate the distinct structural phases [Fig.~\ref{fig:XRDabcPD}(d)].
Noteworthy are the high contributions from the apical oxygen ions in phases~I and~II.
Intriguingly, Tb$_3$Ni$_2$O$_7$ at zero pressure presents a considerably higher DOS at the Fermi level ($\sim 8.5$~States$/$eV) than La$_3$Ni$_2$O$_7$ reaches at around 100 GPa ($\sim 7.2$~States$/$eV) in addition to a sizable involvement of the apical oxygen ions.
(b)~Evolution of the total DOS as a function of the external pressure for selected compounds.
An overall pressure-induced broadening of the O~$2p$-derived valence band can be observed,
as well as of the Ni~$e_g$ manifold located at the Fermi energy.
Characteristic peaks clearly shift to lower energies,
e.g., the Ni~$t_{2g}$ peak visible between $-2$ and $-1$~eV at zero pressure,
or the oxygen peak around $-5.5$~eV ($A=$~La-Tb).
The feature directly below the Fermi energy (marked by the arrow) corresponds to the lower set of Ni~$3d_{z^2}$-O~$2p_{z}$ hybrid states [Fig.~\ref{fig:BandsFS}(a)].
}
\end{center}
\end{figure*}

Motivated by the superconducting infinite-layer nickelates,
which exhibit \textit{in-plane} Ni-O-Ni bond angles of $180^{\circ}$~\cite{Li-Supercond-Inf-NNO-STO:19, GeislerPentcheva-InfNNO:20, GeislerPentcheva-NNOCCOSTO:21, GoodgeGeisler-NNO-IF:22},
we additionally explore these quantities for the bilayer nickelates. %
They directly impact the basal Ni~$3d_{x^2-y^2}$-O~$2p_{x,y}$ hybridization %
and thus the electronic structure at the Fermi energy [see also Fig.~\ref{fig:ElStr}(a) below]
including the Ni~$e_g$ orbital polarization
as well as possible superexchange mechanisms.
Intriguingly, the comprehensive perspective provided in Fig.~\ref{fig:OctRot}(a), top-left panel, unveils an overall maximum for $A=$~La at 20~GPa, %
i.e., close to the experimental onset of superconductivity. %
For higher pressure, the in-plane bond angles are enhanced again,
which we find to surprisingly correlate with the experimentally observed reduction of $T_c$ \cite{Sun-327-Nickelate-SC:23, Hou-LNO327-ExpConfirm:23, Zhang-LNO327-ZeroResistance:23}.
It is important to note here that the in-plane Ni-O-Ni bond angles can deviate from $180^{\circ}$ \textit{despite} $\theta=\phi=0$,
since the basal oxygen ions show an increasing tendency towards a uniform inwards relaxation under high pressure
that does not correspond to finite octahedral rotations.
These observations suggest that the \textit{in-plane} Ni-O-Ni bond angles are an important aspect in understanding  the superconducting phase in bilayer rare-earth nickelates.

Figure~\ref{fig:OctRot}(b) analyzes the structural impact of the in-plane octahedral rotations~$\phi$ in $A_3$Ni$_2$O$_7$ at zero pressure, %
i.e., highlighting the chemical effect of $A$-site variation. %
While absent for $A=$~La-Pr, ferrodistortive $c^+$ rotations stabilize for $A=$~Nd-Lu, including Y and Sc,
with a considerable energy difference $\Delta E \sim -0.6$~eV$/$Ni relative to the metastable $c^0$ case.
We see that they are strongly coupled to the $b/a$ ratio and induce a reversal from $b/a>1$ to $b/a<1$, %
which for  antiferrodistortive $c^-$ rotations  occurs only for Lu and Sc %
and without octahedral rotations ($c^0$) only for Sc.
Moreover, they couple strongly to the octahedral tilts~$\theta$
and enhance them considerably for the later rare-earth nickelates (phase IV, $t<0.82$),
an effect that is clearly absent for $c^-$ and $c^0$. %
This demonstrates that the correct description of the octahedral rotations is key for obtaining accurate structural properties
for $A=$~Nd-Lu, Y, Sc.

\begin{figure}
\begin{center}
\includegraphics[width=\linewidth]{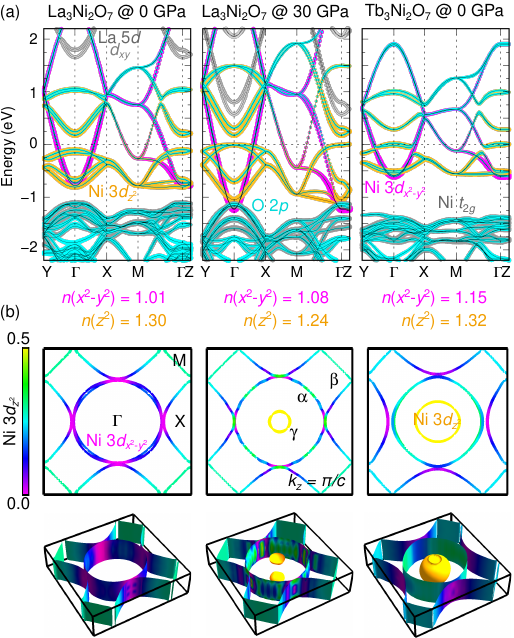}
\caption{\label{fig:BandsFS}(a)~Band structures of La$_3$Ni$_2$O$_7$ at $p = 0$ ($Cmcm$) and $30$~GPa ($I4/mmm$) as well as Tb$_3$Ni$_2$O$_7$ at $0$~GPa ($Cmc2_1$) from DFT+$U$.
The band character is represented by the different colors, and the Ni~$e_g$ occupation numbers~$n$ are explicitly provided.
(b,c)~The corresponding Fermi surfaces unravel an increase in Ni~$3d_{z^2}$ orbital weight (color bar) in pressurized La$_3$Ni$_2$O$_7$, specifically for sheet $\alpha$, and an emerging hole pocket~($\gamma$) around the $\Gamma$ point.
Tb$_3$Ni$_2$O$_7$ presents a similar Fermi surface topology at zero pressure, despite finite octahedral rotations,
in conjunction with a considerably enhanced DOS at the Fermi energy [Fig.~\ref{fig:ElStr}(a)].
}
\end{center}
\end{figure}

\section{\boldmath Electronic structure of $A_3$N\lowercase{i}$_2$O$_7$ rare-earth nickelates under high pressure}

Finally, we explore how the structural observations we discussed above relate to the electronic properties of $A_3$Ni$_2$O$_7$ rare-earth nickelates.
Figure~\ref{fig:ElStr}(a) displays the total density of states (DOS) at the Fermi energy
as a function of the $A$-site element and the external pressure,
which is an important indicator for enhanced superconducting properties.
Furthermore, Fig.~\ref{fig:ElStr}(a) disentangles the relative contributions from the Ni, basal oxygen, and apical oxygen ions.
This allows us to track  the composition of the Fermi surface, %
particularly variations in the basal versus apical oxygen involvement,
which also reflects the Ni~$3d_{x^2-y^2}$ versus $3d_{z^2}$ contributions, respectively.

Substantial correlations of all four panels with the distinct structural phases established above can be identified. %
Along the transition from phase III to II, as well as for $A=$~La-Pr at $\sim 10$~GPa,
we observe a sudden enhancement in the total DOS from roughly $3.5$ to $6.5$~States$/$eV.
This is accompanied by a reduction of the basal oxygen contributions from $\sim 15\%$ to $10\%$,
corresponding to the Ni~$3d_{x^2-y^2}$-O~$2p_{x,y}$ system,
and a concomitant increase of the apical oxygen contributions from $\sim 5\%$ to $11\%$,
related to the Ni~$3d_{z^2}$-O~$2p_{z}$ hybrid states.
Simultaneously, the \textit{absolute} Ni contributions increase as well, even though we see a slight reduction of the \textit{relative} Ni contributions,
which implies that the overall relevance of oxygen increases.

It is surprising that the enhanced involvement of apical oxygen
clearly correlates with phase~II ($\phi \sim 0$) rather than with phase~I ($\theta \sim 0$). %
Nevertheless, within phase~I, the apical oxygen contribution is generally further boosted and reaches its maximum $> 12\%$ for Nd at 150~GPa.
In sharp contrast,
in phase~III and particularly IV the \textit{basal} ions dominate the oxygen contributions to the Fermi surface.
An exception are the unexpectedly high apical oxygen contributions arising along the transition from phase~III to IV,
i.e., at the verge to the strongly bond-disproportionated state
which is accompanied by in-plane Ni~$3d_{z^2}$ orbital order and vertically compressed octahedra discussed above [Fig.~\ref{fig:StructTrends}(b,c)].

The total DOS in La$_3$Ni$_2$O$_7$ ranges from $\sim 3.5$~States$/$eV at 0~GPa to $\sim 7.2$~States$/$eV at 100~GPa.
By screening the $A$-site elements at $p=0$~GPa, we identify even higher values particularly for Tb$_3$Ni$_2$O$_7$ ($\sim 8.5$~States$/$eV).
This compound also exhibits the characteristic reversal from basal to apical oxygen involvement
that usually appears exclusively under finite pressure.
The maximum of the total DOS is observed for Tm$_3$Ni$_2$O$_7$ at 0~GPa.
However, the contributions from the Ni and the apical oxygen ions are relatively low for this compound,
and its band structure (not shown) differs considerably from that of e.g.\ La$_3$Ni$_2$O$_7$ due to the Ni~$e_g$ electronic reconstruction.

Figure~\ref{fig:ElStr}(b) tracks the evolution of the total DOS as a function of the external pressure for a selection of representative compounds.
We observe an overall pressure-induced broadening of the O~$2p$-derived valence band.
Characteristic peaks are clearly shifted to lower energies with increasing pressure,
e.g., the Ni~$t_{2g}$ peak visible between $-2$ and $-1$~eV at 0~GPa,
and, even more strongly, the oxygen peak visible between $-4.5$ and $-5.5$~eV at 0~GPa.

The Ni~$e_g$ manifold is located around the Fermi energy.
We see that its band width is considerably reduced across the rare-earth series,
whereas finite external pressure broadens it.
This observation reflects once again the counteracting nature of these two control parameters.
The feature directly below the Fermi level has Ni~$3d_{z^2}$-O~$2p_{z}$ character [see also Fig.~\ref{fig:BandsFS}(a)]. %
Figure~\ref{fig:ElStr}(b) demonstrates the pressure-induced occupation of these states for $A=$~La, Pr, and Nd.
Surprisingly, for Tb$_3$Ni$_2$O$_7$, we observe that they are already partly occupied at 0~GPa.

Finally, the DOS of Y$_3$Ni$_2$O$_7$ and Lu$_3$Ni$_2$O$_7$ has a distinct shape around the Fermi energy since the Ni~$3d_{z^2}$ orbital is oriented in the plane at 0~GPa (phase~IV),
as we discussed above in the context of Fig.~\ref{fig:StructTrends}(c).
It is interesting to follow the evolution of these curves as pressure drives the structural transition to phase~III,
which uncovers a concomitant electronic reconstruction in the $e_g$ manifold,
i.e., a re-alignment of the Ni~$3d_{z^2}$ orbital with the vertical axis.
The high apical oxygen values at the phase boundary render this transition even more compelling,
and $A=$~Ho at $\sim 10$~GPa as well as $A=$~Y-Tm at $\sim 20$~GPa emerge as further candidates for future in-depth exploration.

Figure~\ref{fig:BandsFS}(a) shows for the prototypical La$_3$Ni$_2$O$_7$ system
that the Ni~$3d_{z^2}$ bands are split into an occupied lower and an empty upper set,
separated by a band gap of $\sim 0.25$~eV at 0~GPa.
Simultaneously, the system is metallic due to highly dispersed Ni~$3d_{x^2-y^2}$ states.
Both Ni~$e_g$ orbitals hybridize substantially with each other around the $M$~point~\cite{Luo-LNO327:23}
and, in addition, present an overall hybridization with the O~$2p$ states.

Application of external pressure enhances the energy difference between the two Ni~$3d_{z^2}$ sets,
but simultaneously also the band width of the $3d_{x^2-y^2}$ and $3d_{z^2}$ states,
which results in a reduction of the band gap.
Moreover, the lower $3d_{z^2}$ set crosses the Fermi energy,
and a fraction of its charge is transferred to the $3d_{x^2-y^2}$ bands.
Thus, a 'self-doping' hole pocket around the $\Gamma$ point appears
[Fig.~\ref{fig:BandsFS}(b); folded back from the $M$~point of the primitive cell] \cite{Sun-327-Nickelate-SC:23, Luo-LNO327:23, Gu-LNO327:23, Yang-LNO327:23, Lechermann-LNO327:23, Liu-LNO327-OxVacDestructive:23, ZhangDagotto-LNO327:23},
which plays a key role in the suggested $s^{\pm}$ superconductivity mechanism~\cite{Gu-LNO327:23, Yang-LNO327:23, Liu-LNO327-OxVacDestructive:23, Lu-LNO327-InterlayerAFM:23, ZhangDagotto-LNO327:23}.
We furthermore see an increase in Ni~$3d_{z^2}$ orbital weight specifically in the $\alpha$ Fermi sheet.
Notably, the Fermi surface of pressurized $I4/mmm$ La$_3$Ni$_2$O$_7$ strongly resembles previous reports based on the $Fmmm$ geometry.
Thus, moderate orthorhombic distortions rather impact details of the electronic structure.

The reduction of the octahedral anisotropy~$R$ with pressure discussed above [Fig.~\ref{fig:StructTrends}(b)]
lowers the energy of the Ni~$3d_{x^2-y^2}$ orbital relative to the Ni~$3d_{z^2}$ states [Fig.~\ref{fig:BandsFS}(a)].
This results in enhanced (reduced) DFT$+U$ occupation numbers of the Ni~$3d_{x^2-y^2}$ ($3d_{z^2}$) orbitals.  %
These findings are consistent with recent reports based on a model study~\cite{ZhangDagotto-LNO327:23}
and demonstrate explicitly that the orbital polarization can be tuned by external pressure.
While the formal Ni$^{2.5+}$ ($d^{7.5}$) configuration implies that $1.5$ electrons  per Ni ion occupy the $e_g$ manifold,
the consistently higher values observed here highlight the involvement of the oxygen system,
reminiscent of the $d^{8}\underline{L}$ configuration characteristic of rare-earth nickelates~\cite{RENickelateReview:16}.

Finally, we find a Fermi surface with a quasi identical topology  as in pressurized La$_3$Ni$_2$O$_7$ %
in Tb$_3$Ni$_2$O$_7$ at zero pressure [Fig.~\ref{fig:BandsFS}(b)],
intriguingly \textit{despite} the presence of pronounced octahedral rotations (space group $Cmc2_1$).
We highlighted this compound  already above
due to its considerably enhanced DOS at the Fermi energy and a concomitant strong involvement of the apical oxygen ions [Fig.~\ref{fig:ElStr}(a)].
Therefore, we suggest this bilayer nickelate as a candidate for superconductivity at ambient pressure.
Investigations of possible superconducting pairing in this system would be highly interesting. %

\section{Summary}

The structural and electronic properties of
$A_3$Ni$_2$O$_7$ bilayer nickelates ($A=$~La-Lu, Y, Sc)
and their pressure dependence (0-150~GPa) were investigated
by performing first-principles simulations including a Coulomb repulsion term,
with particular emphasis on the role of orthorhombic distortion ($b/a$ ratio), octahedral anisotropy, and octahedral rotations.

The lattice parameters were found to exhibit a highly anisotropic response to external pressure.
Surprisingly, in La$_3$Ni$_2$O$_7$ at $\sim 20$~GPa,
we observed an orthorhombic-to-tetragonal transition to an $I4/mmm$ phase at variance with recent x-ray diffraction data,
which points to yet unresolved complexities near the onset of superconductivity, e.g., electron- or hole-doped samples owing to variations in the oxygen stoichiometry.
Due to the sensitivity of particularly the Ni~$3d_{z^2}$-derived states and their energy relative to the Fermi level,
this calls for a careful reassessment of the superconductivity mechanisms  so-far proposed for the undoped  compound. %
We showed that this transition coincides universally with vanishing octahedral tilts.
The associated critical pressure increases quadratically over the rare-earth series,
which we traced back to the enhancement of octahedral rotations with reducing $A$-site ionic radius.
This establishes chemical and external pressure as two distinct control parameters in these complex oxides
that independently and non-interchangeably allow the designing of their quantum state.

We compiled an \textit{ab initio} structural phase diagram,
which unveils two novel phases for $A=$~Nd-Lu, Y, Sc at ambient conditions
that are characterized by the emergence of significant in-plane $c^+$ octahedral rotations as well as in-plane bond disproportionations
and exhibit a surprising pressure-driven electronic reconstruction involving the rotation of the Ni~$3d_{z^2}$ orbital.
The consecutive quenching of the distinct rotational degrees of freedom by pressure %
demonstrates that the potential energy landscape of the in-plane octahedral rotations is significantly shallower than for the octahedral tilts.

Moreover, we found unexpected correlations between $T_c$ and  the \textit{in-plane} Ni-O-Ni bond angles for La$_3$Ni$_2$O$_7$,
which determine the basal Ni~$3d_{x^2-y^2}$-O~$2p_{x,y}$ hybridization.
An overall maximum  near the onset of superconductivity indicates enhanced superexchange interactions
and promotes the in-plane bond angles as an important aspect in understanding the superconductivity mechanism in bilayer rare-earth nickelates.

Finally, by disentangling the contributions of basal versus apical oxygen states at the Fermi level,
we identified Tb$_3$Ni$_2$O$_7$ as an interesting candidate for superconductivity at ambient pressure,
with a considerably higher density of states at the Fermi energy and a Fermi surface similar to pressurized La$_3$Ni$_2$O$_7$.
This is even more astonishing since our results show that the perspectives of conventional chemical precompression are limited in this system.
We suggest further exploration of the superconducting properties of Tb$_3$Ni$_2$O$_7$ at zero pressure.

This comprehensive study uncovers a profound tunability of the structural and electronic phases in this novel materials class.
The richness of the structural phase diagram and the unexpected transitions identified specifically for the later rare-earth nickelates
provide additional opportunities for the discovery of e.g.\ superconducting phases.
It also emphasizes that future work needs to carefully assess the impact of defects, especially oxygen excess (hole doping) or oxygen vacancies (electron doping).

\begin{acknowledgments}
PJH is grateful for discussions with I.~Eremin, A.~Kreisel, and F.~Lechermann.
This work was supported by the National Science Foundation, Grant No.~NSF-DMR-2118718. 
\end{acknowledgments}

\section*{COMPETING INTERESTS}

The authors declare no competing interests.

\section*{Contributions}

BG and PJH conceived of the project. JJH, GRS, RGH, and PJH supervised the research. BG performed the theoretical simulations and corresponding analysis. BG and PJH wrote the manuscript. All authors discussed the results and revised the manuscript.

\section*{Data availability}

The data is available upon reasonable request to the authors.

\appendix

\section{\boldmath Magnetic interactions in La$_3$Ni$_2$O$_7$ under pressure: Ferromagnetic in-plane coupling and proximity of an insulating site-disproportionate state}

\begin{figure}
\begin{center}
\includegraphics[width=\linewidth]{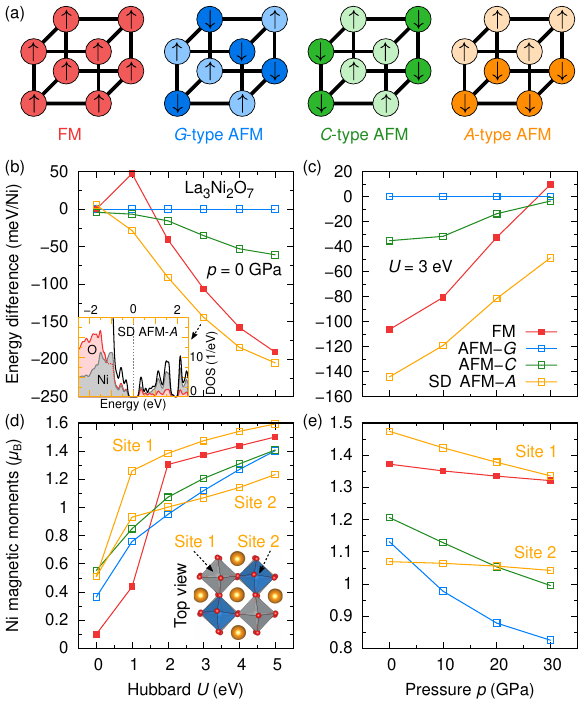}
\caption{\label{fig:LNO-Mag}(a)~Illustration of the four considered magnetic states at the bilayer Ni sites in La$_3$Ni$_2$O$_7$.
(b,c)~Energies of the magnetic phases as a function of the on-site Coulomb repulsion parameter~$U$ and the external pressure~$p$, using $G$-type AFM as reference.
(d,e)~Corresponding Ni magnetic moments, together with a visualization of the site disproportionation (SD) and the concomitant breathing-mode structural distortions
identified in the $A$-type AFM state.
The DOS inset in panel~(b) uncovers this phase to be insulating (black: total DOS; red: oxygen; gray: Ni).
}
\end{center}
\end{figure}

The experimental evidence on the magnetic properties of La$_3$Ni$_2$O$_7$ at ambient pressure is so far ambivalent. %
While recent measurements suggested the presence of a spin density wave under appropriate sample conditions~\cite{Liu-LNO327-SDW:22, Chen-LNO327-SDW:23},
earlier neutron diffraction and nuclear magnetic resonance studies found no long-range magnetic order~\cite{LNO-327-Diffraction-Ling:00, Fukamachi-LNO327-NMR:01}.

To shed some light into this issue,
we explore the magnetic properties of La$_3$Ni$_2$O$_7$ as a function of the on-site Coulomb repulsion parameter~$U$ and the external pressure~$p$ (Fig.~\ref{fig:LNO-Mag}).
At $U=0$~eV, the energy differences between the considered magnetic orders are negligible ($< \pm 5$~meV/Ni),
reflecting the possibility of substantial magnetic fluctuations [Fig.~\ref{fig:LNO-Mag}(b)].
With increasing $U$, ferromagnetic (FM) Ni-Ni interactions stabilize in the NiO$_2$ planes (orange and red curve).
This observation is in sharp contrast to the cupratelike in-plane antiferromagnetic (AFM) interactions (blue and green curve) that have been reported for the infinite-layer nickelates~\cite{Lu-MagExNdNiO2:21, SahinovicGeislerPentcheva:23}.
Simultaneously, the Ni magnetic moments are strongly enhanced with $U$ [Fig.~\ref{fig:LNO-Mag}(d)].
Moreover, we observe that the magnetic interactions in the bilayer nickelate are highly susceptible to external pressure,
which counteracts these trends [Fig.~\ref{fig:LNO-Mag}(c,e)].

Interestingly, we identify an $A$-type AFM ground state  for $U \geq 1$~eV, %
characterized by FM in-plane interactions and AFM-coupled NiO$_2$ layers [Fig.~\ref{fig:LNO-Mag}(a,b)].
Moreover, we find this state to be superimposed by a checkerboard site disproportionation, %
expressed in the emergence of two distinct Ni magnetic moments [Fig.~\ref{fig:LNO-Mag}(d,e); denoted as Site~1 and Site~2].
This phenomenon is accompanied by breathing-mode structural distortions, i.e., larger and smaller NiO$_6$ octahedra,
as reported for (LaNiO$_3$)$_1$/(LaAlO$_3$)$_1$(001) superlattices~\cite{ABR:11, GeislerPentcheva-LNOLAO:18, Geisler-VO-LNOLAO:22}.
A band gap opens around the Fermi energy, rendering an insulating state [Fig.~\ref{fig:LNO-Mag}(b), inset].

The energy differences $E(\text{AFM-$G$}) - E(\text{AFM-$C$})$ and $E(\text{FM}) - E(\text{AFM-$A$})$ are of the order of $35$-$40$~meV/Ni at $p=0$~GPa and $U=3$~eV,
which shows that the NiO$_2$ planes in each bilayer are strongly coupled by superexchange [Fig.~\ref{fig:LNO-Mag}(b,c)].
Simultaneously, $E(\text{AFM-$C$}) - E(\text{FM}) \sim 70$~meV/Ni and $E(\text{AFM-$G$}) - E(\text{AFM-$A$}) \sim 145$~meV/Ni
reflect the in-plane magnetic coupling,
which we find to unambiguously exceed the inter-layer coupling.
Hence, the magnetic interactions are highly anisotropic.
At the same time, these values show that the in-plane magnetic coupling is strongly affected by the inter-layer coupling.
This interplay is also mirrored in the Ni magnetic moments, which are highly distinct in all four considered magnetic phases [Fig.~\ref{fig:LNO-Mag}(d,e)].

\section{\boldmath Role of the Tb~$4f$ states in Tb$_3$Ni$_2$O$_7$}

\begin{figure}
\begin{center}
\includegraphics[width=\linewidth]{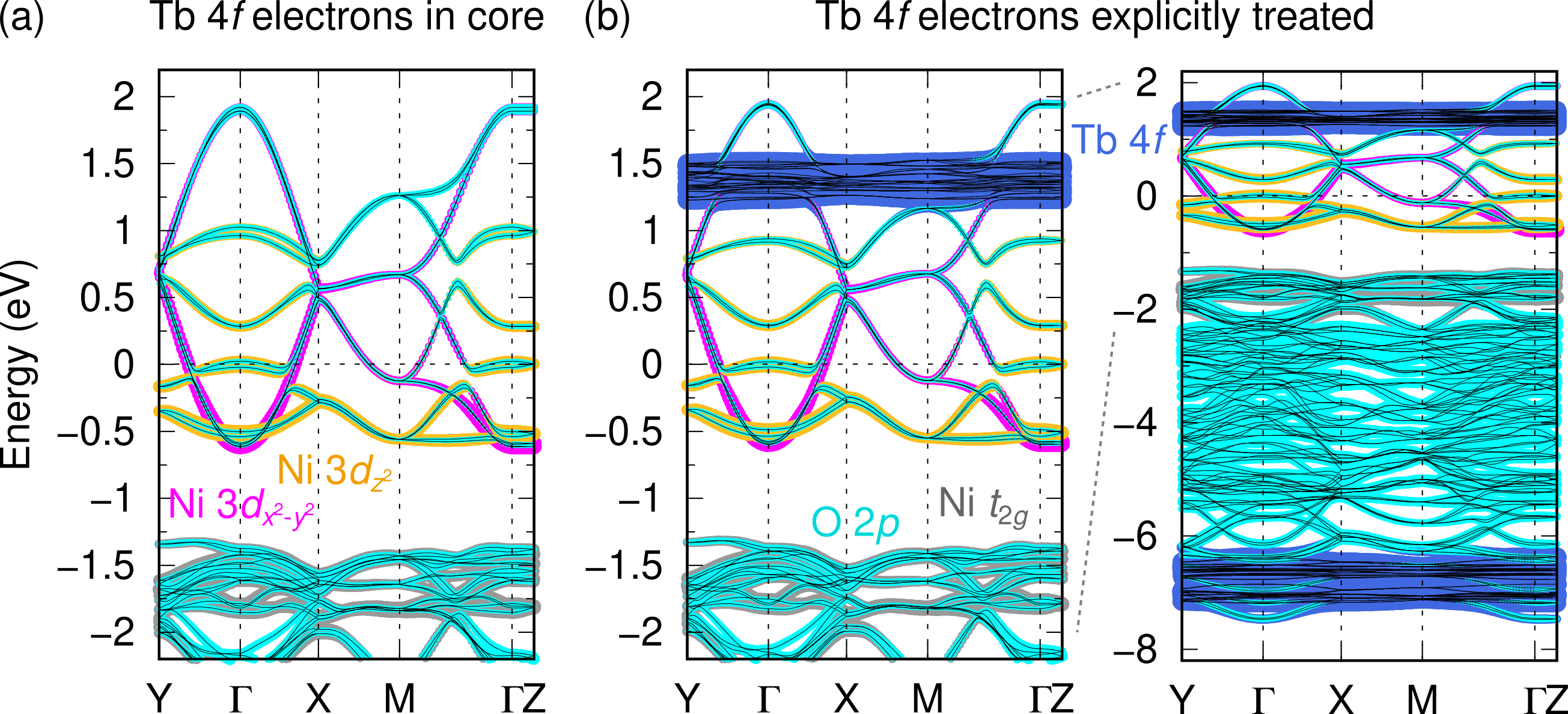}
\caption{\label{fig:TbBands}(a)~Band structure of Tb$_3$Ni$_2$O$_7$ at ambient pressure, keeping the Tb~$4f$ electrons frozen in the core [see Fig.~\ref{fig:BandsFS}(a)].
(b)~The explicit treatment of the Tb~$4f$ electrons unveils an identical electronic structure near the Fermi energy.
}
\end{center}
\end{figure}

Moreover, we investigated the role of the Tb~$4f$ states in Tb$_3$Ni$_2$O$_7$ (Fig.~\ref{fig:TbBands}) as a representative example.
We apply $U = 8$~eV at the Tb sites, similar to previous work on infinite-layer nickelates~\cite{Choi-Lee-Pickett-4fNNO:20, SahinovicGeislerPentcheva:23}.
The Tb~$4f$ states can be observed around $\sim -6.7$ and $1.3$~eV and do not contribute to the Fermi surface.
They are highly localized and thus exhibit almost no dispersion.
Comparison with the frozen-core result unveils an identical electronic structure near the Fermi energy.


\begin{thebibliography}{97}%
\makeatletter
\providecommand \@ifxundefined [1]{%
 \@ifx{#1\undefined}
}%
\providecommand \@ifnum [1]{%
 \ifnum #1\expandafter \@firstoftwo
 \else \expandafter \@secondoftwo
 \fi
}%
\providecommand \@ifx [1]{%
 \ifx #1\expandafter \@firstoftwo
 \else \expandafter \@secondoftwo
 \fi
}%
\providecommand \natexlab [1]{#1}%
\providecommand \enquote  [1]{``#1''}%
\providecommand \bibnamefont  [1]{#1}%
\providecommand \bibfnamefont [1]{#1}%
\providecommand \citenamefont [1]{#1}%
\providecommand \href@noop [0]{\@secondoftwo}%
\providecommand \href [0]{\begingroup \@sanitize@url \@href}%
\providecommand \@href[1]{\@@startlink{#1}\@@href}%
\providecommand \@@href[1]{\endgroup#1\@@endlink}%
\providecommand \@sanitize@url [0]{\catcode `\\12\catcode `\$12\catcode
  `\&12\catcode `\#12\catcode `\^12\catcode `\_12\catcode `\%12\relax}%
\providecommand \@@startlink[1]{}%
\providecommand \@@endlink[0]{}%
\providecommand \url  [0]{\begingroup\@sanitize@url \@url }%
\providecommand \@url [1]{\endgroup\@href {#1}{\urlprefix }}%
\providecommand \urlprefix  [0]{URL }%
\providecommand \Eprint [0]{\href }%
\providecommand \doibase [0]{https://doi.org/}%
\providecommand \selectlanguage [0]{\@gobble}%
\providecommand \bibinfo  [0]{\@secondoftwo}%
\providecommand \bibfield  [0]{\@secondoftwo}%
\providecommand \translation [1]{[#1]}%
\providecommand \BibitemOpen [0]{}%
\providecommand \bibitemStop [0]{}%
\providecommand \bibitemNoStop [0]{.\EOS\space}%
\providecommand \EOS [0]{\spacefactor3000\relax}%
\providecommand \BibitemShut  [1]{\csname bibitem#1\endcsname}%
\let\auto@bib@innerbib\@empty
\bibitem [{\citenamefont {Sun}\ \emph {et~al.}(2023)\citenamefont {Sun},
  \citenamefont {Huo}, \citenamefont {Hu}, \citenamefont {Li}, \citenamefont
  {Liu}, \citenamefont {Han}, \citenamefont {Tang}, \citenamefont {Mao},
  \citenamefont {Yang}, \citenamefont {Wang}, \citenamefont {Cheng},
  \citenamefont {Yao}, \citenamefont {Zhang},\ and\ \citenamefont
  {Wang}}]{Sun-327-Nickelate-SC:23}%
  \BibitemOpen
  \bibfield  {author} {\bibinfo {author} {\bibfnamefont {H.}~\bibnamefont
  {Sun}}, \bibinfo {author} {\bibfnamefont {M.}~\bibnamefont {Huo}}, \bibinfo
  {author} {\bibfnamefont {X.}~\bibnamefont {Hu}}, \bibinfo {author}
  {\bibfnamefont {J.}~\bibnamefont {Li}}, \bibinfo {author} {\bibfnamefont
  {Z.}~\bibnamefont {Liu}}, \bibinfo {author} {\bibfnamefont {Y.}~\bibnamefont
  {Han}}, \bibinfo {author} {\bibfnamefont {L.}~\bibnamefont {Tang}}, \bibinfo
  {author} {\bibfnamefont {Z.}~\bibnamefont {Mao}}, \bibinfo {author}
  {\bibfnamefont {P.}~\bibnamefont {Yang}}, \bibinfo {author} {\bibfnamefont
  {B.}~\bibnamefont {Wang}}, \bibinfo {author} {\bibfnamefont {J.}~\bibnamefont
  {Cheng}}, \bibinfo {author} {\bibfnamefont {D.-X.}\ \bibnamefont {Yao}},
  \bibinfo {author} {\bibfnamefont {G.-M.}\ \bibnamefont {Zhang}},\ and\
  \bibinfo {author} {\bibfnamefont {M.}~\bibnamefont {Wang}},\ }\bibfield
  {title} {\bibinfo {title} {Signatures of superconductivity near {80 K} in a
  nickelate under high pressure},\ }\href
  {https://doi.org/10.1038/s41586-023-06408-7} {\bibfield  {journal} {\bibinfo
  {journal} {Nature}\ }\textbf {\bibinfo {volume} {621}},\ \bibinfo {pages}
  {493} (\bibinfo {year} {2023})}\BibitemShut {NoStop}%
\bibitem [{\citenamefont {Li}\ \emph {et~al.}(2019)\citenamefont {Li},
  \citenamefont {Lee}, \citenamefont {Wang}, \citenamefont {Osada},
  \citenamefont {Crossley}, \citenamefont {Lee}, \citenamefont {Cui},
  \citenamefont {Hikita},\ and\ \citenamefont
  {Hwang}}]{Li-Supercond-Inf-NNO-STO:19}%
  \BibitemOpen
  \bibfield  {author} {\bibinfo {author} {\bibfnamefont {D.}~\bibnamefont
  {Li}}, \bibinfo {author} {\bibfnamefont {K.}~\bibnamefont {Lee}}, \bibinfo
  {author} {\bibfnamefont {B.~Y.}\ \bibnamefont {Wang}}, \bibinfo {author}
  {\bibfnamefont {M.}~\bibnamefont {Osada}}, \bibinfo {author} {\bibfnamefont
  {S.}~\bibnamefont {Crossley}}, \bibinfo {author} {\bibfnamefont {H.~R.}\
  \bibnamefont {Lee}}, \bibinfo {author} {\bibfnamefont {Y.}~\bibnamefont
  {Cui}}, \bibinfo {author} {\bibfnamefont {Y.}~\bibnamefont {Hikita}},\ and\
  \bibinfo {author} {\bibfnamefont {H.~Y.}\ \bibnamefont {Hwang}},\ }\bibfield
  {title} {\bibinfo {title} {Superconductivity in an infinite-layer
  nickelate},\ }\href {https://doi.org/10.1038/s41586-019-1496-5} {\bibfield
  {journal} {\bibinfo  {journal} {Nature}\ }\textbf {\bibinfo {volume} {572}},\
  \bibinfo {pages} {624} (\bibinfo {year} {2019})}\BibitemShut {NoStop}%
\bibitem [{\citenamefont {Li}\ \emph {et~al.}(2020)\citenamefont {Li},
  \citenamefont {Wang}, \citenamefont {Lee}, \citenamefont {Harvey},
  \citenamefont {Osada}, \citenamefont {Goodge}, \citenamefont {Kourkoutis},\
  and\ \citenamefont {Hwang}}]{Li-Supercond-Dome-Inf-NNO-STO:20}%
  \BibitemOpen
  \bibfield  {author} {\bibinfo {author} {\bibfnamefont {D.}~\bibnamefont
  {Li}}, \bibinfo {author} {\bibfnamefont {B.~Y.}\ \bibnamefont {Wang}},
  \bibinfo {author} {\bibfnamefont {K.}~\bibnamefont {Lee}}, \bibinfo {author}
  {\bibfnamefont {S.~P.}\ \bibnamefont {Harvey}}, \bibinfo {author}
  {\bibfnamefont {M.}~\bibnamefont {Osada}}, \bibinfo {author} {\bibfnamefont
  {B.~H.}\ \bibnamefont {Goodge}}, \bibinfo {author} {\bibfnamefont {L.~F.}\
  \bibnamefont {Kourkoutis}},\ and\ \bibinfo {author} {\bibfnamefont {H.~Y.}\
  \bibnamefont {Hwang}},\ }\bibfield  {title} {\bibinfo {title}
  {Superconducting dome in
  {${\mathrm{Nd}}_{1\ensuremath{-}x}{\mathrm{Sr}}_{x}{\mathrm{NiO}}_{2}$}
  infinite layer films},\ }\href
  {https://doi.org/10.1103/PhysRevLett.125.027001} {\bibfield  {journal}
  {\bibinfo  {journal} {Phys. Rev. Lett.}\ }\textbf {\bibinfo {volume} {125}},\
  \bibinfo {pages} {027001} (\bibinfo {year} {2020})}\BibitemShut {NoStop}%
\bibitem [{\citenamefont {Zeng}\ \emph {et~al.}(2020)\citenamefont {Zeng},
  \citenamefont {Tang}, \citenamefont {Yin}, \citenamefont {Li}, \citenamefont
  {Li}, \citenamefont {Huang}, \citenamefont {Hu}, \citenamefont {Liu},
  \citenamefont {Omar}, \citenamefont {Jani}, \citenamefont {Lim},
  \citenamefont {Han}, \citenamefont {Wan}, \citenamefont {Yang}, \citenamefont
  {Pennycook}, \citenamefont {Wee},\ and\ \citenamefont
  {Ariando}}]{Zeng-Inf-NNO:20}%
  \BibitemOpen
  \bibfield  {author} {\bibinfo {author} {\bibfnamefont {S.}~\bibnamefont
  {Zeng}}, \bibinfo {author} {\bibfnamefont {C.~S.}\ \bibnamefont {Tang}},
  \bibinfo {author} {\bibfnamefont {X.}~\bibnamefont {Yin}}, \bibinfo {author}
  {\bibfnamefont {C.}~\bibnamefont {Li}}, \bibinfo {author} {\bibfnamefont
  {M.}~\bibnamefont {Li}}, \bibinfo {author} {\bibfnamefont {Z.}~\bibnamefont
  {Huang}}, \bibinfo {author} {\bibfnamefont {J.}~\bibnamefont {Hu}}, \bibinfo
  {author} {\bibfnamefont {W.}~\bibnamefont {Liu}}, \bibinfo {author}
  {\bibfnamefont {G.~J.}\ \bibnamefont {Omar}}, \bibinfo {author}
  {\bibfnamefont {H.}~\bibnamefont {Jani}}, \bibinfo {author} {\bibfnamefont
  {Z.~S.}\ \bibnamefont {Lim}}, \bibinfo {author} {\bibfnamefont
  {K.}~\bibnamefont {Han}}, \bibinfo {author} {\bibfnamefont {D.}~\bibnamefont
  {Wan}}, \bibinfo {author} {\bibfnamefont {P.}~\bibnamefont {Yang}}, \bibinfo
  {author} {\bibfnamefont {S.~J.}\ \bibnamefont {Pennycook}}, \bibinfo {author}
  {\bibfnamefont {A.~T.~S.}\ \bibnamefont {Wee}},\ and\ \bibinfo {author}
  {\bibfnamefont {A.}~\bibnamefont {Ariando}},\ }\bibfield  {title} {\bibinfo
  {title} {Phase diagram and superconducting dome of infinite-layer
  {${\mathrm{Nd}}_{1\ensuremath{-}x}{\mathrm{Sr}}_{x}{\mathrm{NiO}}_{2}$} thin
  films},\ }\href {https://doi.org/10.1103/PhysRevLett.125.147003} {\bibfield
  {journal} {\bibinfo  {journal} {Phys. Rev. Lett.}\ }\textbf {\bibinfo
  {volume} {125}},\ \bibinfo {pages} {147003} (\bibinfo {year}
  {2020})}\BibitemShut {NoStop}%
\bibitem [{\citenamefont {Nomura}\ \emph {et~al.}(2019)\citenamefont {Nomura},
  \citenamefont {Hirayama}, \citenamefont {Tadano}, \citenamefont {Yoshimoto},
  \citenamefont {Nakamura},\ and\ \citenamefont {Arita}}]{Nomura-Inf-NNO:19}%
  \BibitemOpen
  \bibfield  {author} {\bibinfo {author} {\bibfnamefont {Y.}~\bibnamefont
  {Nomura}}, \bibinfo {author} {\bibfnamefont {M.}~\bibnamefont {Hirayama}},
  \bibinfo {author} {\bibfnamefont {T.}~\bibnamefont {Tadano}}, \bibinfo
  {author} {\bibfnamefont {Y.}~\bibnamefont {Yoshimoto}}, \bibinfo {author}
  {\bibfnamefont {K.}~\bibnamefont {Nakamura}},\ and\ \bibinfo {author}
  {\bibfnamefont {R.}~\bibnamefont {Arita}},\ }\bibfield  {title} {\bibinfo
  {title} {Formation of a two-dimensional single-component correlated electron
  system and band engineering in the nickelate superconductor
  {${\mathrm{NdNiO}}_{2}$}},\ }\href
  {https://doi.org/10.1103/PhysRevB.100.205138} {\bibfield  {journal} {\bibinfo
   {journal} {Phys. Rev. B}\ }\textbf {\bibinfo {volume} {100}},\ \bibinfo
  {pages} {205138} (\bibinfo {year} {2019})}\BibitemShut {NoStop}%
\bibitem [{\citenamefont {Jiang}\ \emph {et~al.}(2019)\citenamefont {Jiang},
  \citenamefont {Si}, \citenamefont {Liao},\ and\ \citenamefont
  {Zhong}}]{JiangZhong-InfNickelates:19}%
  \BibitemOpen
  \bibfield  {author} {\bibinfo {author} {\bibfnamefont {P.}~\bibnamefont
  {Jiang}}, \bibinfo {author} {\bibfnamefont {L.}~\bibnamefont {Si}}, \bibinfo
  {author} {\bibfnamefont {Z.}~\bibnamefont {Liao}},\ and\ \bibinfo {author}
  {\bibfnamefont {Z.}~\bibnamefont {Zhong}},\ }\bibfield  {title} {\bibinfo
  {title} {Electronic structure of rare-earth infinite-layer {$R$NiO$_{2}$}
  {($R=$~La, Nd)}},\ }\href {https://doi.org/10.1103/PhysRevB.100.201106}
  {\bibfield  {journal} {\bibinfo  {journal} {Phys. Rev. B}\ }\textbf {\bibinfo
  {volume} {100}},\ \bibinfo {pages} {201106} (\bibinfo {year}
  {2019})}\BibitemShut {NoStop}%
\bibitem [{\citenamefont {Sakakibara}\ \emph {et~al.}(2020)\citenamefont
  {Sakakibara}, \citenamefont {Usui}, \citenamefont {Suzuki}, \citenamefont
  {Kotani}, \citenamefont {Aoki},\ and\ \citenamefont
  {Kuroki}}]{Sakakibara:20}%
  \BibitemOpen
  \bibfield  {author} {\bibinfo {author} {\bibfnamefont {H.}~\bibnamefont
  {Sakakibara}}, \bibinfo {author} {\bibfnamefont {H.}~\bibnamefont {Usui}},
  \bibinfo {author} {\bibfnamefont {K.}~\bibnamefont {Suzuki}}, \bibinfo
  {author} {\bibfnamefont {T.}~\bibnamefont {Kotani}}, \bibinfo {author}
  {\bibfnamefont {H.}~\bibnamefont {Aoki}},\ and\ \bibinfo {author}
  {\bibfnamefont {K.}~\bibnamefont {Kuroki}},\ }\bibfield  {title} {\bibinfo
  {title} {Model construction and a possibility of cupratelike pairing in a new
  {$d^9$} nickelate superconductor {(Nd,Sr)NiO$_{2}$}},\ }\href
  {https://doi.org/10.1103/PhysRevLett.125.077003} {\bibfield  {journal}
  {\bibinfo  {journal} {Phys. Rev. Lett.}\ }\textbf {\bibinfo {volume} {125}},\
  \bibinfo {pages} {077003} (\bibinfo {year} {2020})}\BibitemShut {NoStop}%
\bibitem [{\citenamefont {Jiang}\ \emph {et~al.}(2020)\citenamefont {Jiang},
  \citenamefont {Berciu},\ and\ \citenamefont
  {Sawatzky}}]{JiangBerciuSawatzky:19}%
  \BibitemOpen
  \bibfield  {author} {\bibinfo {author} {\bibfnamefont {M.}~\bibnamefont
  {Jiang}}, \bibinfo {author} {\bibfnamefont {M.}~\bibnamefont {Berciu}},\ and\
  \bibinfo {author} {\bibfnamefont {G.~A.}\ \bibnamefont {Sawatzky}},\
  }\bibfield  {title} {\bibinfo {title} {Critical nature of the {Ni} spin state
  in doped {${\mathrm{NdNiO}}_{2}$}},\ }\href
  {https://doi.org/10.1103/PhysRevLett.124.207004} {\bibfield  {journal}
  {\bibinfo  {journal} {Phys. Rev. Lett.}\ }\textbf {\bibinfo {volume} {124}},\
  \bibinfo {pages} {207004} (\bibinfo {year} {2020})}\BibitemShut {NoStop}%
\bibitem [{\citenamefont {Botana}\ and\ \citenamefont
  {Norman}(2020)}]{Botana-Inf-Nickelates:19}%
  \BibitemOpen
  \bibfield  {author} {\bibinfo {author} {\bibfnamefont {A.~S.}\ \bibnamefont
  {Botana}}\ and\ \bibinfo {author} {\bibfnamefont {M.~R.}\ \bibnamefont
  {Norman}},\ }\bibfield  {title} {\bibinfo {title} {Similarities and
  differences between {${\mathrm{LaNiO}}_{2}$ and ${\mathrm{CaCuO}}_{2}$} and
  implications for superconductivity},\ }\href
  {https://doi.org/10.1103/PhysRevX.10.011024} {\bibfield  {journal} {\bibinfo
  {journal} {Phys. Rev. X}\ }\textbf {\bibinfo {volume} {10}},\ \bibinfo
  {pages} {011024} (\bibinfo {year} {2020})}\BibitemShut {NoStop}%
\bibitem [{\citenamefont {Geisler}\ and\ \citenamefont
  {Pentcheva}(2020{\natexlab{a}})}]{GeislerPentcheva-InfNNO:20}%
  \BibitemOpen
  \bibfield  {author} {\bibinfo {author} {\bibfnamefont {B.}~\bibnamefont
  {Geisler}}\ and\ \bibinfo {author} {\bibfnamefont {R.}~\bibnamefont
  {Pentcheva}},\ }\bibfield  {title} {\bibinfo {title} {Fundamental difference
  in the electronic reconstruction of infinite-layer versus perovskite
  neodymium nickelate films on {${\mathrm{SrTiO}}_{3}$(001)}},\ }\href
  {https://doi.org/10.1103/PhysRevB.102.020502} {\bibfield  {journal} {\bibinfo
   {journal} {Phys. Rev. B}\ }\textbf {\bibinfo {volume} {102}},\ \bibinfo
  {pages} {020502(R)} (\bibinfo {year} {2020}{\natexlab{a}})}\BibitemShut
  {NoStop}%
\bibitem [{\citenamefont {Lechermann}(2020)}]{Lechermann-Inf:20}%
  \BibitemOpen
  \bibfield  {author} {\bibinfo {author} {\bibfnamefont {F.}~\bibnamefont
  {Lechermann}},\ }\bibfield  {title} {\bibinfo {title} {Late transition metal
  oxides with infinite-layer structure: Nickelates versus cuprates},\ }\href
  {https://doi.org/10.1103/PhysRevB.101.081110} {\bibfield  {journal} {\bibinfo
   {journal} {Phys. Rev. B}\ }\textbf {\bibinfo {volume} {101}},\ \bibinfo
  {pages} {081110} (\bibinfo {year} {2020})}\BibitemShut {NoStop}%
\bibitem [{\citenamefont {Wu}\ \emph {et~al.}(2020)\citenamefont {Wu},
  \citenamefont {Di~Sante}, \citenamefont {Schwemmer}, \citenamefont {Hanke},
  \citenamefont {Hwang}, \citenamefont {Raghu},\ and\ \citenamefont
  {Thomale}}]{NNO-SC-Thomale:20}%
  \BibitemOpen
  \bibfield  {author} {\bibinfo {author} {\bibfnamefont {X.}~\bibnamefont
  {Wu}}, \bibinfo {author} {\bibfnamefont {D.}~\bibnamefont {Di~Sante}},
  \bibinfo {author} {\bibfnamefont {T.}~\bibnamefont {Schwemmer}}, \bibinfo
  {author} {\bibfnamefont {W.}~\bibnamefont {Hanke}}, \bibinfo {author}
  {\bibfnamefont {H.~Y.}\ \bibnamefont {Hwang}}, \bibinfo {author}
  {\bibfnamefont {S.}~\bibnamefont {Raghu}},\ and\ \bibinfo {author}
  {\bibfnamefont {R.}~\bibnamefont {Thomale}},\ }\bibfield  {title} {\bibinfo
  {title} {Robust {$d_{x^2-y^2}$}-wave superconductivity of infinite-layer
  nickelates},\ }\href {https://doi.org/10.1103/PhysRevB.101.060504} {\bibfield
   {journal} {\bibinfo  {journal} {Phys. Rev. B}\ }\textbf {\bibinfo {volume}
  {101}},\ \bibinfo {pages} {060504} (\bibinfo {year} {2020})}\BibitemShut
  {NoStop}%
\bibitem [{\citenamefont {Gu}\ \emph {et~al.}(2020)\citenamefont {Gu},
  \citenamefont {Li}, \citenamefont {Wan}, \citenamefont {Li}, \citenamefont
  {Guo}, \citenamefont {Yang}, \citenamefont {Li}, \citenamefont {Zhu},
  \citenamefont {Pan}, \citenamefont {Nie},\ and\ \citenamefont
  {Wen}}]{Gu-NNO2:20}%
  \BibitemOpen
  \bibfield  {author} {\bibinfo {author} {\bibfnamefont {Q.}~\bibnamefont
  {Gu}}, \bibinfo {author} {\bibfnamefont {Y.}~\bibnamefont {Li}}, \bibinfo
  {author} {\bibfnamefont {S.}~\bibnamefont {Wan}}, \bibinfo {author}
  {\bibfnamefont {H.}~\bibnamefont {Li}}, \bibinfo {author} {\bibfnamefont
  {W.}~\bibnamefont {Guo}}, \bibinfo {author} {\bibfnamefont {H.}~\bibnamefont
  {Yang}}, \bibinfo {author} {\bibfnamefont {Q.}~\bibnamefont {Li}}, \bibinfo
  {author} {\bibfnamefont {X.}~\bibnamefont {Zhu}}, \bibinfo {author}
  {\bibfnamefont {X.}~\bibnamefont {Pan}}, \bibinfo {author} {\bibfnamefont
  {Y.}~\bibnamefont {Nie}},\ and\ \bibinfo {author} {\bibfnamefont {H.-H.}\
  \bibnamefont {Wen}},\ }\bibfield  {title} {\bibinfo {title} {Single particle
  tunneling spectrum of superconducting {Nd$_{1-x}$Sr$_x$NiO$_2$} thin films},\
  }\href {https://doi.org/10.1038/s41467-020-19908-1} {\bibfield  {journal}
  {\bibinfo  {journal} {Nat. Commun.}\ }\textbf {\bibinfo {volume} {11}},\
  \bibinfo {pages} {6027} (\bibinfo {year} {2020})}\BibitemShut {NoStop}%
\bibitem [{\citenamefont {Lu}\ \emph {et~al.}(2021)\citenamefont {Lu},
  \citenamefont {Rossi}, \citenamefont {Nag}, \citenamefont {Osada},
  \citenamefont {Li}, \citenamefont {Lee}, \citenamefont {Wang}, \citenamefont
  {Garcia-Fernandez}, \citenamefont {Agrestini}, \citenamefont {Shen},
  \citenamefont {Been}, \citenamefont {Moritz}, \citenamefont {Devereaux},
  \citenamefont {Zaanen}, \citenamefont {Hwang}, \citenamefont {Zhou},\ and\
  \citenamefont {Lee}}]{Lu-MagExNdNiO2:21}%
  \BibitemOpen
  \bibfield  {author} {\bibinfo {author} {\bibfnamefont {H.}~\bibnamefont
  {Lu}}, \bibinfo {author} {\bibfnamefont {M.}~\bibnamefont {Rossi}}, \bibinfo
  {author} {\bibfnamefont {A.}~\bibnamefont {Nag}}, \bibinfo {author}
  {\bibfnamefont {M.}~\bibnamefont {Osada}}, \bibinfo {author} {\bibfnamefont
  {D.~F.}\ \bibnamefont {Li}}, \bibinfo {author} {\bibfnamefont
  {K.}~\bibnamefont {Lee}}, \bibinfo {author} {\bibfnamefont {B.~Y.}\
  \bibnamefont {Wang}}, \bibinfo {author} {\bibfnamefont {M.}~\bibnamefont
  {Garcia-Fernandez}}, \bibinfo {author} {\bibfnamefont {S.}~\bibnamefont
  {Agrestini}}, \bibinfo {author} {\bibfnamefont {Z.~X.}\ \bibnamefont {Shen}},
  \bibinfo {author} {\bibfnamefont {E.~M.}\ \bibnamefont {Been}}, \bibinfo
  {author} {\bibfnamefont {B.}~\bibnamefont {Moritz}}, \bibinfo {author}
  {\bibfnamefont {T.~P.}\ \bibnamefont {Devereaux}}, \bibinfo {author}
  {\bibfnamefont {J.}~\bibnamefont {Zaanen}}, \bibinfo {author} {\bibfnamefont
  {H.~Y.}\ \bibnamefont {Hwang}}, \bibinfo {author} {\bibfnamefont {K.-J.}\
  \bibnamefont {Zhou}},\ and\ \bibinfo {author} {\bibfnamefont {W.~S.}\
  \bibnamefont {Lee}},\ }\bibfield  {title} {\bibinfo {title} {Magnetic
  excitations in infinite-layer nickelates},\ }\href
  {https://doi.org/10.1126/science.abd7726} {\bibfield  {journal} {\bibinfo
  {journal} {Science}\ }\textbf {\bibinfo {volume} {373}},\ \bibinfo {pages}
  {213} (\bibinfo {year} {2021})}\BibitemShut {NoStop}%
\bibitem [{\citenamefont {Ortiz}\ \emph {et~al.}(2021)\citenamefont {Ortiz},
  \citenamefont {Menke}, \citenamefont {Misj\'ak}, \citenamefont {Mantadakis},
  \citenamefont {F\"ursich}, \citenamefont {Schierle}, \citenamefont
  {Logvenov}, \citenamefont {Kaiser}, \citenamefont {Keimer}, \citenamefont
  {Hansmann},\ and\ \citenamefont {Benckiser}}]{Ortiz-NNO:21}%
  \BibitemOpen
  \bibfield  {author} {\bibinfo {author} {\bibfnamefont {R.~A.}\ \bibnamefont
  {Ortiz}}, \bibinfo {author} {\bibfnamefont {H.}~\bibnamefont {Menke}},
  \bibinfo {author} {\bibfnamefont {F.}~\bibnamefont {Misj\'ak}}, \bibinfo
  {author} {\bibfnamefont {D.~T.}\ \bibnamefont {Mantadakis}}, \bibinfo
  {author} {\bibfnamefont {K.}~\bibnamefont {F\"ursich}}, \bibinfo {author}
  {\bibfnamefont {E.}~\bibnamefont {Schierle}}, \bibinfo {author}
  {\bibfnamefont {G.}~\bibnamefont {Logvenov}}, \bibinfo {author}
  {\bibfnamefont {U.}~\bibnamefont {Kaiser}}, \bibinfo {author} {\bibfnamefont
  {B.}~\bibnamefont {Keimer}}, \bibinfo {author} {\bibfnamefont
  {P.}~\bibnamefont {Hansmann}},\ and\ \bibinfo {author} {\bibfnamefont
  {E.}~\bibnamefont {Benckiser}},\ }\bibfield  {title} {\bibinfo {title}
  {Superlattice approach to doping infinite-layer nickelates},\ }\href
  {https://doi.org/10.1103/PhysRevB.104.165137} {\bibfield  {journal} {\bibinfo
   {journal} {Phys. Rev. B}\ }\textbf {\bibinfo {volume} {104}},\ \bibinfo
  {pages} {165137} (\bibinfo {year} {2021})}\BibitemShut {NoStop}%
\bibitem [{\citenamefont {Lechermann}(2021)}]{Lechermann-Inf:21}%
  \BibitemOpen
  \bibfield  {author} {\bibinfo {author} {\bibfnamefont {F.}~\bibnamefont
  {Lechermann}},\ }\bibfield  {title} {\bibinfo {title} {Doping-dependent
  character and possible magnetic ordering of {${\mathrm{NdNiO}}_{2}$}},\
  }\href {https://doi.org/10.1103/PhysRevMaterials.5.044803} {\bibfield
  {journal} {\bibinfo  {journal} {Phys. Rev. Mater.}\ }\textbf {\bibinfo
  {volume} {5}},\ \bibinfo {pages} {044803} (\bibinfo {year}
  {2021})}\BibitemShut {NoStop}%
\bibitem [{\citenamefont {Sahinovic}\ and\ \citenamefont
  {Geisler}(2021)}]{SahinovicGeisler:21}%
  \BibitemOpen
  \bibfield  {author} {\bibinfo {author} {\bibfnamefont {A.}~\bibnamefont
  {Sahinovic}}\ and\ \bibinfo {author} {\bibfnamefont {B.}~\bibnamefont
  {Geisler}},\ }\bibfield  {title} {\bibinfo {title} {Active learning and
  element-embedding approach in neural networks for infinite-layer versus
  perovskite oxides},\ }\href
  {https://doi.org/10.1103/PhysRevResearch.3.L042022} {\bibfield  {journal}
  {\bibinfo  {journal} {Phys. Rev. Research}\ }\textbf {\bibinfo {volume}
  {3}},\ \bibinfo {pages} {L042022} (\bibinfo {year} {2021})}\BibitemShut
  {NoStop}%
\bibitem [{\citenamefont {Wang}\ \emph {et~al.}(2021)\citenamefont {Wang},
  \citenamefont {Li}, \citenamefont {Goodge}, \citenamefont {Lee},
  \citenamefont {Osada}, \citenamefont {Harvey}, \citenamefont {Kourkoutis},
  \citenamefont {Beasley},\ and\ \citenamefont {Hwang}}]{Wang-IL-Pauli:21}%
  \BibitemOpen
  \bibfield  {author} {\bibinfo {author} {\bibfnamefont {B.~Y.}\ \bibnamefont
  {Wang}}, \bibinfo {author} {\bibfnamefont {D.}~\bibnamefont {Li}}, \bibinfo
  {author} {\bibfnamefont {B.~H.}\ \bibnamefont {Goodge}}, \bibinfo {author}
  {\bibfnamefont {K.}~\bibnamefont {Lee}}, \bibinfo {author} {\bibfnamefont
  {M.}~\bibnamefont {Osada}}, \bibinfo {author} {\bibfnamefont {S.~P.}\
  \bibnamefont {Harvey}}, \bibinfo {author} {\bibfnamefont {L.~F.}\
  \bibnamefont {Kourkoutis}}, \bibinfo {author} {\bibfnamefont {M.~R.}\
  \bibnamefont {Beasley}},\ and\ \bibinfo {author} {\bibfnamefont {H.~Y.}\
  \bibnamefont {Hwang}},\ }\bibfield  {title} {\bibinfo {title} {Isotropic
  {Pauli-limited} superconductivity in the infinite-layer nickelate
  {Nd$_{0.775}$Sr$_{0.225}$NiO$_2$}},\ }\href
  {https://doi.org/10.1038/s41567-020-01128-5} {\bibfield  {journal} {\bibinfo
  {journal} {Nat. Phys.}\ }\textbf {\bibinfo {volume} {17}},\ \bibinfo {pages}
  {473} (\bibinfo {year} {2021})}\BibitemShut {NoStop}%
\bibitem [{\citenamefont {Geisler}\ and\ \citenamefont
  {Pentcheva}(2021)}]{GeislerPentcheva-NNOCCOSTO:21}%
  \BibitemOpen
  \bibfield  {author} {\bibinfo {author} {\bibfnamefont {B.}~\bibnamefont
  {Geisler}}\ and\ \bibinfo {author} {\bibfnamefont {R.}~\bibnamefont
  {Pentcheva}},\ }\bibfield  {title} {\bibinfo {title} {Correlated interface
  electron gas in infinite-layer nickelate versus cuprate films on
  {${\mathrm{SrTiO}}_{3}(001)$}},\ }\href
  {https://doi.org/10.1103/PhysRevResearch.3.013261} {\bibfield  {journal}
  {\bibinfo  {journal} {Phys. Rev. Research}\ }\textbf {\bibinfo {volume}
  {3}},\ \bibinfo {pages} {013261} (\bibinfo {year} {2021})}\BibitemShut
  {NoStop}%
\bibitem [{\citenamefont {Zeng}\ \emph {et~al.}(2022)\citenamefont {Zeng},
  \citenamefont {Yin}, \citenamefont {Li}, \citenamefont {Chow}, \citenamefont
  {Tang}, \citenamefont {Han}, \citenamefont {Huang}, \citenamefont {Cao},
  \citenamefont {Wan}, \citenamefont {Zhang}, \citenamefont {Lim},
  \citenamefont {Diao}, \citenamefont {Yang}, \citenamefont {Wee},
  \citenamefont {Pennycook},\ and\ \citenamefont {Ariando}}]{Zeng-Inf-NNO:22}%
  \BibitemOpen
  \bibfield  {author} {\bibinfo {author} {\bibfnamefont {S.~W.}\ \bibnamefont
  {Zeng}}, \bibinfo {author} {\bibfnamefont {X.~M.}\ \bibnamefont {Yin}},
  \bibinfo {author} {\bibfnamefont {C.~J.}\ \bibnamefont {Li}}, \bibinfo
  {author} {\bibfnamefont {L.~E.}\ \bibnamefont {Chow}}, \bibinfo {author}
  {\bibfnamefont {C.~S.}\ \bibnamefont {Tang}}, \bibinfo {author}
  {\bibfnamefont {K.}~\bibnamefont {Han}}, \bibinfo {author} {\bibfnamefont
  {Z.}~\bibnamefont {Huang}}, \bibinfo {author} {\bibfnamefont
  {Y.}~\bibnamefont {Cao}}, \bibinfo {author} {\bibfnamefont {D.~Y.}\
  \bibnamefont {Wan}}, \bibinfo {author} {\bibfnamefont {Z.~T.}\ \bibnamefont
  {Zhang}}, \bibinfo {author} {\bibfnamefont {Z.~S.}\ \bibnamefont {Lim}},
  \bibinfo {author} {\bibfnamefont {C.~Z.}\ \bibnamefont {Diao}}, \bibinfo
  {author} {\bibfnamefont {P.}~\bibnamefont {Yang}}, \bibinfo {author}
  {\bibfnamefont {A.~T.~S.}\ \bibnamefont {Wee}}, \bibinfo {author}
  {\bibfnamefont {S.~J.}\ \bibnamefont {Pennycook}},\ and\ \bibinfo {author}
  {\bibfnamefont {A.}~\bibnamefont {Ariando}},\ }\bibfield  {title} {\bibinfo
  {title} {Observation of perfect diamagnetism and interfacial effect on the
  electronic structures in infinite layer {Nd$_{0.8}$Sr$_{0.2}$NiO$_2$}
  superconductors},\ }\href {https://doi.org/10.1038/s41467-022-28390-w}
  {\bibfield  {journal} {\bibinfo  {journal} {Nature Communications}\ }\textbf
  {\bibinfo {volume} {13}},\ \bibinfo {pages} {743} (\bibinfo {year}
  {2022})}\BibitemShut {NoStop}%
\bibitem [{\citenamefont {Goodge}\ \emph {et~al.}(2023)\citenamefont {Goodge},
  \citenamefont {Geisler}, \citenamefont {Lee}, \citenamefont {Osada},
  \citenamefont {Wang}, \citenamefont {Li}, \citenamefont {Hwang},
  \citenamefont {Pentcheva},\ and\ \citenamefont
  {Kourkoutis}}]{GoodgeGeisler-NNO-IF:22}%
  \BibitemOpen
  \bibfield  {author} {\bibinfo {author} {\bibfnamefont {B.~H.}\ \bibnamefont
  {Goodge}}, \bibinfo {author} {\bibfnamefont {B.}~\bibnamefont {Geisler}},
  \bibinfo {author} {\bibfnamefont {K.}~\bibnamefont {Lee}}, \bibinfo {author}
  {\bibfnamefont {M.}~\bibnamefont {Osada}}, \bibinfo {author} {\bibfnamefont
  {B.~Y.}\ \bibnamefont {Wang}}, \bibinfo {author} {\bibfnamefont
  {D.}~\bibnamefont {Li}}, \bibinfo {author} {\bibfnamefont {H.~Y.}\
  \bibnamefont {Hwang}}, \bibinfo {author} {\bibfnamefont {R.}~\bibnamefont
  {Pentcheva}},\ and\ \bibinfo {author} {\bibfnamefont {L.~F.}\ \bibnamefont
  {Kourkoutis}},\ }\bibfield  {title} {\bibinfo {title} {Resolving the polar
  interface of infinite-layer nickelate thin films},\ }\href
  {https://doi.org/10.1038/s41563-023-01510-7} {\bibfield  {journal} {\bibinfo
  {journal} {Nature Materials}\ }\textbf {\bibinfo {volume} {22}},\ \bibinfo
  {pages} {466} (\bibinfo {year} {2023})}\BibitemShut {NoStop}%
\bibitem [{\citenamefont {Kreisel}\ \emph {et~al.}(2022)\citenamefont
  {Kreisel}, \citenamefont {Andersen}, \citenamefont {R\o{}mer}, \citenamefont
  {Eremin},\ and\ \citenamefont {Lechermann}}]{KreiselLechermann-IL:22}%
  \BibitemOpen
  \bibfield  {author} {\bibinfo {author} {\bibfnamefont {A.}~\bibnamefont
  {Kreisel}}, \bibinfo {author} {\bibfnamefont {B.~M.}\ \bibnamefont
  {Andersen}}, \bibinfo {author} {\bibfnamefont {A.~T.}\ \bibnamefont
  {R\o{}mer}}, \bibinfo {author} {\bibfnamefont {I.~M.}\ \bibnamefont
  {Eremin}},\ and\ \bibinfo {author} {\bibfnamefont {F.}~\bibnamefont
  {Lechermann}},\ }\bibfield  {title} {\bibinfo {title} {Superconducting
  instabilities in strongly correlated infinite-layer nickelates},\ }\href
  {https://doi.org/10.1103/PhysRevLett.129.077002} {\bibfield  {journal}
  {\bibinfo  {journal} {Phys. Rev. Lett.}\ }\textbf {\bibinfo {volume} {129}},\
  \bibinfo {pages} {077002} (\bibinfo {year} {2022})}\BibitemShut {NoStop}%
\bibitem [{\citenamefont {Rossi}\ \emph {et~al.}(2022)\citenamefont {Rossi},
  \citenamefont {Osada}, \citenamefont {Choi}, \citenamefont {Agrestini},
  \citenamefont {Jost}, \citenamefont {Lee}, \citenamefont {Lu}, \citenamefont
  {Wang}, \citenamefont {Lee}, \citenamefont {Nag}, \citenamefont {Chuang},
  \citenamefont {Kuo}, \citenamefont {Lee}, \citenamefont {Moritz},
  \citenamefont {Devereaux}, \citenamefont {Shen}, \citenamefont {Lee},
  \citenamefont {Zhou}, \citenamefont {Hwang},\ and\ \citenamefont
  {Lee}}]{Rossi-IL-CO:22}%
  \BibitemOpen
  \bibfield  {author} {\bibinfo {author} {\bibfnamefont {M.}~\bibnamefont
  {Rossi}}, \bibinfo {author} {\bibfnamefont {M.}~\bibnamefont {Osada}},
  \bibinfo {author} {\bibfnamefont {J.}~\bibnamefont {Choi}}, \bibinfo {author}
  {\bibfnamefont {S.}~\bibnamefont {Agrestini}}, \bibinfo {author}
  {\bibfnamefont {D.}~\bibnamefont {Jost}}, \bibinfo {author} {\bibfnamefont
  {Y.}~\bibnamefont {Lee}}, \bibinfo {author} {\bibfnamefont {H.}~\bibnamefont
  {Lu}}, \bibinfo {author} {\bibfnamefont {B.~Y.}\ \bibnamefont {Wang}},
  \bibinfo {author} {\bibfnamefont {K.}~\bibnamefont {Lee}}, \bibinfo {author}
  {\bibfnamefont {A.}~\bibnamefont {Nag}}, \bibinfo {author} {\bibfnamefont
  {Y.-D.}\ \bibnamefont {Chuang}}, \bibinfo {author} {\bibfnamefont {C.-T.}\
  \bibnamefont {Kuo}}, \bibinfo {author} {\bibfnamefont {S.-J.}\ \bibnamefont
  {Lee}}, \bibinfo {author} {\bibfnamefont {B.}~\bibnamefont {Moritz}},
  \bibinfo {author} {\bibfnamefont {T.~P.}\ \bibnamefont {Devereaux}}, \bibinfo
  {author} {\bibfnamefont {Z.-X.}\ \bibnamefont {Shen}}, \bibinfo {author}
  {\bibfnamefont {J.-S.}\ \bibnamefont {Lee}}, \bibinfo {author} {\bibfnamefont
  {K.-J.}\ \bibnamefont {Zhou}}, \bibinfo {author} {\bibfnamefont {H.~Y.}\
  \bibnamefont {Hwang}},\ and\ \bibinfo {author} {\bibfnamefont {W.-S.}\
  \bibnamefont {Lee}},\ }\bibfield  {title} {\bibinfo {title} {A broken
  translational symmetry state in an infinite-layer nickelate},\ }\href
  {https://doi.org/10.1038/s41567-022-01660-6} {\bibfield  {journal} {\bibinfo
  {journal} {Nature Physics}\ }\textbf {\bibinfo {volume} {18}},\ \bibinfo
  {pages} {869} (\bibinfo {year} {2022})}\BibitemShut {NoStop}%
\bibitem [{\citenamefont {Fowlie}\ \emph {et~al.}(2022)\citenamefont {Fowlie},
  \citenamefont {Hadjimichael}, \citenamefont {Martins}, \citenamefont {Li},
  \citenamefont {Osada}, \citenamefont {Wang}, \citenamefont {Lee},
  \citenamefont {Lee}, \citenamefont {Salman}, \citenamefont {Prokscha},
  \citenamefont {Triscone}, \citenamefont {Hwang},\ and\ \citenamefont
  {Suter}}]{Fowlie-IL-IntrinsicMag:22}%
  \BibitemOpen
  \bibfield  {author} {\bibinfo {author} {\bibfnamefont {J.}~\bibnamefont
  {Fowlie}}, \bibinfo {author} {\bibfnamefont {M.}~\bibnamefont
  {Hadjimichael}}, \bibinfo {author} {\bibfnamefont {M.~M.}\ \bibnamefont
  {Martins}}, \bibinfo {author} {\bibfnamefont {D.}~\bibnamefont {Li}},
  \bibinfo {author} {\bibfnamefont {M.}~\bibnamefont {Osada}}, \bibinfo
  {author} {\bibfnamefont {B.~Y.}\ \bibnamefont {Wang}}, \bibinfo {author}
  {\bibfnamefont {K.}~\bibnamefont {Lee}}, \bibinfo {author} {\bibfnamefont
  {Y.}~\bibnamefont {Lee}}, \bibinfo {author} {\bibfnamefont {Z.}~\bibnamefont
  {Salman}}, \bibinfo {author} {\bibfnamefont {T.}~\bibnamefont {Prokscha}},
  \bibinfo {author} {\bibfnamefont {J.-M.}\ \bibnamefont {Triscone}}, \bibinfo
  {author} {\bibfnamefont {H.~Y.}\ \bibnamefont {Hwang}},\ and\ \bibinfo
  {author} {\bibfnamefont {A.}~\bibnamefont {Suter}},\ }\bibfield  {title}
  {\bibinfo {title} {Intrinsic magnetism in superconducting infinite-layer
  nickelates},\ }\href {https://doi.org/10.1038/s41567-022-01684-y} {\bibfield
  {journal} {\bibinfo  {journal} {Nature Physics}\ }\textbf {\bibinfo {volume}
  {18}},\ \bibinfo {pages} {1043} (\bibinfo {year} {2022})}\BibitemShut
  {NoStop}%
\bibitem [{\citenamefont {Sahinovic}\ and\ \citenamefont
  {Geisler}(2022)}]{SahinovicGeisler:22}%
  \BibitemOpen
  \bibfield  {author} {\bibinfo {author} {\bibfnamefont {A.}~\bibnamefont
  {Sahinovic}}\ and\ \bibinfo {author} {\bibfnamefont {B.}~\bibnamefont
  {Geisler}},\ }\bibfield  {title} {\bibinfo {title} {Quantifying transfer
  learning synergies in infinite-layer and perovskite nitrides, oxides, and
  fluorides},\ }\href {https://doi.org/10.1088/1361-648x/ac5995} {\bibfield
  {journal} {\bibinfo  {journal} {J. Phys.: Condens. Matter}\ }\textbf
  {\bibinfo {volume} {34}},\ \bibinfo {pages} {214003} (\bibinfo {year}
  {2022})}\BibitemShut {NoStop}%
\bibitem [{\citenamefont {Sahinovic}\ \emph {et~al.}(2023)\citenamefont
  {Sahinovic}, \citenamefont {Geisler},\ and\ \citenamefont
  {Pentcheva}}]{SahinovicGeislerPentcheva:23}%
  \BibitemOpen
  \bibfield  {author} {\bibinfo {author} {\bibfnamefont {A.}~\bibnamefont
  {Sahinovic}}, \bibinfo {author} {\bibfnamefont {B.}~\bibnamefont {Geisler}},\
  and\ \bibinfo {author} {\bibfnamefont {R.}~\bibnamefont {Pentcheva}},\
  }\href@noop {} {\bibinfo {title} {Nature of the magnetic coupling in
  infinite-layer nickelates versus cuprates}} (\bibinfo {year} {2023}),\
  \Eprint {https://arxiv.org/abs/2309.12840} {arXiv:2309.12840
  [cond-mat.supr-con]} \BibitemShut {NoStop}%
\bibitem [{\citenamefont {Geisler}(2023)}]{Geisler-Rashba-NNOSTOKTO:23}%
  \BibitemOpen
  \bibfield  {author} {\bibinfo {author} {\bibfnamefont {B.}~\bibnamefont
  {Geisler}},\ }\href@noop {} {\bibinfo {title} {Rashba spin-orbit coupling in
  infinite-layer nickelate films on {SrTiO$_3$(001)} and {KTaO$_3$(001)}}}
  (\bibinfo {year} {2023}),\ \Eprint {https://arxiv.org/abs/2303.00717}
  {arXiv:2303.00717 [cond-mat.supr-con]} \BibitemShut {NoStop}%
\bibitem [{\citenamefont {Osada}\ \emph {et~al.}(2020)\citenamefont {Osada},
  \citenamefont {Wang}, \citenamefont {Goodge}, \citenamefont {Lee},
  \citenamefont {Yoon}, \citenamefont {Sakuma}, \citenamefont {Li},
  \citenamefont {Miura}, \citenamefont {Kourkoutis},\ and\ \citenamefont
  {Hwang}}]{Osada-PrNiO2-SC:20}%
  \BibitemOpen
  \bibfield  {author} {\bibinfo {author} {\bibfnamefont {M.}~\bibnamefont
  {Osada}}, \bibinfo {author} {\bibfnamefont {B.~Y.}\ \bibnamefont {Wang}},
  \bibinfo {author} {\bibfnamefont {B.~H.}\ \bibnamefont {Goodge}}, \bibinfo
  {author} {\bibfnamefont {K.}~\bibnamefont {Lee}}, \bibinfo {author}
  {\bibfnamefont {H.}~\bibnamefont {Yoon}}, \bibinfo {author} {\bibfnamefont
  {K.}~\bibnamefont {Sakuma}}, \bibinfo {author} {\bibfnamefont
  {D.}~\bibnamefont {Li}}, \bibinfo {author} {\bibfnamefont {M.}~\bibnamefont
  {Miura}}, \bibinfo {author} {\bibfnamefont {L.~F.}\ \bibnamefont
  {Kourkoutis}},\ and\ \bibinfo {author} {\bibfnamefont {H.~Y.}\ \bibnamefont
  {Hwang}},\ }\bibfield  {title} {\bibinfo {title} {A superconducting
  praseodymium nickelate with infinite layer structure},\ }\href
  {https://doi.org/10.1021/acs.nanolett.0c01392} {\bibfield  {journal}
  {\bibinfo  {journal} {Nano Lett.}\ }\textbf {\bibinfo {volume} {20}},\
  \bibinfo {pages} {5735} (\bibinfo {year} {2020})}\BibitemShut {NoStop}%
\bibitem [{\citenamefont {Osada}\ \emph {et~al.}(2021)\citenamefont {Osada},
  \citenamefont {Wang}, \citenamefont {Goodge}, \citenamefont {Harvey},
  \citenamefont {Lee}, \citenamefont {Li}, \citenamefont {Kourkoutis},\ and\
  \citenamefont {Hwang}}]{Osada-LaNiO2-SC:21}%
  \BibitemOpen
  \bibfield  {author} {\bibinfo {author} {\bibfnamefont {M.}~\bibnamefont
  {Osada}}, \bibinfo {author} {\bibfnamefont {B.~Y.}\ \bibnamefont {Wang}},
  \bibinfo {author} {\bibfnamefont {B.~H.}\ \bibnamefont {Goodge}}, \bibinfo
  {author} {\bibfnamefont {S.~P.}\ \bibnamefont {Harvey}}, \bibinfo {author}
  {\bibfnamefont {K.}~\bibnamefont {Lee}}, \bibinfo {author} {\bibfnamefont
  {D.}~\bibnamefont {Li}}, \bibinfo {author} {\bibfnamefont {L.~F.}\
  \bibnamefont {Kourkoutis}},\ and\ \bibinfo {author} {\bibfnamefont {H.~Y.}\
  \bibnamefont {Hwang}},\ }\bibfield  {title} {\bibinfo {title} {Nickelate
  superconductivity without rare-earth magnetism: {(La,Sr)NiO$_2$}},\ }\href
  {https://doi.org/https://doi.org/10.1002/adma.202104083} {\bibfield
  {journal} {\bibinfo  {journal} {Adv. Mater.}\ }\textbf {\bibinfo {volume}
  {33}},\ \bibinfo {pages} {2104083} (\bibinfo {year} {2021})}\BibitemShut
  {NoStop}%
\bibitem [{\citenamefont {Pan}\ \emph {et~al.}(2022)\citenamefont {Pan},
  \citenamefont {Ferenc~Segedin}, \citenamefont {LaBollita}, \citenamefont
  {Song}, \citenamefont {Nica}, \citenamefont {Goodge}, \citenamefont {Pierce},
  \citenamefont {Doyle}, \citenamefont {Novakov}, \citenamefont
  {C{\'o}rdova~Carrizales}, \citenamefont {N'Diaye}, \citenamefont {Shafer},
  \citenamefont {Paik}, \citenamefont {Heron}, \citenamefont {Mason},
  \citenamefont {Yacoby}, \citenamefont {Kourkoutis}, \citenamefont {Erten},
  \citenamefont {Brooks}, \citenamefont {Botana},\ and\ \citenamefont
  {Mundy}}]{Pan-ILSC:22}%
  \BibitemOpen
  \bibfield  {author} {\bibinfo {author} {\bibfnamefont {G.~A.}\ \bibnamefont
  {Pan}}, \bibinfo {author} {\bibfnamefont {D.}~\bibnamefont {Ferenc~Segedin}},
  \bibinfo {author} {\bibfnamefont {H.}~\bibnamefont {LaBollita}}, \bibinfo
  {author} {\bibfnamefont {Q.}~\bibnamefont {Song}}, \bibinfo {author}
  {\bibfnamefont {E.~M.}\ \bibnamefont {Nica}}, \bibinfo {author}
  {\bibfnamefont {B.~H.}\ \bibnamefont {Goodge}}, \bibinfo {author}
  {\bibfnamefont {A.~T.}\ \bibnamefont {Pierce}}, \bibinfo {author}
  {\bibfnamefont {S.}~\bibnamefont {Doyle}}, \bibinfo {author} {\bibfnamefont
  {S.}~\bibnamefont {Novakov}}, \bibinfo {author} {\bibfnamefont
  {D.}~\bibnamefont {C{\'o}rdova~Carrizales}}, \bibinfo {author} {\bibfnamefont
  {A.~T.}\ \bibnamefont {N'Diaye}}, \bibinfo {author} {\bibfnamefont
  {P.}~\bibnamefont {Shafer}}, \bibinfo {author} {\bibfnamefont
  {H.}~\bibnamefont {Paik}}, \bibinfo {author} {\bibfnamefont {J.~T.}\
  \bibnamefont {Heron}}, \bibinfo {author} {\bibfnamefont {J.~A.}\ \bibnamefont
  {Mason}}, \bibinfo {author} {\bibfnamefont {A.}~\bibnamefont {Yacoby}},
  \bibinfo {author} {\bibfnamefont {L.~F.}\ \bibnamefont {Kourkoutis}},
  \bibinfo {author} {\bibfnamefont {O.}~\bibnamefont {Erten}}, \bibinfo
  {author} {\bibfnamefont {C.~M.}\ \bibnamefont {Brooks}}, \bibinfo {author}
  {\bibfnamefont {A.~S.}\ \bibnamefont {Botana}},\ and\ \bibinfo {author}
  {\bibfnamefont {J.~A.}\ \bibnamefont {Mundy}},\ }\bibfield  {title} {\bibinfo
  {title} {Superconductivity in a quintuple-layer square-planar nickelate},\
  }\href {https://doi.org/10.1038/s41563-021-01142-9} {\bibfield  {journal}
  {\bibinfo  {journal} {Nat. Mater.}\ }\textbf {\bibinfo {volume} {21}},\
  \bibinfo {pages} {160} (\bibinfo {year} {2022})}\BibitemShut {NoStop}%
\bibitem [{\citenamefont {Lorenz}\ \emph {et~al.}(2016)\citenamefont {Lorenz},
  \citenamefont {Rao}, \citenamefont {Venkatesan}, \citenamefont {Fortunato},
  \citenamefont {Barquinha}, \citenamefont {Branquinho}, \citenamefont
  {Salgueiro}, \citenamefont {Martins}, \citenamefont {Carlos}, \citenamefont
  {Liu}, \citenamefont {Shan}, \citenamefont {Grundmann}, \citenamefont
  {Boschker}, \citenamefont {Mukherjee}, \citenamefont {Priyadarshini},
  \citenamefont {DasGupta}, \citenamefont {Rogers}, \citenamefont {Teherani},
  \citenamefont {Sandana}, \citenamefont {Bove}, \citenamefont {Rietwyk},
  \citenamefont {Zaban}, \citenamefont {Veziridis}, \citenamefont {Weidenkaff},
  \citenamefont {Muralidhar}, \citenamefont {Murakami}, \citenamefont {Abel},
  \citenamefont {Fompeyrine}, \citenamefont {Zuniga-Perez}, \citenamefont
  {Ramesh}, \citenamefont {Spaldin}, \citenamefont {Ostanin}, \citenamefont
  {Borisov}, \citenamefont {Mertig}, \citenamefont {Lazenka}, \citenamefont
  {Srinivasan}, \citenamefont {Prellier}, \citenamefont {Uchida}, \citenamefont
  {Kawasaki}, \citenamefont {Pentcheva}, \citenamefont {Gegenwart},
  \citenamefont {Granozio}, \citenamefont {Fontcuberta},\ and\ \citenamefont
  {Pryds}}]{OxideRoadmap:16}%
  \BibitemOpen
  \bibfield  {author} {\bibinfo {author} {\bibfnamefont {M.}~\bibnamefont
  {Lorenz}}, \bibinfo {author} {\bibfnamefont {M.~S.~R.}\ \bibnamefont {Rao}},
  \bibinfo {author} {\bibfnamefont {T.}~\bibnamefont {Venkatesan}}, \bibinfo
  {author} {\bibfnamefont {E.}~\bibnamefont {Fortunato}}, \bibinfo {author}
  {\bibfnamefont {P.}~\bibnamefont {Barquinha}}, \bibinfo {author}
  {\bibfnamefont {R.}~\bibnamefont {Branquinho}}, \bibinfo {author}
  {\bibfnamefont {D.}~\bibnamefont {Salgueiro}}, \bibinfo {author}
  {\bibfnamefont {R.}~\bibnamefont {Martins}}, \bibinfo {author} {\bibfnamefont
  {E.}~\bibnamefont {Carlos}}, \bibinfo {author} {\bibfnamefont
  {A.}~\bibnamefont {Liu}}, \bibinfo {author} {\bibfnamefont {F.~K.}\
  \bibnamefont {Shan}}, \bibinfo {author} {\bibfnamefont {M.}~\bibnamefont
  {Grundmann}}, \bibinfo {author} {\bibfnamefont {H.}~\bibnamefont {Boschker}},
  \bibinfo {author} {\bibfnamefont {J.}~\bibnamefont {Mukherjee}}, \bibinfo
  {author} {\bibfnamefont {M.}~\bibnamefont {Priyadarshini}}, \bibinfo {author}
  {\bibfnamefont {N.}~\bibnamefont {DasGupta}}, \bibinfo {author}
  {\bibfnamefont {D.~J.}\ \bibnamefont {Rogers}}, \bibinfo {author}
  {\bibfnamefont {F.~H.}\ \bibnamefont {Teherani}}, \bibinfo {author}
  {\bibfnamefont {E.~V.}\ \bibnamefont {Sandana}}, \bibinfo {author}
  {\bibfnamefont {P.}~\bibnamefont {Bove}}, \bibinfo {author} {\bibfnamefont
  {K.}~\bibnamefont {Rietwyk}}, \bibinfo {author} {\bibfnamefont
  {A.}~\bibnamefont {Zaban}}, \bibinfo {author} {\bibfnamefont
  {A.}~\bibnamefont {Veziridis}}, \bibinfo {author} {\bibfnamefont
  {A.}~\bibnamefont {Weidenkaff}}, \bibinfo {author} {\bibfnamefont
  {M.}~\bibnamefont {Muralidhar}}, \bibinfo {author} {\bibfnamefont
  {M.}~\bibnamefont {Murakami}}, \bibinfo {author} {\bibfnamefont
  {S.}~\bibnamefont {Abel}}, \bibinfo {author} {\bibfnamefont {J.}~\bibnamefont
  {Fompeyrine}}, \bibinfo {author} {\bibfnamefont {J.}~\bibnamefont
  {Zuniga-Perez}}, \bibinfo {author} {\bibfnamefont {R.}~\bibnamefont
  {Ramesh}}, \bibinfo {author} {\bibfnamefont {N.~A.}\ \bibnamefont {Spaldin}},
  \bibinfo {author} {\bibfnamefont {S.}~\bibnamefont {Ostanin}}, \bibinfo
  {author} {\bibfnamefont {V.}~\bibnamefont {Borisov}}, \bibinfo {author}
  {\bibfnamefont {I.}~\bibnamefont {Mertig}}, \bibinfo {author} {\bibfnamefont
  {V.}~\bibnamefont {Lazenka}}, \bibinfo {author} {\bibfnamefont
  {G.}~\bibnamefont {Srinivasan}}, \bibinfo {author} {\bibfnamefont
  {W.}~\bibnamefont {Prellier}}, \bibinfo {author} {\bibfnamefont
  {M.}~\bibnamefont {Uchida}}, \bibinfo {author} {\bibfnamefont
  {M.}~\bibnamefont {Kawasaki}}, \bibinfo {author} {\bibfnamefont
  {R.}~\bibnamefont {Pentcheva}}, \bibinfo {author} {\bibfnamefont
  {P.}~\bibnamefont {Gegenwart}}, \bibinfo {author} {\bibfnamefont {F.~M.}\
  \bibnamefont {Granozio}}, \bibinfo {author} {\bibfnamefont {J.}~\bibnamefont
  {Fontcuberta}},\ and\ \bibinfo {author} {\bibfnamefont {N.}~\bibnamefont
  {Pryds}},\ }\bibfield  {title} {\bibinfo {title} {The 2016 oxide electronic
  materials and oxide interfaces roadmap},\ }\href
  {http://stacks.iop.org/0022-3727/49/i=43/a=433001} {\bibfield  {journal}
  {\bibinfo  {journal} {J. Phys. D: Appl. Phys.}\ }\textbf {\bibinfo {volume}
  {49}},\ \bibinfo {pages} {433001} (\bibinfo {year} {2016})}\BibitemShut
  {NoStop}%
\bibitem [{\citenamefont {Middey}\ \emph {et~al.}(2016)\citenamefont {Middey},
  \citenamefont {Chakhalian}, \citenamefont {Mahadevan}, \citenamefont
  {Freeland}, \citenamefont {Millis},\ and\ \citenamefont
  {Sarma}}]{RENickelateReview:16}%
  \BibitemOpen
  \bibfield  {author} {\bibinfo {author} {\bibfnamefont {S.}~\bibnamefont
  {Middey}}, \bibinfo {author} {\bibfnamefont {J.}~\bibnamefont {Chakhalian}},
  \bibinfo {author} {\bibfnamefont {P.}~\bibnamefont {Mahadevan}}, \bibinfo
  {author} {\bibfnamefont {J.}~\bibnamefont {Freeland}}, \bibinfo {author}
  {\bibfnamefont {A.}~\bibnamefont {Millis}},\ and\ \bibinfo {author}
  {\bibfnamefont {D.}~\bibnamefont {Sarma}},\ }\bibfield  {title} {\bibinfo
  {title} {Physics of ultrathin films and heterostructures of rare-earth
  nickelates},\ }\href {https://doi.org/10.1146/annurev-matsci-070115-032057}
  {\bibfield  {journal} {\bibinfo  {journal} {Annu. Rev. Mater. Res.}\ }\textbf
  {\bibinfo {volume} {46}},\ \bibinfo {pages} {305} (\bibinfo {year}
  {2016})}\BibitemShut {NoStop}%
\bibitem [{\citenamefont {Belviso}\ \emph {et~al.}(2019)\citenamefont
  {Belviso}, \citenamefont {Claerbout}, \citenamefont {Comas-Vives},
  \citenamefont {Dalal}, \citenamefont {Fan}, \citenamefont {Filippetti},
  \citenamefont {Fiorentini}, \citenamefont {Foppa}, \citenamefont {Franchini},
  \citenamefont {Geisler}, \citenamefont {Ghiringhelli}, \citenamefont
  {Gro{\ss}}, \citenamefont {Hu}, \citenamefont {{\'I}{\~n}iguez},
  \citenamefont {Kauwe}, \citenamefont {Musfeldt}, \citenamefont {Nicolini},
  \citenamefont {Pentcheva}, \citenamefont {Polcar}, \citenamefont {Ren},
  \citenamefont {Ricci}, \citenamefont {Ricci}, \citenamefont {Sen},
  \citenamefont {Skelton}, \citenamefont {Sparks}, \citenamefont {Stroppa},
  \citenamefont {Urru}, \citenamefont {Vandichel}, \citenamefont {Vavassori},
  \citenamefont {Wu}, \citenamefont {Yang}, \citenamefont {Zhao}, \citenamefont
  {Puggioni}, \citenamefont {Cortese},\ and\ \citenamefont
  {Cammarata}}]{Viewpoint:19}%
  \BibitemOpen
  \bibfield  {author} {\bibinfo {author} {\bibfnamefont {F.}~\bibnamefont
  {Belviso}}, \bibinfo {author} {\bibfnamefont {V.~E.~P.}\ \bibnamefont
  {Claerbout}}, \bibinfo {author} {\bibfnamefont {A.}~\bibnamefont
  {Comas-Vives}}, \bibinfo {author} {\bibfnamefont {N.~S.}\ \bibnamefont
  {Dalal}}, \bibinfo {author} {\bibfnamefont {F.-R.}\ \bibnamefont {Fan}},
  \bibinfo {author} {\bibfnamefont {A.}~\bibnamefont {Filippetti}}, \bibinfo
  {author} {\bibfnamefont {V.}~\bibnamefont {Fiorentini}}, \bibinfo {author}
  {\bibfnamefont {L.}~\bibnamefont {Foppa}}, \bibinfo {author} {\bibfnamefont
  {C.}~\bibnamefont {Franchini}}, \bibinfo {author} {\bibfnamefont
  {B.}~\bibnamefont {Geisler}}, \bibinfo {author} {\bibfnamefont {L.~M.}\
  \bibnamefont {Ghiringhelli}}, \bibinfo {author} {\bibfnamefont
  {A.}~\bibnamefont {Gro{\ss}}}, \bibinfo {author} {\bibfnamefont
  {S.}~\bibnamefont {Hu}}, \bibinfo {author} {\bibfnamefont {J.}~\bibnamefont
  {{\'I}{\~n}iguez}}, \bibinfo {author} {\bibfnamefont {S.~K.}\ \bibnamefont
  {Kauwe}}, \bibinfo {author} {\bibfnamefont {J.~L.}\ \bibnamefont {Musfeldt}},
  \bibinfo {author} {\bibfnamefont {P.}~\bibnamefont {Nicolini}}, \bibinfo
  {author} {\bibfnamefont {R.}~\bibnamefont {Pentcheva}}, \bibinfo {author}
  {\bibfnamefont {T.}~\bibnamefont {Polcar}}, \bibinfo {author} {\bibfnamefont
  {W.}~\bibnamefont {Ren}}, \bibinfo {author} {\bibfnamefont {F.}~\bibnamefont
  {Ricci}}, \bibinfo {author} {\bibfnamefont {F.}~\bibnamefont {Ricci}},
  \bibinfo {author} {\bibfnamefont {H.~S.}\ \bibnamefont {Sen}}, \bibinfo
  {author} {\bibfnamefont {J.~M.}\ \bibnamefont {Skelton}}, \bibinfo {author}
  {\bibfnamefont {T.~D.}\ \bibnamefont {Sparks}}, \bibinfo {author}
  {\bibfnamefont {A.}~\bibnamefont {Stroppa}}, \bibinfo {author} {\bibfnamefont
  {A.}~\bibnamefont {Urru}}, \bibinfo {author} {\bibfnamefont {M.}~\bibnamefont
  {Vandichel}}, \bibinfo {author} {\bibfnamefont {P.}~\bibnamefont
  {Vavassori}}, \bibinfo {author} {\bibfnamefont {H.}~\bibnamefont {Wu}},
  \bibinfo {author} {\bibfnamefont {K.}~\bibnamefont {Yang}}, \bibinfo {author}
  {\bibfnamefont {H.~J.}\ \bibnamefont {Zhao}}, \bibinfo {author}
  {\bibfnamefont {D.}~\bibnamefont {Puggioni}}, \bibinfo {author}
  {\bibfnamefont {R.}~\bibnamefont {Cortese}},\ and\ \bibinfo {author}
  {\bibfnamefont {A.}~\bibnamefont {Cammarata}},\ }\bibfield  {title} {\bibinfo
  {title} {Viewpoint: Atomic-scale design protocols toward energy, electronic,
  catalysis, and sensing applications},\ }\href
  {https://doi.org/10.1021/acs.inorgchem.9b01785} {\bibfield  {journal}
  {\bibinfo  {journal} {Inorg. Chem.}\ }\textbf {\bibinfo {volume} {58}},\
  \bibinfo {pages} {14939} (\bibinfo {year} {2019})}\BibitemShut {NoStop}%
\bibitem [{\citenamefont {Geisler}\ \emph {et~al.}(2021)\citenamefont
  {Geisler}, \citenamefont {Yordanov}, \citenamefont {Gruner}, \citenamefont
  {Keimer},\ and\ \citenamefont {Pentcheva}}]{TE-Oxides-Review-Geisler:21}%
  \BibitemOpen
  \bibfield  {author} {\bibinfo {author} {\bibfnamefont {B.}~\bibnamefont
  {Geisler}}, \bibinfo {author} {\bibfnamefont {P.}~\bibnamefont {Yordanov}},
  \bibinfo {author} {\bibfnamefont {M.~E.}\ \bibnamefont {Gruner}}, \bibinfo
  {author} {\bibfnamefont {B.}~\bibnamefont {Keimer}},\ and\ \bibinfo {author}
  {\bibfnamefont {R.}~\bibnamefont {Pentcheva}},\ }\bibfield  {title} {\bibinfo
  {title} {Tuning the thermoelectric properties of transition metal oxide thin
  films and superlattices on the quantum scale},\ }\href
  {https://doi.org/https://doi.org/10.1002/pssb.202100270} {\bibfield
  {journal} {\bibinfo  {journal} {physica status solidi (b)}\ ,\ \bibinfo
  {pages} {2100270}} (\bibinfo {year} {2021})}\BibitemShut {NoStop}%
\bibitem [{\citenamefont {Wang}\ \emph {et~al.}(2022)\citenamefont {Wang},
  \citenamefont {Yang}, \citenamefont {Yang}, \citenamefont {Chen},
  \citenamefont {Zhang}, \citenamefont {Zhang}, \citenamefont {Zhu},
  \citenamefont {Uwatoko}, \citenamefont {Gu}, \citenamefont {Dong},
  \citenamefont {Sun}, \citenamefont {Jin},\ and\ \citenamefont
  {Cheng}}]{Wang-Pressure-PNO:22}%
  \BibitemOpen
  \bibfield  {author} {\bibinfo {author} {\bibfnamefont {N.~N.}\ \bibnamefont
  {Wang}}, \bibinfo {author} {\bibfnamefont {M.~W.}\ \bibnamefont {Yang}},
  \bibinfo {author} {\bibfnamefont {Z.}~\bibnamefont {Yang}}, \bibinfo {author}
  {\bibfnamefont {K.~Y.}\ \bibnamefont {Chen}}, \bibinfo {author}
  {\bibfnamefont {H.}~\bibnamefont {Zhang}}, \bibinfo {author} {\bibfnamefont
  {Q.~H.}\ \bibnamefont {Zhang}}, \bibinfo {author} {\bibfnamefont {Z.~H.}\
  \bibnamefont {Zhu}}, \bibinfo {author} {\bibfnamefont {Y.}~\bibnamefont
  {Uwatoko}}, \bibinfo {author} {\bibfnamefont {L.}~\bibnamefont {Gu}},
  \bibinfo {author} {\bibfnamefont {X.~L.}\ \bibnamefont {Dong}}, \bibinfo
  {author} {\bibfnamefont {J.~P.}\ \bibnamefont {Sun}}, \bibinfo {author}
  {\bibfnamefont {K.~J.}\ \bibnamefont {Jin}},\ and\ \bibinfo {author}
  {\bibfnamefont {J.-G.}\ \bibnamefont {Cheng}},\ }\bibfield  {title} {\bibinfo
  {title} {Pressure-induced monotonic enhancement of tc to over 30{\thinspace}k
  in superconducting {Pr$_{0.82}$Sr$_{0.18}$NiO$_2$} thin films},\ }\href
  {https://doi.org/10.1038/s41467-022-32065-x} {\bibfield  {journal} {\bibinfo
  {journal} {Nat. Commun.}\ }\textbf {\bibinfo {volume} {13}},\ \bibinfo
  {pages} {4367} (\bibinfo {year} {2022})}\BibitemShut {NoStop}%
\bibitem [{\citenamefont {Hou}\ \emph {et~al.}(2023)\citenamefont {Hou},
  \citenamefont {Yang}, \citenamefont {Liu}, \citenamefont {Li}, \citenamefont
  {Shan}, \citenamefont {Ma}, \citenamefont {Wang}, \citenamefont {Wang},
  \citenamefont {Guo}, \citenamefont {Sun}, \citenamefont {Uwatoko},
  \citenamefont {Wang}, \citenamefont {Zhang}, \citenamefont {Wang},\ and\
  \citenamefont {Cheng}}]{Hou-LNO327-ExpConfirm:23}%
  \BibitemOpen
  \bibfield  {author} {\bibinfo {author} {\bibfnamefont {J.}~\bibnamefont
  {Hou}}, \bibinfo {author} {\bibfnamefont {P.~T.}\ \bibnamefont {Yang}},
  \bibinfo {author} {\bibfnamefont {Z.~Y.}\ \bibnamefont {Liu}}, \bibinfo
  {author} {\bibfnamefont {J.~Y.}\ \bibnamefont {Li}}, \bibinfo {author}
  {\bibfnamefont {P.~F.}\ \bibnamefont {Shan}}, \bibinfo {author}
  {\bibfnamefont {L.}~\bibnamefont {Ma}}, \bibinfo {author} {\bibfnamefont
  {G.}~\bibnamefont {Wang}}, \bibinfo {author} {\bibfnamefont {N.~N.}\
  \bibnamefont {Wang}}, \bibinfo {author} {\bibfnamefont {H.~Z.}\ \bibnamefont
  {Guo}}, \bibinfo {author} {\bibfnamefont {J.~P.}\ \bibnamefont {Sun}},
  \bibinfo {author} {\bibfnamefont {Y.}~\bibnamefont {Uwatoko}}, \bibinfo
  {author} {\bibfnamefont {M.}~\bibnamefont {Wang}}, \bibinfo {author}
  {\bibfnamefont {G.~M.}\ \bibnamefont {Zhang}}, \bibinfo {author}
  {\bibfnamefont {B.~S.}\ \bibnamefont {Wang}},\ and\ \bibinfo {author}
  {\bibfnamefont {J.~G.}\ \bibnamefont {Cheng}},\ }\href@noop {} {\bibinfo
  {title} {Emergence of high-temperature superconducting phase in the
  pressurized {La$_3$Ni$_2$O$_7$} crystals}} (\bibinfo {year} {2023}),\ \Eprint
  {https://arxiv.org/abs/2307.09865} {arXiv:2307.09865 [cond-mat.supr-con]}
  \BibitemShut {NoStop}%
\bibitem [{\citenamefont {Zhang}\ \emph
  {et~al.}(2023{\natexlab{a}})\citenamefont {Zhang}, \citenamefont {Su},
  \citenamefont {Huang}, \citenamefont {Sun}, \citenamefont {Huo},
  \citenamefont {Shan}, \citenamefont {Ye}, \citenamefont {Yang}, \citenamefont
  {Li}, \citenamefont {Smidman}, \citenamefont {Wang}, \citenamefont {Jiao},\
  and\ \citenamefont {Yuan}}]{Zhang-LNO327-ZeroResistance:23}%
  \BibitemOpen
  \bibfield  {author} {\bibinfo {author} {\bibfnamefont {Y.}~\bibnamefont
  {Zhang}}, \bibinfo {author} {\bibfnamefont {D.}~\bibnamefont {Su}}, \bibinfo
  {author} {\bibfnamefont {Y.}~\bibnamefont {Huang}}, \bibinfo {author}
  {\bibfnamefont {H.}~\bibnamefont {Sun}}, \bibinfo {author} {\bibfnamefont
  {M.}~\bibnamefont {Huo}}, \bibinfo {author} {\bibfnamefont {Z.}~\bibnamefont
  {Shan}}, \bibinfo {author} {\bibfnamefont {K.}~\bibnamefont {Ye}}, \bibinfo
  {author} {\bibfnamefont {Z.}~\bibnamefont {Yang}}, \bibinfo {author}
  {\bibfnamefont {R.}~\bibnamefont {Li}}, \bibinfo {author} {\bibfnamefont
  {M.}~\bibnamefont {Smidman}}, \bibinfo {author} {\bibfnamefont
  {M.}~\bibnamefont {Wang}}, \bibinfo {author} {\bibfnamefont {L.}~\bibnamefont
  {Jiao}},\ and\ \bibinfo {author} {\bibfnamefont {H.}~\bibnamefont {Yuan}},\
  }\href@noop {} {\bibinfo {title} {High-temperature superconductivity with
  zero-resistance and strange metal behavior in {La$_3$Ni$_2$O$_7$}}} (\bibinfo
  {year} {2023}{\natexlab{a}}),\ \Eprint {https://arxiv.org/abs/2307.14819}
  {arXiv:2307.14819 [cond-mat.supr-con]} \BibitemShut {NoStop}%
\bibitem [{\citenamefont {Luo}\ \emph {et~al.}(2023)\citenamefont {Luo},
  \citenamefont {Hu}, \citenamefont {Wang}, \citenamefont {W\'u},\ and\
  \citenamefont {Yao}}]{Luo-LNO327:23}%
  \BibitemOpen
  \bibfield  {author} {\bibinfo {author} {\bibfnamefont {Z.}~\bibnamefont
  {Luo}}, \bibinfo {author} {\bibfnamefont {X.}~\bibnamefont {Hu}}, \bibinfo
  {author} {\bibfnamefont {M.}~\bibnamefont {Wang}}, \bibinfo {author}
  {\bibfnamefont {W.}~\bibnamefont {W\'u}},\ and\ \bibinfo {author}
  {\bibfnamefont {D.-X.}\ \bibnamefont {Yao}},\ }\bibfield  {title} {\bibinfo
  {title} {Bilayer two-orbital model of {La$_3$Ni$_2$O$_7$} under pressure},\
  }\href {https://doi.org/10.1103/PhysRevLett.131.126001} {\bibfield  {journal}
  {\bibinfo  {journal} {Phys. Rev. Lett.}\ }\textbf {\bibinfo {volume} {131}},\
  \bibinfo {pages} {126001} (\bibinfo {year} {2023})}\BibitemShut {NoStop}%
\bibitem [{\citenamefont {Gu}\ \emph {et~al.}(2023)\citenamefont {Gu},
  \citenamefont {Le}, \citenamefont {Yang}, \citenamefont {Wu},\ and\
  \citenamefont {Hu}}]{Gu-LNO327:23}%
  \BibitemOpen
  \bibfield  {author} {\bibinfo {author} {\bibfnamefont {Y.}~\bibnamefont
  {Gu}}, \bibinfo {author} {\bibfnamefont {C.}~\bibnamefont {Le}}, \bibinfo
  {author} {\bibfnamefont {Z.}~\bibnamefont {Yang}}, \bibinfo {author}
  {\bibfnamefont {X.}~\bibnamefont {Wu}},\ and\ \bibinfo {author}
  {\bibfnamefont {J.}~\bibnamefont {Hu}},\ }\href@noop {} {\bibinfo {title}
  {Effective model and pairing tendency in bilayer {Ni}-based superconductor
  {La$_3$Ni$_2$O$_7$}}} (\bibinfo {year} {2023}),\ \Eprint
  {https://arxiv.org/abs/2306.07275} {arXiv:2306.07275 [cond-mat.supr-con]}
  \BibitemShut {NoStop}%
\bibitem [{\citenamefont {Yang}\ \emph {et~al.}(2023)\citenamefont {Yang},
  \citenamefont {Wang},\ and\ \citenamefont {Wang}}]{Yang-LNO327:23}%
  \BibitemOpen
  \bibfield  {author} {\bibinfo {author} {\bibfnamefont {Q.-G.}\ \bibnamefont
  {Yang}}, \bibinfo {author} {\bibfnamefont {D.}~\bibnamefont {Wang}},\ and\
  \bibinfo {author} {\bibfnamefont {Q.-H.}\ \bibnamefont {Wang}},\ }\bibfield
  {title} {\bibinfo {title} {Possible ${s}_{\ifmmode\pm\else\textpm\fi{}}$-wave
  superconductivity in {La$_3$Ni$_2$O$_7$}},\ }\href
  {https://doi.org/10.1103/PhysRevB.108.L140505} {\bibfield  {journal}
  {\bibinfo  {journal} {Phys. Rev. B}\ }\textbf {\bibinfo {volume} {108}},\
  \bibinfo {pages} {L140505} (\bibinfo {year} {2023})}\BibitemShut {NoStop}%
\bibitem [{\citenamefont {Lechermann}\ \emph {et~al.}(2023)\citenamefont
  {Lechermann}, \citenamefont {Gondolf}, \citenamefont {B\"otzel},\ and\
  \citenamefont {Eremin}}]{Lechermann-LNO327:23}%
  \BibitemOpen
  \bibfield  {author} {\bibinfo {author} {\bibfnamefont {F.}~\bibnamefont
  {Lechermann}}, \bibinfo {author} {\bibfnamefont {J.}~\bibnamefont {Gondolf}},
  \bibinfo {author} {\bibfnamefont {S.}~\bibnamefont {B\"otzel}},\ and\
  \bibinfo {author} {\bibfnamefont {I.~M.}\ \bibnamefont {Eremin}},\
  }\href@noop {} {\bibinfo {title} {Electronic correlations and superconducting
  instability in {La$_3$Ni$_2$O$_7$} under high pressure}} (\bibinfo {year}
  {2023}),\ \Eprint {https://arxiv.org/abs/2306.05121} {arXiv:2306.05121
  [cond-mat.str-el]} \BibitemShut {NoStop}%
\bibitem [{\citenamefont {Sakakibara}\ \emph {et~al.}(2023)\citenamefont
  {Sakakibara}, \citenamefont {Kitamine}, \citenamefont {Ochi},\ and\
  \citenamefont {Kuroki}}]{Sakakibara-LNO327:23}%
  \BibitemOpen
  \bibfield  {author} {\bibinfo {author} {\bibfnamefont {H.}~\bibnamefont
  {Sakakibara}}, \bibinfo {author} {\bibfnamefont {N.}~\bibnamefont
  {Kitamine}}, \bibinfo {author} {\bibfnamefont {M.}~\bibnamefont {Ochi}},\
  and\ \bibinfo {author} {\bibfnamefont {K.}~\bibnamefont {Kuroki}},\
  }\href@noop {} {\bibinfo {title} {Possible high {$T_c$} superconductivity in
  {La$_3$Ni$_2$O$_7$} under high pressure through manifestation of a
  nearly-half-filled bilayer hubbard model}} (\bibinfo {year} {2023}),\ \Eprint
  {https://arxiv.org/abs/2306.06039} {arXiv:2306.06039 [cond-mat.supr-con]}
  \BibitemShut {NoStop}%
\bibitem [{\citenamefont {Shen}\ \emph {et~al.}(2023)\citenamefont {Shen},
  \citenamefont {Qin},\ and\ \citenamefont {Zhang}}]{Shen-LNO327:23}%
  \BibitemOpen
  \bibfield  {author} {\bibinfo {author} {\bibfnamefont {Y.}~\bibnamefont
  {Shen}}, \bibinfo {author} {\bibfnamefont {M.}~\bibnamefont {Qin}},\ and\
  \bibinfo {author} {\bibfnamefont {G.-M.}\ \bibnamefont {Zhang}},\ }\href@noop
  {} {\bibinfo {title} {Effective bi-layer model {Hamiltonian} and
  density-matrix renormalization group study for the {high-$T_c$}
  superconductivity in {La$_{3}$Ni$_{2}$O$_{7}$} under high pressure}}
  (\bibinfo {year} {2023}),\ \Eprint {https://arxiv.org/abs/2306.07837}
  {arXiv:2306.07837 [cond-mat.str-el]} \BibitemShut {NoStop}%
\bibitem [{\citenamefont {Christiansson}\ \emph {et~al.}(2023)\citenamefont
  {Christiansson}, \citenamefont {Petocchi},\ and\ \citenamefont
  {Werner}}]{Christiansson-LNO327:23}%
  \BibitemOpen
  \bibfield  {author} {\bibinfo {author} {\bibfnamefont {V.}~\bibnamefont
  {Christiansson}}, \bibinfo {author} {\bibfnamefont {F.}~\bibnamefont
  {Petocchi}},\ and\ \bibinfo {author} {\bibfnamefont {P.}~\bibnamefont
  {Werner}},\ }\href@noop {} {\bibinfo {title} {Correlated electronic structure
  of {La$_3$Ni$_2$O$_7$} under pressure}} (\bibinfo {year} {2023}),\ \Eprint
  {https://arxiv.org/abs/2306.07931} {arXiv:2306.07931 [cond-mat.str-el]}
  \BibitemShut {NoStop}%
\bibitem [{\citenamefont {Shilenko}\ and\ \citenamefont
  {Leonov}(2023)}]{Shilenko-LNO327:23}%
  \BibitemOpen
  \bibfield  {author} {\bibinfo {author} {\bibfnamefont {D.~A.}\ \bibnamefont
  {Shilenko}}\ and\ \bibinfo {author} {\bibfnamefont {I.~V.}\ \bibnamefont
  {Leonov}},\ }\href@noop {} {\bibinfo {title} {Correlated electronic
  structure, orbital-selective behavior, and magnetic correlations in
  double-layer {La$_3$Ni$_2$O$_7$} under pressure}} (\bibinfo {year} {2023}),\
  \Eprint {https://arxiv.org/abs/2306.14841} {arXiv:2306.14841
  [cond-mat.str-el]} \BibitemShut {NoStop}%
\bibitem [{\citenamefont {Liu}\ \emph {et~al.}(2023{\natexlab{a}})\citenamefont
  {Liu}, \citenamefont {Huo}, \citenamefont {Li}, \citenamefont {Li},
  \citenamefont {Liu}, \citenamefont {Dai}, \citenamefont {Zhou}, \citenamefont
  {Hao}, \citenamefont {Lu}, \citenamefont {Wang},\ and\ \citenamefont
  {Wen}}]{Liu-LNO327-Optics:23}%
  \BibitemOpen
  \bibfield  {author} {\bibinfo {author} {\bibfnamefont {Z.}~\bibnamefont
  {Liu}}, \bibinfo {author} {\bibfnamefont {M.}~\bibnamefont {Huo}}, \bibinfo
  {author} {\bibfnamefont {J.}~\bibnamefont {Li}}, \bibinfo {author}
  {\bibfnamefont {Q.}~\bibnamefont {Li}}, \bibinfo {author} {\bibfnamefont
  {Y.}~\bibnamefont {Liu}}, \bibinfo {author} {\bibfnamefont {Y.}~\bibnamefont
  {Dai}}, \bibinfo {author} {\bibfnamefont {X.}~\bibnamefont {Zhou}}, \bibinfo
  {author} {\bibfnamefont {J.}~\bibnamefont {Hao}}, \bibinfo {author}
  {\bibfnamefont {Y.}~\bibnamefont {Lu}}, \bibinfo {author} {\bibfnamefont
  {M.}~\bibnamefont {Wang}},\ and\ \bibinfo {author} {\bibfnamefont {H.-H.}\
  \bibnamefont {Wen}},\ }\href@noop {} {\bibinfo {title} {Electronic
  correlations and energy gap in the bilayer nickelate
  {La$_{3}$Ni$_{2}$O$_{7}$}}} (\bibinfo {year} {2023}{\natexlab{a}}),\ \Eprint
  {https://arxiv.org/abs/2307.02950} {arXiv:2307.02950 [cond-mat.supr-con]}
  \BibitemShut {NoStop}%
\bibitem [{\citenamefont {Wu}\ \emph {et~al.}(2023)\citenamefont {Wu},
  \citenamefont {Luo}, \citenamefont {Yao},\ and\ \citenamefont
  {Wang}}]{Wu-LNO327:23}%
  \BibitemOpen
  \bibfield  {author} {\bibinfo {author} {\bibfnamefont {W.}~\bibnamefont
  {Wu}}, \bibinfo {author} {\bibfnamefont {Z.}~\bibnamefont {Luo}}, \bibinfo
  {author} {\bibfnamefont {D.-X.}\ \bibnamefont {Yao}},\ and\ \bibinfo {author}
  {\bibfnamefont {M.}~\bibnamefont {Wang}},\ }\href@noop {} {\bibinfo {title}
  {Charge transfer and {Zhang-Rice} singlet bands in the nickelate
  superconductor $\mathrm{La_3Ni_2O_7}$ under pressure}} (\bibinfo {year}
  {2023}),\ \Eprint {https://arxiv.org/abs/2307.05662} {arXiv:2307.05662
  [cond-mat.str-el]} \BibitemShut {NoStop}%
\bibitem [{\citenamefont {Cao}\ and\ \citenamefont {feng
  Yang}(2023)}]{Cao-LNO327:23}%
  \BibitemOpen
  \bibfield  {author} {\bibinfo {author} {\bibfnamefont {Y.}~\bibnamefont
  {Cao}}\ and\ \bibinfo {author} {\bibfnamefont {Y.}~\bibnamefont {feng
  Yang}},\ }\href@noop {} {\bibinfo {title} {Flat bands promoted by hund's rule
  coupling in the candidate double-layer high-temperature superconductor
  {La$_3$Ni$_2$O$_7$}}} (\bibinfo {year} {2023}),\ \Eprint
  {https://arxiv.org/abs/2307.06806} {arXiv:2307.06806 [cond-mat.supr-con]}
  \BibitemShut {NoStop}%
\bibitem [{\citenamefont {Chen}\ \emph
  {et~al.}(2023{\natexlab{a}})\citenamefont {Chen}, \citenamefont {Jiang},
  \citenamefont {Li}, \citenamefont {Zhong},\ and\ \citenamefont
  {Lu}}]{Chen-LNO327:23}%
  \BibitemOpen
  \bibfield  {author} {\bibinfo {author} {\bibfnamefont {X.}~\bibnamefont
  {Chen}}, \bibinfo {author} {\bibfnamefont {P.}~\bibnamefont {Jiang}},
  \bibinfo {author} {\bibfnamefont {J.}~\bibnamefont {Li}}, \bibinfo {author}
  {\bibfnamefont {Z.}~\bibnamefont {Zhong}},\ and\ \bibinfo {author}
  {\bibfnamefont {Y.}~\bibnamefont {Lu}},\ }\href@noop {} {\bibinfo {title}
  {Critical charge and spin instabilities in superconducting
  {La$_3$Ni$_2$O$_7$}}} (\bibinfo {year} {2023}{\natexlab{a}}),\ \Eprint
  {https://arxiv.org/abs/2307.07154} {arXiv:2307.07154 [cond-mat.supr-con]}
  \BibitemShut {NoStop}%
\bibitem [{\citenamefont {Liu}\ \emph {et~al.}(2023{\natexlab{b}})\citenamefont
  {Liu}, \citenamefont {Mei}, \citenamefont {Ye}, \citenamefont {Chen},\ and\
  \citenamefont {Yang}}]{Liu-LNO327-OxVacDestructive:23}%
  \BibitemOpen
  \bibfield  {author} {\bibinfo {author} {\bibfnamefont {Y.-B.}\ \bibnamefont
  {Liu}}, \bibinfo {author} {\bibfnamefont {J.-W.}\ \bibnamefont {Mei}},
  \bibinfo {author} {\bibfnamefont {F.}~\bibnamefont {Ye}}, \bibinfo {author}
  {\bibfnamefont {W.-Q.}\ \bibnamefont {Chen}},\ and\ \bibinfo {author}
  {\bibfnamefont {F.}~\bibnamefont {Yang}},\ }\href@noop {} {\bibinfo {title}
  {The s$^\pm$-wave pairing and the destructive role of apical-oxygen
  deficiencies in {La$_3$Ni$_2$O$_7$} under pressure}} (\bibinfo {year}
  {2023}{\natexlab{b}}),\ \Eprint {https://arxiv.org/abs/2307.10144}
  {arXiv:2307.10144 [cond-mat.supr-con]} \BibitemShut {NoStop}%
\bibitem [{\citenamefont {Lu}\ \emph {et~al.}(2023)\citenamefont {Lu},
  \citenamefont {Pan}, \citenamefont {Yang},\ and\ \citenamefont
  {Wu}}]{Lu-LNO327-InterlayerAFM:23}%
  \BibitemOpen
  \bibfield  {author} {\bibinfo {author} {\bibfnamefont {C.}~\bibnamefont
  {Lu}}, \bibinfo {author} {\bibfnamefont {Z.}~\bibnamefont {Pan}}, \bibinfo
  {author} {\bibfnamefont {F.}~\bibnamefont {Yang}},\ and\ \bibinfo {author}
  {\bibfnamefont {C.}~\bibnamefont {Wu}},\ }\href@noop {} {\bibinfo {title}
  {Interlayer coupling driven high-temperature superconductivity in
  {La$_3$Ni$_2$O$_7$} under pressure}} (\bibinfo {year} {2023}),\ \Eprint
  {https://arxiv.org/abs/2307.14965} {arXiv:2307.14965 [cond-mat.supr-con]}
  \BibitemShut {NoStop}%
\bibitem [{\citenamefont {Zhang}\ \emph
  {et~al.}(2023{\natexlab{b}})\citenamefont {Zhang}, \citenamefont {Lin},
  \citenamefont {Moreo}, \citenamefont {Maier},\ and\ \citenamefont
  {Dagotto}}]{ZhangDagotto-LNO327:23}%
  \BibitemOpen
  \bibfield  {author} {\bibinfo {author} {\bibfnamefont {Y.}~\bibnamefont
  {Zhang}}, \bibinfo {author} {\bibfnamefont {L.-F.}\ \bibnamefont {Lin}},
  \bibinfo {author} {\bibfnamefont {A.}~\bibnamefont {Moreo}}, \bibinfo
  {author} {\bibfnamefont {T.~A.}\ \bibnamefont {Maier}},\ and\ \bibinfo
  {author} {\bibfnamefont {E.}~\bibnamefont {Dagotto}},\ }\href@noop {}
  {\bibinfo {title} {Structural phase transition, $s_{\pm}$-wave pairing and
  magnetic stripe order in the bilayered nickelate superconductor
  {La$_3$Ni$_2$O$_7$} under pressure}} (\bibinfo {year} {2023}{\natexlab{b}}),\
  \Eprint {https://arxiv.org/abs/2307.15276} {arXiv:2307.15276
  [cond-mat.supr-con]} \BibitemShut {NoStop}%
\bibitem [{\citenamefont {Oh}\ and\ \citenamefont
  {Zhang}(2023)}]{OhZhang-LNO327:23}%
  \BibitemOpen
  \bibfield  {author} {\bibinfo {author} {\bibfnamefont {H.}~\bibnamefont
  {Oh}}\ and\ \bibinfo {author} {\bibfnamefont {Y.-H.}\ \bibnamefont {Zhang}},\
  }\href@noop {} {\bibinfo {title} {Type {II} {t-J} model and shared
  antiferromagnetic spin coupling from {Hund's} rule in superconducting
  {La$_3$Ni$_2$O$_7$}}} (\bibinfo {year} {2023}),\ \Eprint
  {https://arxiv.org/abs/2307.15706} {arXiv:2307.15706 [cond-mat.str-el]}
  \BibitemShut {NoStop}%
\bibitem [{\citenamefont {Liao}\ \emph {et~al.}(2023)\citenamefont {Liao},
  \citenamefont {Chen}, \citenamefont {Duan}, \citenamefont {Wang},
  \citenamefont {Liu}, \citenamefont {Yu},\ and\ \citenamefont
  {Si}}]{Liao-LNO327:23}%
  \BibitemOpen
  \bibfield  {author} {\bibinfo {author} {\bibfnamefont {Z.}~\bibnamefont
  {Liao}}, \bibinfo {author} {\bibfnamefont {L.}~\bibnamefont {Chen}}, \bibinfo
  {author} {\bibfnamefont {G.}~\bibnamefont {Duan}}, \bibinfo {author}
  {\bibfnamefont {Y.}~\bibnamefont {Wang}}, \bibinfo {author} {\bibfnamefont
  {C.}~\bibnamefont {Liu}}, \bibinfo {author} {\bibfnamefont {R.}~\bibnamefont
  {Yu}},\ and\ \bibinfo {author} {\bibfnamefont {Q.}~\bibnamefont {Si}},\
  }\href@noop {} {\bibinfo {title} {Electron correlations and superconductivity
  in {La$_3$Ni$_2$O$_7$} under pressure tuning}} (\bibinfo {year} {2023}),\
  \Eprint {https://arxiv.org/abs/2307.16697} {arXiv:2307.16697
  [cond-mat.supr-con]} \BibitemShut {NoStop}%
\bibitem [{\citenamefont {Qu}\ \emph {et~al.}(2023)\citenamefont {Qu},
  \citenamefont {Qu}, \citenamefont {Chen}, \citenamefont {Wu}, \citenamefont
  {Yang}, \citenamefont {Li},\ and\ \citenamefont {Su}}]{Qu-LNO327:23}%
  \BibitemOpen
  \bibfield  {author} {\bibinfo {author} {\bibfnamefont {X.-Z.}\ \bibnamefont
  {Qu}}, \bibinfo {author} {\bibfnamefont {D.-W.}\ \bibnamefont {Qu}}, \bibinfo
  {author} {\bibfnamefont {J.}~\bibnamefont {Chen}}, \bibinfo {author}
  {\bibfnamefont {C.}~\bibnamefont {Wu}}, \bibinfo {author} {\bibfnamefont
  {F.}~\bibnamefont {Yang}}, \bibinfo {author} {\bibfnamefont {W.}~\bibnamefont
  {Li}},\ and\ \bibinfo {author} {\bibfnamefont {G.}~\bibnamefont {Su}},\
  }\href@noop {} {\bibinfo {title} {Bilayer $t$-$j$-$j_\perp$ model and
  magnetically mediated pairing in the pressurized nickelate
  {La$_3$Ni$_2$O$_7$}}} (\bibinfo {year} {2023}),\ \Eprint
  {https://arxiv.org/abs/2307.16873} {arXiv:2307.16873 [cond-mat.str-el]}
  \BibitemShut {NoStop}%
\bibitem [{\citenamefont {Keimer}\ \emph {et~al.}(2015)\citenamefont {Keimer},
  \citenamefont {Kivelson}, \citenamefont {Norman}, \citenamefont {Uchida},\
  and\ \citenamefont {Zaanen}}]{Keimer:15}%
  \BibitemOpen
  \bibfield  {author} {\bibinfo {author} {\bibfnamefont {B.}~\bibnamefont
  {Keimer}}, \bibinfo {author} {\bibfnamefont {S.~A.}\ \bibnamefont
  {Kivelson}}, \bibinfo {author} {\bibfnamefont {M.~R.}\ \bibnamefont
  {Norman}}, \bibinfo {author} {\bibfnamefont {S.}~\bibnamefont {Uchida}},\
  and\ \bibinfo {author} {\bibfnamefont {J.}~\bibnamefont {Zaanen}},\
  }\bibfield  {title} {\bibinfo {title} {From quantum matter to
  high-temperature superconductivity in copper oxides},\ }\href
  {https://doi.org/10.1038/nature14165} {\bibfield  {journal} {\bibinfo
  {journal} {Nature}\ }\textbf {\bibinfo {volume} {518}},\ \bibinfo {pages}
  {179} (\bibinfo {year} {2015})}\BibitemShut {NoStop}%
\bibitem [{\citenamefont {Chubukov}\ and\ \citenamefont
  {Hirschfeld}(2015)}]{ChubukovHirschfeld-FeSC:15}%
  \BibitemOpen
  \bibfield  {author} {\bibinfo {author} {\bibfnamefont {A.}~\bibnamefont
  {Chubukov}}\ and\ \bibinfo {author} {\bibfnamefont {P.~J.}\ \bibnamefont
  {Hirschfeld}},\ }\bibfield  {title} {\bibinfo {title} {Iron-based
  superconductors, seven years later},\ }\href
  {https://doi.org/10.1063/PT.3.2818} {\bibfield  {journal} {\bibinfo
  {journal} {Physics Today}\ }\textbf {\bibinfo {volume} {68}},\ \bibinfo
  {pages} {46} (\bibinfo {year} {2015})}\BibitemShut {NoStop}%
\bibitem [{\citenamefont {Drozdov}\ \emph {et~al.}(2015)\citenamefont
  {Drozdov}, \citenamefont {Eremets}, \citenamefont {Troyan}, \citenamefont
  {Ksenofontov},\ and\ \citenamefont {Shylin}}]{Drozdov:15}%
  \BibitemOpen
  \bibfield  {author} {\bibinfo {author} {\bibfnamefont {A.~P.}\ \bibnamefont
  {Drozdov}}, \bibinfo {author} {\bibfnamefont {M.~I.}\ \bibnamefont
  {Eremets}}, \bibinfo {author} {\bibfnamefont {I.~A.}\ \bibnamefont {Troyan}},
  \bibinfo {author} {\bibfnamefont {V.}~\bibnamefont {Ksenofontov}},\ and\
  \bibinfo {author} {\bibfnamefont {S.~I.}\ \bibnamefont {Shylin}},\ }\bibfield
   {title} {\bibinfo {title} {Conventional superconductivity at 203 {Kelvin} at
  high pressures in the sulfur hydride system},\ }\href
  {https://doi.org/10.1038/nature14964} {\bibfield  {journal} {\bibinfo
  {journal} {Nature}\ }\textbf {\bibinfo {volume} {525}},\ \bibinfo {pages}
  {73} (\bibinfo {year} {2015})}\BibitemShut {NoStop}%
\bibitem [{\citenamefont {Drozdov}\ \emph {et~al.}(2019)\citenamefont
  {Drozdov}, \citenamefont {Kong}, \citenamefont {Minkov}, \citenamefont
  {Besedin}, \citenamefont {Kuzovnikov}, \citenamefont {Mozaffari},
  \citenamefont {Balicas}, \citenamefont {Balakirev}, \citenamefont {Graf},
  \citenamefont {Prakapenka}, \citenamefont {Greenberg}, \citenamefont
  {Knyazev}, \citenamefont {Tkacz},\ and\ \citenamefont
  {Eremets}}]{Drozdov:19}%
  \BibitemOpen
  \bibfield  {author} {\bibinfo {author} {\bibfnamefont {A.~P.}\ \bibnamefont
  {Drozdov}}, \bibinfo {author} {\bibfnamefont {P.~P.}\ \bibnamefont {Kong}},
  \bibinfo {author} {\bibfnamefont {V.~S.}\ \bibnamefont {Minkov}}, \bibinfo
  {author} {\bibfnamefont {S.~P.}\ \bibnamefont {Besedin}}, \bibinfo {author}
  {\bibfnamefont {M.~A.}\ \bibnamefont {Kuzovnikov}}, \bibinfo {author}
  {\bibfnamefont {S.}~\bibnamefont {Mozaffari}}, \bibinfo {author}
  {\bibfnamefont {L.}~\bibnamefont {Balicas}}, \bibinfo {author} {\bibfnamefont
  {F.~F.}\ \bibnamefont {Balakirev}}, \bibinfo {author} {\bibfnamefont {D.~E.}\
  \bibnamefont {Graf}}, \bibinfo {author} {\bibfnamefont {V.~B.}\ \bibnamefont
  {Prakapenka}}, \bibinfo {author} {\bibfnamefont {E.}~\bibnamefont
  {Greenberg}}, \bibinfo {author} {\bibfnamefont {D.~A.}\ \bibnamefont
  {Knyazev}}, \bibinfo {author} {\bibfnamefont {M.}~\bibnamefont {Tkacz}},\
  and\ \bibinfo {author} {\bibfnamefont {M.~I.}\ \bibnamefont {Eremets}},\
  }\bibfield  {title} {\bibinfo {title} {Superconductivity at 250 {K} in
  lanthanum hydride under high pressures},\ }\href
  {https://doi.org/10.1038/s41586-019-1201-8} {\bibfield  {journal} {\bibinfo
  {journal} {Nature}\ }\textbf {\bibinfo {volume} {569}},\ \bibinfo {pages}
  {528} (\bibinfo {year} {2019})}\BibitemShut {NoStop}%
\bibitem [{\citenamefont {Hu}\ \emph {et~al.}(2011)\citenamefont {Hu},
  \citenamefont {McCandless}, \citenamefont {Garlea}, \citenamefont {Stadler},
  \citenamefont {Xiong}, \citenamefont {Chan}, \citenamefont {Plummer},\ and\
  \citenamefont {Jin}}]{Hu2011}%
  \BibitemOpen
  \bibfield  {author} {\bibinfo {author} {\bibfnamefont {B.}~\bibnamefont
  {Hu}}, \bibinfo {author} {\bibfnamefont {G.~T.}\ \bibnamefont {McCandless}},
  \bibinfo {author} {\bibfnamefont {V.~O.}\ \bibnamefont {Garlea}}, \bibinfo
  {author} {\bibfnamefont {S.}~\bibnamefont {Stadler}}, \bibinfo {author}
  {\bibfnamefont {Y.}~\bibnamefont {Xiong}}, \bibinfo {author} {\bibfnamefont
  {J.~Y.}\ \bibnamefont {Chan}}, \bibinfo {author} {\bibfnamefont {E.~W.}\
  \bibnamefont {Plummer}},\ and\ \bibinfo {author} {\bibfnamefont
  {R.}~\bibnamefont {Jin}},\ }\bibfield  {title} {\bibinfo {title}
  {Structure-property coupling in
  {Sr${}_{3}$(Ru${}_{1\ensuremath{-}x}$Mn${}_{x}$)${}_{2}$O${}_{7}$}},\ }\href
  {https://doi.org/10.1103/PhysRevB.84.174411} {\bibfield  {journal} {\bibinfo
  {journal} {Phys. Rev. B}\ }\textbf {\bibinfo {volume} {84}},\ \bibinfo
  {pages} {174411} (\bibinfo {year} {2011})}\BibitemShut {NoStop}%
\bibitem [{\citenamefont {Puetter}\ \emph {et~al.}(2010)\citenamefont
  {Puetter}, \citenamefont {Rau},\ and\ \citenamefont {Kee}}]{Puetter2010}%
  \BibitemOpen
  \bibfield  {author} {\bibinfo {author} {\bibfnamefont {C.~M.}\ \bibnamefont
  {Puetter}}, \bibinfo {author} {\bibfnamefont {J.~G.}\ \bibnamefont {Rau}},\
  and\ \bibinfo {author} {\bibfnamefont {H.-Y.}\ \bibnamefont {Kee}},\
  }\bibfield  {title} {\bibinfo {title} {Microscopic route to nematicity in
  {${\text{Sr}}_{3}{\text{Ru}}_{2}{\text{O}}_{7}$}},\ }\href
  {https://doi.org/10.1103/PhysRevB.81.081105} {\bibfield  {journal} {\bibinfo
  {journal} {Phys. Rev. B}\ }\textbf {\bibinfo {volume} {81}},\ \bibinfo
  {pages} {081105} (\bibinfo {year} {2010})}\BibitemShut {NoStop}%
\bibitem [{\citenamefont {Mesa}\ \emph {et~al.}(2012)\citenamefont {Mesa},
  \citenamefont {Ye}, \citenamefont {Chi}, \citenamefont {Fernandez-Baca},
  \citenamefont {Tian}, \citenamefont {Hu}, \citenamefont {Jin}, \citenamefont
  {Plummer},\ and\ \citenamefont {Zhang}}]{Mesa2012}%
  \BibitemOpen
  \bibfield  {author} {\bibinfo {author} {\bibfnamefont {D.}~\bibnamefont
  {Mesa}}, \bibinfo {author} {\bibfnamefont {F.}~\bibnamefont {Ye}}, \bibinfo
  {author} {\bibfnamefont {S.}~\bibnamefont {Chi}}, \bibinfo {author}
  {\bibfnamefont {J.~A.}\ \bibnamefont {Fernandez-Baca}}, \bibinfo {author}
  {\bibfnamefont {W.}~\bibnamefont {Tian}}, \bibinfo {author} {\bibfnamefont
  {B.}~\bibnamefont {Hu}}, \bibinfo {author} {\bibfnamefont {R.}~\bibnamefont
  {Jin}}, \bibinfo {author} {\bibfnamefont {E.~W.}\ \bibnamefont {Plummer}},\
  and\ \bibinfo {author} {\bibfnamefont {J.}~\bibnamefont {Zhang}},\ }\bibfield
   {title} {\bibinfo {title} {Single-bilayer {$E$-type} antiferromagnetism in
  {Mn}-substituted {Sr${}_{3}$Ru${}_{2}$O${}_{7}$}: {Neutron} scattering
  study},\ }\href {https://doi.org/10.1103/PhysRevB.85.180410} {\bibfield
  {journal} {\bibinfo  {journal} {Phys. Rev. B}\ }\textbf {\bibinfo {volume}
  {85}},\ \bibinfo {pages} {180410} (\bibinfo {year} {2012})}\BibitemShut
  {NoStop}%
\bibitem [{\citenamefont {Mukherjee}\ and\ \citenamefont
  {Lee}(2016)}]{Mukherjee2016}%
  \BibitemOpen
  \bibfield  {author} {\bibinfo {author} {\bibfnamefont {S.}~\bibnamefont
  {Mukherjee}}\ and\ \bibinfo {author} {\bibfnamefont {W.-C.}\ \bibnamefont
  {Lee}},\ }\bibfield  {title} {\bibinfo {title} {Structural and magnetic field
  effects on spin fluctuations in
  {${\mathrm{Sr}}_{3}{\mathrm{Ru}}_{2}{\mathrm{O}}_{7}$}},\ }\href
  {https://doi.org/10.1103/PhysRevB.94.064407} {\bibfield  {journal} {\bibinfo
  {journal} {Phys. Rev. B}\ }\textbf {\bibinfo {volume} {94}},\ \bibinfo
  {pages} {064407} (\bibinfo {year} {2016})}\BibitemShut {NoStop}%
\bibitem [{\citenamefont {Nakayama}\ \emph {et~al.}(2018)\citenamefont
  {Nakayama}, \citenamefont {Kondo}, \citenamefont {Kuroda}, \citenamefont
  {Bareille}, \citenamefont {Watson}, \citenamefont {Kunisada}, \citenamefont
  {Noguchi}, \citenamefont {Kim}, \citenamefont {Hoesch}, \citenamefont
  {Yoshida},\ and\ \citenamefont {Shin}}]{Nakayam2018}%
  \BibitemOpen
  \bibfield  {author} {\bibinfo {author} {\bibfnamefont {M.}~\bibnamefont
  {Nakayama}}, \bibinfo {author} {\bibfnamefont {T.}~\bibnamefont {Kondo}},
  \bibinfo {author} {\bibfnamefont {K.}~\bibnamefont {Kuroda}}, \bibinfo
  {author} {\bibfnamefont {C.}~\bibnamefont {Bareille}}, \bibinfo {author}
  {\bibfnamefont {M.~D.}\ \bibnamefont {Watson}}, \bibinfo {author}
  {\bibfnamefont {S.}~\bibnamefont {Kunisada}}, \bibinfo {author}
  {\bibfnamefont {R.}~\bibnamefont {Noguchi}}, \bibinfo {author} {\bibfnamefont
  {T.~K.}\ \bibnamefont {Kim}}, \bibinfo {author} {\bibfnamefont
  {M.}~\bibnamefont {Hoesch}}, \bibinfo {author} {\bibfnamefont
  {Y.}~\bibnamefont {Yoshida}},\ and\ \bibinfo {author} {\bibfnamefont
  {S.}~\bibnamefont {Shin}},\ }\bibfield  {title} {\bibinfo {title}
  {Orbital-selective metal-insulator transition lifting the ${t}_{2g}$ band
  hybridization in the hund metal
  {${\mathrm{Sr}}_{3}{({\mathrm{Ru}}_{1\ensuremath{-}x}{\mathrm{Mn}}_{x})}_{2}{\mathrm{O}}_{7}$}},\
  }\href {https://doi.org/10.1103/PhysRevB.98.161102} {\bibfield  {journal}
  {\bibinfo  {journal} {Phys. Rev. B}\ }\textbf {\bibinfo {volume} {98}},\
  \bibinfo {pages} {161102} (\bibinfo {year} {2018})}\BibitemShut {NoStop}%
\bibitem [{\citenamefont {Mathieu}\ \emph {et~al.}(2005)\citenamefont
  {Mathieu}, \citenamefont {Asamitsu}, \citenamefont {Kaneko}, \citenamefont
  {He}, \citenamefont {Yu}, \citenamefont {Kumai}, \citenamefont {Onose},
  \citenamefont {Takeshita}, \citenamefont {Arima}, \citenamefont {Takagi},\
  and\ \citenamefont {Tokura}}]{Mathieu2005}%
  \BibitemOpen
  \bibfield  {author} {\bibinfo {author} {\bibfnamefont {R.}~\bibnamefont
  {Mathieu}}, \bibinfo {author} {\bibfnamefont {A.}~\bibnamefont {Asamitsu}},
  \bibinfo {author} {\bibfnamefont {Y.}~\bibnamefont {Kaneko}}, \bibinfo
  {author} {\bibfnamefont {J.~P.}\ \bibnamefont {He}}, \bibinfo {author}
  {\bibfnamefont {X.~Z.}\ \bibnamefont {Yu}}, \bibinfo {author} {\bibfnamefont
  {R.}~\bibnamefont {Kumai}}, \bibinfo {author} {\bibfnamefont
  {Y.}~\bibnamefont {Onose}}, \bibinfo {author} {\bibfnamefont
  {N.}~\bibnamefont {Takeshita}}, \bibinfo {author} {\bibfnamefont
  {T.}~\bibnamefont {Arima}}, \bibinfo {author} {\bibfnamefont
  {H.}~\bibnamefont {Takagi}},\ and\ \bibinfo {author} {\bibfnamefont
  {Y.}~\bibnamefont {Tokura}},\ }\bibfield  {title} {\bibinfo {title}
  {Impurity-induced transition to a mott insulator in
  {${\mathrm{Sr}}_{3}{\mathrm{Ru}}_{2}{\mathrm{O}}_{7}$}},\ }\href
  {https://doi.org/10.1103/PhysRevB.72.092404} {\bibfield  {journal} {\bibinfo
  {journal} {Phys. Rev. B}\ }\textbf {\bibinfo {volume} {72}},\ \bibinfo
  {pages} {092404} (\bibinfo {year} {2005})}\BibitemShut {NoStop}%
\bibitem [{\citenamefont {Kohn}\ and\ \citenamefont {Sham}(1965)}]{KoSh65}%
  \BibitemOpen
  \bibfield  {author} {\bibinfo {author} {\bibfnamefont {W.}~\bibnamefont
  {Kohn}}\ and\ \bibinfo {author} {\bibfnamefont {L.~J.}\ \bibnamefont
  {Sham}},\ }\bibfield  {title} {\bibinfo {title} {Self-consistent equations
  including exchange and correlation effects},\ }\href
  {https://doi.org/10.1103/PhysRev.140.A1133} {\bibfield  {journal} {\bibinfo
  {journal} {Phys. Rev.}\ }\textbf {\bibinfo {volume} {140}},\ \bibinfo {pages}
  {A1133} (\bibinfo {year} {1965})}\BibitemShut {NoStop}%
\bibitem [{\citenamefont {Kresse}\ and\ \citenamefont
  {Joubert}(1999)}]{USPP-PAW:99}%
  \BibitemOpen
  \bibfield  {author} {\bibinfo {author} {\bibfnamefont {G.}~\bibnamefont
  {Kresse}}\ and\ \bibinfo {author} {\bibfnamefont {D.}~\bibnamefont
  {Joubert}},\ }\bibfield  {title} {\bibinfo {title} {From ultrasoft
  pseudopotentials to the projector augmented-wave method},\ }\href
  {https://doi.org/10.1103/PhysRevB.59.1758} {\bibfield  {journal} {\bibinfo
  {journal} {Phys. Rev. B}\ }\textbf {\bibinfo {volume} {59}},\ \bibinfo
  {pages} {1758} (\bibinfo {year} {1999})}\BibitemShut {NoStop}%
\bibitem [{\citenamefont {Bl\"ochl}(1994)}]{PAW:94}%
  \BibitemOpen
  \bibfield  {author} {\bibinfo {author} {\bibfnamefont {P.~E.}\ \bibnamefont
  {Bl\"ochl}},\ }\bibfield  {title} {\bibinfo {title} {Projector augmented-wave
  method},\ }\href {https://doi.org/10.1103/PhysRevB.50.17953} {\bibfield
  {journal} {\bibinfo  {journal} {Phys. Rev. B}\ }\textbf {\bibinfo {volume}
  {50}},\ \bibinfo {pages} {17953} (\bibinfo {year} {1994})}\BibitemShut
  {NoStop}%
\bibitem [{\citenamefont {Perdew}\ \emph {et~al.}(1996)\citenamefont {Perdew},
  \citenamefont {Burke},\ and\ \citenamefont {Ernzerhof}}]{PeBu96}%
  \BibitemOpen
  \bibfield  {author} {\bibinfo {author} {\bibfnamefont {J.~P.}\ \bibnamefont
  {Perdew}}, \bibinfo {author} {\bibfnamefont {K.}~\bibnamefont {Burke}},\ and\
  \bibinfo {author} {\bibfnamefont {M.}~\bibnamefont {Ernzerhof}},\ }\bibfield
  {title} {\bibinfo {title} {Generalized gradient approximation made simple},\
  }\href {https://doi.org/10.1103/PhysRevLett.77.3865} {\bibfield  {journal}
  {\bibinfo  {journal} {Phys. Rev. Lett.}\ }\textbf {\bibinfo {volume} {77}},\
  \bibinfo {pages} {3865} (\bibinfo {year} {1996})}\BibitemShut {NoStop}%
\bibitem [{\citenamefont {Liechtenstein}\ \emph {et~al.}(1995)\citenamefont
  {Liechtenstein}, \citenamefont {Anisimov},\ and\ \citenamefont
  {Zaanen}}]{LiechtensteinAnisimov:95}%
  \BibitemOpen
  \bibfield  {author} {\bibinfo {author} {\bibfnamefont {A.~I.}\ \bibnamefont
  {Liechtenstein}}, \bibinfo {author} {\bibfnamefont {V.~I.}\ \bibnamefont
  {Anisimov}},\ and\ \bibinfo {author} {\bibfnamefont {J.}~\bibnamefont
  {Zaanen}},\ }\bibfield  {title} {\bibinfo {title} {Density-functional theory
  and strong interactions: {Orbital} ordering in {Mott-Hubbard} insulators},\
  }\href {https://doi.org/10.1103/PhysRevB.52.R5467} {\bibfield  {journal}
  {\bibinfo  {journal} {Phys. Rev. B}\ }\textbf {\bibinfo {volume} {52}},\
  \bibinfo {pages} {R5467} (\bibinfo {year} {1995})}\BibitemShut {NoStop}%
\bibitem [{\citenamefont {Dudarev}\ \emph {et~al.}(1998)\citenamefont
  {Dudarev}, \citenamefont {Botton}, \citenamefont {Savrasov}, \citenamefont
  {Humphreys},\ and\ \citenamefont {Sutton}}]{Dudarev:98}%
  \BibitemOpen
  \bibfield  {author} {\bibinfo {author} {\bibfnamefont {S.~L.}\ \bibnamefont
  {Dudarev}}, \bibinfo {author} {\bibfnamefont {G.~A.}\ \bibnamefont {Botton}},
  \bibinfo {author} {\bibfnamefont {S.~Y.}\ \bibnamefont {Savrasov}}, \bibinfo
  {author} {\bibfnamefont {C.~J.}\ \bibnamefont {Humphreys}},\ and\ \bibinfo
  {author} {\bibfnamefont {A.~P.}\ \bibnamefont {Sutton}},\ }\bibfield  {title}
  {\bibinfo {title} {Electron-energy-loss spectra and the structural stability
  of nickel oxide: An {LSDA$+U$} study},\ }\href
  {https://doi.org/10.1103/PhysRevB.57.1505} {\bibfield  {journal} {\bibinfo
  {journal} {Phys. Rev. B}\ }\textbf {\bibinfo {volume} {57}},\ \bibinfo
  {pages} {1505} (\bibinfo {year} {1998})}\BibitemShut {NoStop}%
\bibitem [{\citenamefont {Liu}\ \emph {et~al.}(2013)\citenamefont {Liu},
  \citenamefont {Kargarian}, \citenamefont {Kareev}, \citenamefont {Gray},
  \citenamefont {Ryan}, \citenamefont {Cruz}, \citenamefont {Tahir},
  \citenamefont {Chuang}, \citenamefont {Guo}, \citenamefont {Rondinelli},
  \citenamefont {Freeland}, \citenamefont {Fiete},\ and\ \citenamefont
  {Chakhalian}}]{Liu-NNO:13}%
  \BibitemOpen
  \bibfield  {author} {\bibinfo {author} {\bibfnamefont {J.}~\bibnamefont
  {Liu}}, \bibinfo {author} {\bibfnamefont {M.}~\bibnamefont {Kargarian}},
  \bibinfo {author} {\bibfnamefont {M.}~\bibnamefont {Kareev}}, \bibinfo
  {author} {\bibfnamefont {B.}~\bibnamefont {Gray}}, \bibinfo {author}
  {\bibfnamefont {P.~J.}\ \bibnamefont {Ryan}}, \bibinfo {author}
  {\bibfnamefont {A.}~\bibnamefont {Cruz}}, \bibinfo {author} {\bibfnamefont
  {N.}~\bibnamefont {Tahir}}, \bibinfo {author} {\bibfnamefont {Y.-D.}\
  \bibnamefont {Chuang}}, \bibinfo {author} {\bibfnamefont {J.}~\bibnamefont
  {Guo}}, \bibinfo {author} {\bibfnamefont {J.~M.}\ \bibnamefont {Rondinelli}},
  \bibinfo {author} {\bibfnamefont {J.~W.}\ \bibnamefont {Freeland}}, \bibinfo
  {author} {\bibfnamefont {G.~A.}\ \bibnamefont {Fiete}},\ and\ \bibinfo
  {author} {\bibfnamefont {J.}~\bibnamefont {Chakhalian}},\ }\bibfield  {title}
  {\bibinfo {title} {Heterointerface engineered electronic and magnetic phases
  of {NdNiO$_3$} thin films},\ }\href {https://doi.org/10.1038/ncomms3714}
  {\bibfield  {journal} {\bibinfo  {journal} {Nat. Commun.}\ }\textbf {\bibinfo
  {volume} {4}},\ \bibinfo {pages} {2714} (\bibinfo {year} {2013})}\BibitemShut
  {NoStop}%
\bibitem [{\citenamefont {Geisler}\ \emph {et~al.}(2017)\citenamefont
  {Geisler}, \citenamefont {Blanca-Romero},\ and\ \citenamefont
  {Pentcheva}}]{Geisler-LNOSTO:17}%
  \BibitemOpen
  \bibfield  {author} {\bibinfo {author} {\bibfnamefont {B.}~\bibnamefont
  {Geisler}}, \bibinfo {author} {\bibfnamefont {A.}~\bibnamefont
  {Blanca-Romero}},\ and\ \bibinfo {author} {\bibfnamefont {R.}~\bibnamefont
  {Pentcheva}},\ }\bibfield  {title} {\bibinfo {title} {Design of $n$- and
  $p$-type oxide thermoelectrics in {LaNiO$_{3}$/SrTiO$_{3}(001)$}
  superlattices exploiting interface polarity},\ }\href
  {https://doi.org/10.1103/PhysRevB.95.125301} {\bibfield  {journal} {\bibinfo
  {journal} {Phys. Rev. B}\ }\textbf {\bibinfo {volume} {95}},\ \bibinfo
  {pages} {125301} (\bibinfo {year} {2017})}\BibitemShut {NoStop}%
\bibitem [{\citenamefont {Wrobel}\ \emph {et~al.}(2018)\citenamefont {Wrobel},
  \citenamefont {Geisler}, \citenamefont {Wang}, \citenamefont {Christiani},
  \citenamefont {Logvenov}, \citenamefont {Bluschke}, \citenamefont {Schierle},
  \citenamefont {van Aken}, \citenamefont {Keimer}, \citenamefont {Pentcheva},\
  and\ \citenamefont {Benckiser}}]{WrobelGeisler:18}%
  \BibitemOpen
  \bibfield  {author} {\bibinfo {author} {\bibfnamefont {F.}~\bibnamefont
  {Wrobel}}, \bibinfo {author} {\bibfnamefont {B.}~\bibnamefont {Geisler}},
  \bibinfo {author} {\bibfnamefont {Y.}~\bibnamefont {Wang}}, \bibinfo {author}
  {\bibfnamefont {G.}~\bibnamefont {Christiani}}, \bibinfo {author}
  {\bibfnamefont {G.}~\bibnamefont {Logvenov}}, \bibinfo {author}
  {\bibfnamefont {M.}~\bibnamefont {Bluschke}}, \bibinfo {author}
  {\bibfnamefont {E.}~\bibnamefont {Schierle}}, \bibinfo {author}
  {\bibfnamefont {P.~A.}\ \bibnamefont {van Aken}}, \bibinfo {author}
  {\bibfnamefont {B.}~\bibnamefont {Keimer}}, \bibinfo {author} {\bibfnamefont
  {R.}~\bibnamefont {Pentcheva}},\ and\ \bibinfo {author} {\bibfnamefont
  {E.}~\bibnamefont {Benckiser}},\ }\bibfield  {title} {\bibinfo {title}
  {Digital modulation of the nickel valence state in a cuprate-nickelate
  heterostructure},\ }\href {https://doi.org/10.1103/PhysRevMaterials.2.035001}
  {\bibfield  {journal} {\bibinfo  {journal} {Phys. Rev. Materials}\ }\textbf
  {\bibinfo {volume} {2}},\ \bibinfo {pages} {035001} (\bibinfo {year}
  {2018})}\BibitemShut {NoStop}%
\bibitem [{\citenamefont {Geisler}\ and\ \citenamefont
  {Pentcheva}(2018)}]{GeislerPentcheva-LNOLAO:18}%
  \BibitemOpen
  \bibfield  {author} {\bibinfo {author} {\bibfnamefont {B.}~\bibnamefont
  {Geisler}}\ and\ \bibinfo {author} {\bibfnamefont {R.}~\bibnamefont
  {Pentcheva}},\ }\bibfield  {title} {\bibinfo {title} {Confinement- and
  strain-induced enhancement of thermoelectric properties in
  {LaNiO$_{3}$}/{LaAlO$_{3}(001)$} superlattices},\ }\href
  {https://doi.org/10.1103/PhysRevMaterials.2.055403} {\bibfield  {journal}
  {\bibinfo  {journal} {Phys. Rev. Materials}\ }\textbf {\bibinfo {volume}
  {2}},\ \bibinfo {pages} {055403} (\bibinfo {year} {2018})}\BibitemShut
  {NoStop}%
\bibitem [{\citenamefont {Geisler}\ and\ \citenamefont
  {Pentcheva}(2019)}]{GeislerPentcheva-LNOLAO-Resonances:19}%
  \BibitemOpen
  \bibfield  {author} {\bibinfo {author} {\bibfnamefont {B.}~\bibnamefont
  {Geisler}}\ and\ \bibinfo {author} {\bibfnamefont {R.}~\bibnamefont
  {Pentcheva}},\ }\bibfield  {title} {\bibinfo {title} {Inducing $n$- and
  $p$-type thermoelectricity in oxide superlattices by strain tuning of
  orbital-selective transport resonances},\ }\href
  {https://doi.org/10.1103/PhysRevApplied.11.044047} {\bibfield  {journal}
  {\bibinfo  {journal} {Phys. Rev. Applied}\ }\textbf {\bibinfo {volume}
  {11}},\ \bibinfo {pages} {044047} (\bibinfo {year} {2019})}\BibitemShut
  {NoStop}%
\bibitem [{\citenamefont {Geisler}\ \emph {et~al.}(2022)\citenamefont
  {Geisler}, \citenamefont {Follmann},\ and\ \citenamefont
  {Pentcheva}}]{Geisler-VO-LNOLAO:22}%
  \BibitemOpen
  \bibfield  {author} {\bibinfo {author} {\bibfnamefont {B.}~\bibnamefont
  {Geisler}}, \bibinfo {author} {\bibfnamefont {S.}~\bibnamefont {Follmann}},\
  and\ \bibinfo {author} {\bibfnamefont {R.}~\bibnamefont {Pentcheva}},\
  }\bibfield  {title} {\bibinfo {title} {Oxygen vacancy formation and
  electronic reconstruction in strained {${\mathrm{LaNiO}}_{3}$} and
  {${\mathrm{LaNiO}}_{3}/{\mathrm{LaAlO}}_{3}$} superlattices},\ }\href
  {https://doi.org/10.1103/PhysRevB.106.155139} {\bibfield  {journal} {\bibinfo
   {journal} {Phys. Rev. B}\ }\textbf {\bibinfo {volume} {106}},\ \bibinfo
  {pages} {155139} (\bibinfo {year} {2022})}\BibitemShut {NoStop}%
\bibitem [{\citenamefont {Geisler}\ \emph {et~al.}(2024)\citenamefont
  {Geisler}, \citenamefont {Fanfarillo}, \citenamefont {Hamlin}, \citenamefont
  {Stewart}, \citenamefont {Hennig},\ and\ \citenamefont
  {Hirschfeld}}]{Geisler-LNO327-Optical:24}%
  \BibitemOpen
  \bibfield  {author} {\bibinfo {author} {\bibfnamefont {B.}~\bibnamefont
  {Geisler}}, \bibinfo {author} {\bibfnamefont {L.}~\bibnamefont {Fanfarillo}},
  \bibinfo {author} {\bibfnamefont {J.~J.}\ \bibnamefont {Hamlin}}, \bibinfo
  {author} {\bibfnamefont {G.~R.}\ \bibnamefont {Stewart}}, \bibinfo {author}
  {\bibfnamefont {R.~G.}\ \bibnamefont {Hennig}},\ and\ \bibinfo {author}
  {\bibfnamefont {P.~J.}\ \bibnamefont {Hirschfeld}},\ }\href@noop {} {\bibinfo
  {title} {Optical properties and electronic correlations in
  {La$_3$Ni$_2$O$_{7-\delta}$} bilayer nickelates under high pressure}}
  (\bibinfo {year} {2024}),\ \Eprint {https://arxiv.org/abs/2401.04258}
  {arXiv:2401.04258 [cond-mat.supr-con]} \BibitemShut {NoStop}%
\bibitem [{\citenamefont {Monkhorst}\ and\ \citenamefont
  {Pack}(1976)}]{MoPa76}%
  \BibitemOpen
  \bibfield  {author} {\bibinfo {author} {\bibfnamefont {H.~J.}\ \bibnamefont
  {Monkhorst}}\ and\ \bibinfo {author} {\bibfnamefont {J.~D.}\ \bibnamefont
  {Pack}},\ }\bibfield  {title} {\bibinfo {title} {Special points for
  brillouin-zone integrations},\ }\href
  {https://doi.org/10.1103/PhysRevB.13.5188} {\bibfield  {journal} {\bibinfo
  {journal} {Phys. Rev. B}\ }\textbf {\bibinfo {volume} {13}},\ \bibinfo
  {pages} {5188} (\bibinfo {year} {1976})}\BibitemShut {NoStop}%
\bibitem [{\citenamefont {Zhang}\ \emph {et~al.}(1994)\citenamefont {Zhang},
  \citenamefont {Greenblatt},\ and\ \citenamefont
  {Goodenough}}]{LNO-327-Diffraction-Zhang:94}%
  \BibitemOpen
  \bibfield  {author} {\bibinfo {author} {\bibfnamefont {Z.}~\bibnamefont
  {Zhang}}, \bibinfo {author} {\bibfnamefont {M.}~\bibnamefont {Greenblatt}},\
  and\ \bibinfo {author} {\bibfnamefont {J.}~\bibnamefont {Goodenough}},\
  }\bibfield  {title} {\bibinfo {title} {Synthesis, structure, and properties
  of the layered perovskite {La$_3$Ni$_2$O$_{7-\delta}$}},\ }\href
  {https://doi.org/https://doi.org/10.1006/jssc.1994.1059} {\bibfield
  {journal} {\bibinfo  {journal} {Journal of Solid State Chemistry}\ }\textbf
  {\bibinfo {volume} {108}},\ \bibinfo {pages} {402} (\bibinfo {year}
  {1994})}\BibitemShut {NoStop}%
\bibitem [{\citenamefont {Ling}\ \emph {et~al.}(2000)\citenamefont {Ling},
  \citenamefont {Argyriou}, \citenamefont {Wu},\ and\ \citenamefont
  {Neumeier}}]{LNO-327-Diffraction-Ling:00}%
  \BibitemOpen
  \bibfield  {author} {\bibinfo {author} {\bibfnamefont {C.~D.}\ \bibnamefont
  {Ling}}, \bibinfo {author} {\bibfnamefont {D.~N.}\ \bibnamefont {Argyriou}},
  \bibinfo {author} {\bibfnamefont {G.}~\bibnamefont {Wu}},\ and\ \bibinfo
  {author} {\bibfnamefont {J.}~\bibnamefont {Neumeier}},\ }\bibfield  {title}
  {\bibinfo {title} {Neutron diffraction study of {La$_3$Ni$_2$O$_7$}:
  {Structural} relationships among $n=$1, 2, and 3 phases
  {La$_{n+1}$Ni$_n$O$_{3n+1}$}},\ }\href
  {https://doi.org/https://doi.org/10.1006/jssc.2000.8721} {\bibfield
  {journal} {\bibinfo  {journal} {Journal of Solid State Chemistry}\ }\textbf
  {\bibinfo {volume} {152}},\ \bibinfo {pages} {517} (\bibinfo {year}
  {2000})}\BibitemShut {NoStop}%
\bibitem [{\citenamefont {Wang}\ \emph {et~al.}(2023)\citenamefont {Wang},
  \citenamefont {Li}, \citenamefont {Xie}, \citenamefont {Liu}, \citenamefont
  {Sun}, \citenamefont {Huang}, \citenamefont {Gao}, \citenamefont {Nakagawa},
  \citenamefont {Fu}, \citenamefont {Dong}, \citenamefont {Cao}, \citenamefont
  {Yu}, \citenamefont {Kawaguchi}, \citenamefont {Kadobayashi}, \citenamefont
  {Wang}, \citenamefont {Jin}, \citenamefont {kwang Mao},\ and\ \citenamefont
  {Liu}}]{Wang-LNO327-I4mmm:23}%
  \BibitemOpen
  \bibfield  {author} {\bibinfo {author} {\bibfnamefont {L.}~\bibnamefont
  {Wang}}, \bibinfo {author} {\bibfnamefont {Y.}~\bibnamefont {Li}}, \bibinfo
  {author} {\bibfnamefont {S.}~\bibnamefont {Xie}}, \bibinfo {author}
  {\bibfnamefont {F.}~\bibnamefont {Liu}}, \bibinfo {author} {\bibfnamefont
  {H.}~\bibnamefont {Sun}}, \bibinfo {author} {\bibfnamefont {C.}~\bibnamefont
  {Huang}}, \bibinfo {author} {\bibfnamefont {Y.}~\bibnamefont {Gao}}, \bibinfo
  {author} {\bibfnamefont {T.}~\bibnamefont {Nakagawa}}, \bibinfo {author}
  {\bibfnamefont {B.}~\bibnamefont {Fu}}, \bibinfo {author} {\bibfnamefont
  {B.}~\bibnamefont {Dong}}, \bibinfo {author} {\bibfnamefont {Z.}~\bibnamefont
  {Cao}}, \bibinfo {author} {\bibfnamefont {R.}~\bibnamefont {Yu}}, \bibinfo
  {author} {\bibfnamefont {S.~I.}\ \bibnamefont {Kawaguchi}}, \bibinfo {author}
  {\bibfnamefont {H.}~\bibnamefont {Kadobayashi}}, \bibinfo {author}
  {\bibfnamefont {M.}~\bibnamefont {Wang}}, \bibinfo {author} {\bibfnamefont
  {C.}~\bibnamefont {Jin}}, \bibinfo {author} {\bibfnamefont {H.}~\bibnamefont
  {kwang Mao}},\ and\ \bibinfo {author} {\bibfnamefont {H.}~\bibnamefont
  {Liu}},\ }\href@noop {} {\bibinfo {title} {Structure responsible for the
  superconducting state in {La$_3$Ni$_2$O$_7$} at high pressure and low
  temperature conditions}} (\bibinfo {year} {2023}),\ \Eprint
  {https://arxiv.org/abs/2311.09186} {arXiv:2311.09186 [cond-mat.supr-con]}
  \BibitemShut {NoStop}%
\bibitem [{\citenamefont {Taniguchi}\ \emph {et~al.}(1995)\citenamefont
  {Taniguchi}, \citenamefont {Nishikawa}, \citenamefont {Yasui}, \citenamefont
  {Kobayashi}, \citenamefont {Takeda}, \citenamefont {Shamoto},\ and\
  \citenamefont {Sato}}]{Taniguchi-LNO327:95}%
  \BibitemOpen
  \bibfield  {author} {\bibinfo {author} {\bibfnamefont {S.}~\bibnamefont
  {Taniguchi}}, \bibinfo {author} {\bibfnamefont {T.}~\bibnamefont
  {Nishikawa}}, \bibinfo {author} {\bibfnamefont {Y.}~\bibnamefont {Yasui}},
  \bibinfo {author} {\bibfnamefont {Y.}~\bibnamefont {Kobayashi}}, \bibinfo
  {author} {\bibfnamefont {J.}~\bibnamefont {Takeda}}, \bibinfo {author}
  {\bibfnamefont {S.-i.}\ \bibnamefont {Shamoto}},\ and\ \bibinfo {author}
  {\bibfnamefont {M.}~\bibnamefont {Sato}},\ }\bibfield  {title} {\bibinfo
  {title} {Transport, magnetic and thermal properties of
  {La$_3$Ni$_2$O$_{7-\delta}$}},\ }\href {https://doi.org/10.1143/JPSJ.64.1644}
  {\bibfield  {journal} {\bibinfo  {journal} {Journal of the Physical Society
  of Japan}\ }\textbf {\bibinfo {volume} {64}},\ \bibinfo {pages} {1644}
  (\bibinfo {year} {1995})}\BibitemShut {NoStop}%
\bibitem [{\citenamefont {Varignon}\ \emph {et~al.}(2017)\citenamefont
  {Varignon}, \citenamefont {Grisolia}, \citenamefont {{\'I}{\~n}iguez},
  \citenamefont {Barth{\'e}l{\'e}my},\ and\ \citenamefont
  {Bibes}}]{Varignon:17}%
  \BibitemOpen
  \bibfield  {author} {\bibinfo {author} {\bibfnamefont {J.}~\bibnamefont
  {Varignon}}, \bibinfo {author} {\bibfnamefont {M.~N.}\ \bibnamefont
  {Grisolia}}, \bibinfo {author} {\bibfnamefont {J.}~\bibnamefont
  {{\'I}{\~n}iguez}}, \bibinfo {author} {\bibfnamefont {A.}~\bibnamefont
  {Barth{\'e}l{\'e}my}},\ and\ \bibinfo {author} {\bibfnamefont
  {M.}~\bibnamefont {Bibes}},\ }\bibfield  {title} {\bibinfo {title} {Complete
  phase diagram of rare-earth nickelates from first-principles},\ }\href
  {https://doi.org/10.1038/s41535-017-0024-9} {\bibfield  {journal} {\bibinfo
  {journal} {npj Quantum Materials}\ }\textbf {\bibinfo {volume} {2}},\
  \bibinfo {pages} {21} (\bibinfo {year} {2017})}\BibitemShut {NoStop}%
\bibitem [{\citenamefont {Catalano}\ \emph {et~al.}(2018)\citenamefont
  {Catalano}, \citenamefont {Gibert}, \citenamefont {Fowlie}, \citenamefont
  {Iniguez}, \citenamefont {Triscone},\ and\ \citenamefont
  {Kreisel}}]{Catalano:18}%
  \BibitemOpen
  \bibfield  {author} {\bibinfo {author} {\bibfnamefont {S.}~\bibnamefont
  {Catalano}}, \bibinfo {author} {\bibfnamefont {M.}~\bibnamefont {Gibert}},
  \bibinfo {author} {\bibfnamefont {J.}~\bibnamefont {Fowlie}}, \bibinfo
  {author} {\bibfnamefont {J.}~\bibnamefont {Iniguez}}, \bibinfo {author}
  {\bibfnamefont {J.-M.}\ \bibnamefont {Triscone}},\ and\ \bibinfo {author}
  {\bibfnamefont {J.}~\bibnamefont {Kreisel}},\ }\bibfield  {title} {\bibinfo
  {title} {Rare-earth nickelates {$R$NiO$_3$}: thin films and
  heterostructures},\ }\href {https://doi.org/10.1088/1361-6633/aaa37a}
  {\bibfield  {journal} {\bibinfo  {journal} {Reports on Progress in Physics}\
  }\textbf {\bibinfo {volume} {81}},\ \bibinfo {pages} {046501} (\bibinfo
  {year} {2018})}\BibitemShut {NoStop}%
\bibitem [{\citenamefont {Geisler}\ and\ \citenamefont
  {Pentcheva}(2020{\natexlab{b}})}]{GeislerPentcheva-LCO:20}%
  \BibitemOpen
  \bibfield  {author} {\bibinfo {author} {\bibfnamefont {B.}~\bibnamefont
  {Geisler}}\ and\ \bibinfo {author} {\bibfnamefont {R.}~\bibnamefont
  {Pentcheva}},\ }\bibfield  {title} {\bibinfo {title} {Competition of defect
  ordering and site disproportionation in strained {${\mathrm{LaCoO}}_{3}$ on
  ${\mathrm{SrTiO}}_{3}$(001)}},\ }\href
  {https://doi.org/10.1103/PhysRevB.101.165108} {\bibfield  {journal} {\bibinfo
   {journal} {Phys. Rev. B}\ }\textbf {\bibinfo {volume} {101}},\ \bibinfo
  {pages} {165108} (\bibinfo {year} {2020}{\natexlab{b}})}\BibitemShut
  {NoStop}%
\bibitem [{\citenamefont {Radhakrishnan}\ \emph {et~al.}(2021)\citenamefont
  {Radhakrishnan}, \citenamefont {Geisler}, \citenamefont {F\"ursich},
  \citenamefont {Putzky}, \citenamefont {Wang}, \citenamefont {Ilse},
  \citenamefont {Christiani}, \citenamefont {Logvenov}, \citenamefont
  {Wochner}, \citenamefont {van Aken}, \citenamefont {Goering}, \citenamefont
  {Pentcheva},\ and\ \citenamefont
  {Benckiser}}]{Radhakrishnan-Geisler-YVOLAO:21}%
  \BibitemOpen
  \bibfield  {author} {\bibinfo {author} {\bibfnamefont {P.}~\bibnamefont
  {Radhakrishnan}}, \bibinfo {author} {\bibfnamefont {B.}~\bibnamefont
  {Geisler}}, \bibinfo {author} {\bibfnamefont {K.}~\bibnamefont {F\"ursich}},
  \bibinfo {author} {\bibfnamefont {D.}~\bibnamefont {Putzky}}, \bibinfo
  {author} {\bibfnamefont {Y.}~\bibnamefont {Wang}}, \bibinfo {author}
  {\bibfnamefont {S.~E.}\ \bibnamefont {Ilse}}, \bibinfo {author}
  {\bibfnamefont {G.}~\bibnamefont {Christiani}}, \bibinfo {author}
  {\bibfnamefont {G.}~\bibnamefont {Logvenov}}, \bibinfo {author}
  {\bibfnamefont {P.}~\bibnamefont {Wochner}}, \bibinfo {author} {\bibfnamefont
  {P.~A.}\ \bibnamefont {van Aken}}, \bibinfo {author} {\bibfnamefont
  {E.}~\bibnamefont {Goering}}, \bibinfo {author} {\bibfnamefont
  {R.}~\bibnamefont {Pentcheva}},\ and\ \bibinfo {author} {\bibfnamefont
  {E.}~\bibnamefont {Benckiser}},\ }\bibfield  {title} {\bibinfo {title}
  {Orbital engineering in
  {${\mathrm{YVO}}_{3}$-$\mathrm{La}\mathrm{Al}{\mathrm{O}}_{3}$}
  superlattices},\ }\href {https://doi.org/10.1103/PhysRevB.104.L121102}
  {\bibfield  {journal} {\bibinfo  {journal} {Phys. Rev. B}\ }\textbf {\bibinfo
  {volume} {104}},\ \bibinfo {pages} {L121102} (\bibinfo {year}
  {2021})}\BibitemShut {NoStop}%
\bibitem [{\citenamefont {Radhakrishnan}\ \emph {et~al.}(2022)\citenamefont
  {Radhakrishnan}, \citenamefont {Geisler}, \citenamefont {F\"ursich},
  \citenamefont {Putzky}, \citenamefont {Wang}, \citenamefont {Christiani},
  \citenamefont {Logvenov}, \citenamefont {Wochner}, \citenamefont {van Aken},
  \citenamefont {Pentcheva},\ and\ \citenamefont
  {Benckiser}}]{Radhakrishnan-Geisler-YVOLAO:22}%
  \BibitemOpen
  \bibfield  {author} {\bibinfo {author} {\bibfnamefont {P.}~\bibnamefont
  {Radhakrishnan}}, \bibinfo {author} {\bibfnamefont {B.}~\bibnamefont
  {Geisler}}, \bibinfo {author} {\bibfnamefont {K.}~\bibnamefont {F\"ursich}},
  \bibinfo {author} {\bibfnamefont {D.}~\bibnamefont {Putzky}}, \bibinfo
  {author} {\bibfnamefont {Y.}~\bibnamefont {Wang}}, \bibinfo {author}
  {\bibfnamefont {G.}~\bibnamefont {Christiani}}, \bibinfo {author}
  {\bibfnamefont {G.}~\bibnamefont {Logvenov}}, \bibinfo {author}
  {\bibfnamefont {P.}~\bibnamefont {Wochner}}, \bibinfo {author} {\bibfnamefont
  {P.~A.}\ \bibnamefont {van Aken}}, \bibinfo {author} {\bibfnamefont
  {R.}~\bibnamefont {Pentcheva}},\ and\ \bibinfo {author} {\bibfnamefont
  {E.}~\bibnamefont {Benckiser}},\ }\bibfield  {title} {\bibinfo {title}
  {Coupling of electronic and structural degrees of freedom in vanadate
  superlattices},\ }\href {https://doi.org/10.1103/PhysRevB.105.165117}
  {\bibfield  {journal} {\bibinfo  {journal} {Phys. Rev. B}\ }\textbf {\bibinfo
  {volume} {105}},\ \bibinfo {pages} {165117} (\bibinfo {year}
  {2022})}\BibitemShut {NoStop}%
\bibitem [{\citenamefont {Zhang}\ \emph
  {et~al.}(2023{\natexlab{c}})\citenamefont {Zhang}, \citenamefont {Lin},
  \citenamefont {Moreo}, \citenamefont {Maier},\ and\ \citenamefont
  {Dagotto}}]{ZhangDagotto-RE-LNO327:23}%
  \BibitemOpen
  \bibfield  {author} {\bibinfo {author} {\bibfnamefont {Y.}~\bibnamefont
  {Zhang}}, \bibinfo {author} {\bibfnamefont {L.-F.}\ \bibnamefont {Lin}},
  \bibinfo {author} {\bibfnamefont {A.}~\bibnamefont {Moreo}}, \bibinfo
  {author} {\bibfnamefont {T.~A.}\ \bibnamefont {Maier}},\ and\ \bibinfo
  {author} {\bibfnamefont {E.}~\bibnamefont {Dagotto}},\ }\href@noop {}
  {\bibinfo {title} {Trends of electronic structures and $s_{\pm}$-wave pairing
  for the rare-earth series in bilayer nickelate superconductor
  {$R_3$Ni$_2$O$_7$}}} (\bibinfo {year} {2023}{\natexlab{c}}),\ \Eprint
  {https://arxiv.org/abs/2308.07386} {arXiv:2308.07386 [cond-mat.supr-con]}
  \BibitemShut {NoStop}%
\bibitem [{\citenamefont {Wu}\ \emph {et~al.}(2013)\citenamefont {Wu},
  \citenamefont {Benckiser}, \citenamefont {Haverkort}, \citenamefont {Frano},
  \citenamefont {Lu}, \citenamefont {Nwankwo}, \citenamefont {Br\"uck},
  \citenamefont {Audehm}, \citenamefont {Goering}, \citenamefont {Macke},
  \citenamefont {Hinkov}, \citenamefont {Wochner}, \citenamefont {Christiani},
  \citenamefont {Heinze}, \citenamefont {Logvenov}, \citenamefont
  {Habermeier},\ and\ \citenamefont {Keimer}}]{WuBenckiser:13}%
  \BibitemOpen
  \bibfield  {author} {\bibinfo {author} {\bibfnamefont {M.}~\bibnamefont
  {Wu}}, \bibinfo {author} {\bibfnamefont {E.}~\bibnamefont {Benckiser}},
  \bibinfo {author} {\bibfnamefont {M.~W.}\ \bibnamefont {Haverkort}}, \bibinfo
  {author} {\bibfnamefont {A.}~\bibnamefont {Frano}}, \bibinfo {author}
  {\bibfnamefont {Y.}~\bibnamefont {Lu}}, \bibinfo {author} {\bibfnamefont
  {U.}~\bibnamefont {Nwankwo}}, \bibinfo {author} {\bibfnamefont
  {S.}~\bibnamefont {Br\"uck}}, \bibinfo {author} {\bibfnamefont
  {P.}~\bibnamefont {Audehm}}, \bibinfo {author} {\bibfnamefont
  {E.}~\bibnamefont {Goering}}, \bibinfo {author} {\bibfnamefont
  {S.}~\bibnamefont {Macke}}, \bibinfo {author} {\bibfnamefont
  {V.}~\bibnamefont {Hinkov}}, \bibinfo {author} {\bibfnamefont
  {P.}~\bibnamefont {Wochner}}, \bibinfo {author} {\bibfnamefont
  {G.}~\bibnamefont {Christiani}}, \bibinfo {author} {\bibfnamefont
  {S.}~\bibnamefont {Heinze}}, \bibinfo {author} {\bibfnamefont
  {G.}~\bibnamefont {Logvenov}}, \bibinfo {author} {\bibfnamefont {H.-U.}\
  \bibnamefont {Habermeier}},\ and\ \bibinfo {author} {\bibfnamefont
  {B.}~\bibnamefont {Keimer}},\ }\bibfield  {title} {\bibinfo {title} {Strain
  and composition dependence of orbital polarization in nickel oxide
  superlattices},\ }\href {https://doi.org/10.1103/PhysRevB.88.125124}
  {\bibfield  {journal} {\bibinfo  {journal} {Phys. Rev. B}\ }\textbf {\bibinfo
  {volume} {88}},\ \bibinfo {pages} {125124} (\bibinfo {year}
  {2013})}\BibitemShut {NoStop}%
\bibitem [{\citenamefont {Iliev}\ \emph {et~al.}(1998)\citenamefont {Iliev},
  \citenamefont {Abrashev}, \citenamefont {Lee}, \citenamefont {Popov},
  \citenamefont {Sun}, \citenamefont {Thomsen}, \citenamefont {Meng},\ and\
  \citenamefont {Chu}}]{Raman-Manganates-Iliev:98}%
  \BibitemOpen
  \bibfield  {author} {\bibinfo {author} {\bibfnamefont {M.~N.}\ \bibnamefont
  {Iliev}}, \bibinfo {author} {\bibfnamefont {M.~V.}\ \bibnamefont {Abrashev}},
  \bibinfo {author} {\bibfnamefont {H.-G.}\ \bibnamefont {Lee}}, \bibinfo
  {author} {\bibfnamefont {V.~N.}\ \bibnamefont {Popov}}, \bibinfo {author}
  {\bibfnamefont {Y.~Y.}\ \bibnamefont {Sun}}, \bibinfo {author} {\bibfnamefont
  {C.}~\bibnamefont {Thomsen}}, \bibinfo {author} {\bibfnamefont {R.~L.}\
  \bibnamefont {Meng}},\ and\ \bibinfo {author} {\bibfnamefont {C.~W.}\
  \bibnamefont {Chu}},\ }\bibfield  {title} {\bibinfo {title} {Raman
  spectroscopy of orthorhombic perovskitelike {${\mathrm{YMnO}}_{3}$} and
  {${\mathrm{LaMnO}}_{3}$}},\ }\href {https://doi.org/10.1103/PhysRevB.57.2872}
  {\bibfield  {journal} {\bibinfo  {journal} {Phys. Rev. B}\ }\textbf {\bibinfo
  {volume} {57}},\ \bibinfo {pages} {2872} (\bibinfo {year}
  {1998})}\BibitemShut {NoStop}%
\bibitem [{\citenamefont {Baldini}\ \emph {et~al.}(2015)\citenamefont
  {Baldini}, \citenamefont {Muramatsu}, \citenamefont {Sherafati},
  \citenamefont {Mao}, \citenamefont {Malavasi}, \citenamefont {Postorino},
  \citenamefont {Satpathy},\ and\ \citenamefont {Struzhkin}}]{Baldini-LMO:15}%
  \BibitemOpen
  \bibfield  {author} {\bibinfo {author} {\bibfnamefont {M.}~\bibnamefont
  {Baldini}}, \bibinfo {author} {\bibfnamefont {T.}~\bibnamefont {Muramatsu}},
  \bibinfo {author} {\bibfnamefont {M.}~\bibnamefont {Sherafati}}, \bibinfo
  {author} {\bibfnamefont {H.-k.}\ \bibnamefont {Mao}}, \bibinfo {author}
  {\bibfnamefont {L.}~\bibnamefont {Malavasi}}, \bibinfo {author}
  {\bibfnamefont {P.}~\bibnamefont {Postorino}}, \bibinfo {author}
  {\bibfnamefont {S.}~\bibnamefont {Satpathy}},\ and\ \bibinfo {author}
  {\bibfnamefont {V.~V.}\ \bibnamefont {Struzhkin}},\ }\bibfield  {title}
  {\bibinfo {title} {Origin of colossal magnetoresistance in {LaMnO$_3$}
  manganite},\ }\href {https://doi.org/10.1073/pnas.1424866112} {\bibfield
  {journal} {\bibinfo  {journal} {Proc. Natl. Acad. Sci. USA}\ }\textbf
  {\bibinfo {volume} {112}},\ \bibinfo {pages} {10869} (\bibinfo {year}
  {2015})}\BibitemShut {NoStop}%
\bibitem [{\citenamefont {Liu}\ \emph {et~al.}(2022)\citenamefont {Liu},
  \citenamefont {Sun}, \citenamefont {Huo}, \citenamefont {Ma}, \citenamefont
  {Ji}, \citenamefont {Yi}, \citenamefont {Li}, \citenamefont {Liu},
  \citenamefont {Yu}, \citenamefont {Zhang}, \citenamefont {Chen},
  \citenamefont {Liang}, \citenamefont {Dong}, \citenamefont {Guo},
  \citenamefont {Zhong}, \citenamefont {Shen}, \citenamefont {Li},\ and\
  \citenamefont {Wang}}]{Liu-LNO327-SDW:22}%
  \BibitemOpen
  \bibfield  {author} {\bibinfo {author} {\bibfnamefont {Z.}~\bibnamefont
  {Liu}}, \bibinfo {author} {\bibfnamefont {H.}~\bibnamefont {Sun}}, \bibinfo
  {author} {\bibfnamefont {M.}~\bibnamefont {Huo}}, \bibinfo {author}
  {\bibfnamefont {X.}~\bibnamefont {Ma}}, \bibinfo {author} {\bibfnamefont
  {Y.}~\bibnamefont {Ji}}, \bibinfo {author} {\bibfnamefont {E.}~\bibnamefont
  {Yi}}, \bibinfo {author} {\bibfnamefont {L.}~\bibnamefont {Li}}, \bibinfo
  {author} {\bibfnamefont {H.}~\bibnamefont {Liu}}, \bibinfo {author}
  {\bibfnamefont {J.}~\bibnamefont {Yu}}, \bibinfo {author} {\bibfnamefont
  {Z.}~\bibnamefont {Zhang}}, \bibinfo {author} {\bibfnamefont
  {Z.}~\bibnamefont {Chen}}, \bibinfo {author} {\bibfnamefont {F.}~\bibnamefont
  {Liang}}, \bibinfo {author} {\bibfnamefont {H.}~\bibnamefont {Dong}},
  \bibinfo {author} {\bibfnamefont {H.}~\bibnamefont {Guo}}, \bibinfo {author}
  {\bibfnamefont {D.}~\bibnamefont {Zhong}}, \bibinfo {author} {\bibfnamefont
  {B.}~\bibnamefont {Shen}}, \bibinfo {author} {\bibfnamefont {S.}~\bibnamefont
  {Li}},\ and\ \bibinfo {author} {\bibfnamefont {M.}~\bibnamefont {Wang}},\
  }\bibfield  {title} {\bibinfo {title} {Evidence for charge and spin density
  waves in single crystals of {La$_3$Ni$_2$O$_7$} and {La$_3$Ni$_2$O$_6$}},\
  }\href {https://doi.org/10.1007/s11433-022-1962-4} {\bibfield  {journal}
  {\bibinfo  {journal} {Science China Physics, Mechanics {\&} Astronomy}\
  }\textbf {\bibinfo {volume} {66}},\ \bibinfo {pages} {217411} (\bibinfo
  {year} {2022})}\BibitemShut {NoStop}%
\bibitem [{\citenamefont {Chen}\ \emph
  {et~al.}(2023{\natexlab{b}})\citenamefont {Chen}, \citenamefont {Liu},
  \citenamefont {Jiao}, \citenamefont {Zou}, \citenamefont {Luo}, \citenamefont
  {Wu}, \citenamefont {Zhang}, \citenamefont {Guo},\ and\ \citenamefont
  {Shu}}]{Chen-LNO327-SDW:23}%
  \BibitemOpen
  \bibfield  {author} {\bibinfo {author} {\bibfnamefont {K.}~\bibnamefont
  {Chen}}, \bibinfo {author} {\bibfnamefont {X.}~\bibnamefont {Liu}}, \bibinfo
  {author} {\bibfnamefont {J.}~\bibnamefont {Jiao}}, \bibinfo {author}
  {\bibfnamefont {M.}~\bibnamefont {Zou}}, \bibinfo {author} {\bibfnamefont
  {Y.}~\bibnamefont {Luo}}, \bibinfo {author} {\bibfnamefont {Q.}~\bibnamefont
  {Wu}}, \bibinfo {author} {\bibfnamefont {N.}~\bibnamefont {Zhang}}, \bibinfo
  {author} {\bibfnamefont {Y.}~\bibnamefont {Guo}},\ and\ \bibinfo {author}
  {\bibfnamefont {L.}~\bibnamefont {Shu}},\ }\href@noop {} {\bibinfo {title}
  {Evidence of spin density waves in {La$_3$Ni$_2$O$_{7-\delta}$}}} (\bibinfo
  {year} {2023}{\natexlab{b}}),\ \Eprint {https://arxiv.org/abs/2311.15717}
  {arXiv:2311.15717 [cond-mat.str-el]} \BibitemShut {NoStop}%
\bibitem [{\citenamefont {Fukamachi}\ \emph {et~al.}(2001)\citenamefont
  {Fukamachi}, \citenamefont {Kobayashi}, \citenamefont {Miyashita},\ and\
  \citenamefont {Sato}}]{Fukamachi-LNO327-NMR:01}%
  \BibitemOpen
  \bibfield  {author} {\bibinfo {author} {\bibfnamefont {T.}~\bibnamefont
  {Fukamachi}}, \bibinfo {author} {\bibfnamefont {Y.}~\bibnamefont
  {Kobayashi}}, \bibinfo {author} {\bibfnamefont {T.}~\bibnamefont
  {Miyashita}},\ and\ \bibinfo {author} {\bibfnamefont {M.}~\bibnamefont
  {Sato}},\ }\bibfield  {title} {\bibinfo {title} {${}^{139}${La} {NMR} studies
  of layered perovskite systems {La$_3$Ni$_2$O$_{7-\delta}$} and
  {La$_4$Ni$_3$O$_{10}$}},\ }\href
  {https://doi.org/https://doi.org/10.1016/S0022-3697(00)00127-X} {\bibfield
  {journal} {\bibinfo  {journal} {Journal of Physics and Chemistry of Solids}\
  }\textbf {\bibinfo {volume} {62}},\ \bibinfo {pages} {195} (\bibinfo {year}
  {2001})}\BibitemShut {NoStop}%
\bibitem [{\citenamefont {Blanca-Romero}\ and\ \citenamefont
  {Pentcheva}(2011)}]{ABR:11}%
  \BibitemOpen
  \bibfield  {author} {\bibinfo {author} {\bibfnamefont {A.}~\bibnamefont
  {Blanca-Romero}}\ and\ \bibinfo {author} {\bibfnamefont {R.}~\bibnamefont
  {Pentcheva}},\ }\bibfield  {title} {\bibinfo {title} {Confinement-induced
  metal-to-insulator transition in strained {LaNiO$_{3}$}/{LaAlO$_{3}$}
  superlattices},\ }\href {https://doi.org/10.1103/PhysRevB.84.195450}
  {\bibfield  {journal} {\bibinfo  {journal} {Phys. Rev. B}\ }\textbf {\bibinfo
  {volume} {84}},\ \bibinfo {pages} {195450} (\bibinfo {year}
  {2011})}\BibitemShut {NoStop}%
\bibitem [{\citenamefont {Choi}\ \emph {et~al.}(2020)\citenamefont {Choi},
  \citenamefont {Lee},\ and\ \citenamefont
  {Pickett}}]{Choi-Lee-Pickett-4fNNO:20}%
  \BibitemOpen
  \bibfield  {author} {\bibinfo {author} {\bibfnamefont {M.-Y.}\ \bibnamefont
  {Choi}}, \bibinfo {author} {\bibfnamefont {K.-W.}\ \bibnamefont {Lee}},\ and\
  \bibinfo {author} {\bibfnamefont {W.~E.}\ \bibnamefont {Pickett}},\
  }\bibfield  {title} {\bibinfo {title} {Role of $4f$ states in infinite-layer
  {${\mathrm{NdNiO}}_{2}$}},\ }\href
  {https://doi.org/10.1103/PhysRevB.101.020503} {\bibfield  {journal} {\bibinfo
   {journal} {Phys. Rev. B}\ }\textbf {\bibinfo {volume} {101}},\ \bibinfo
  {pages} {020503} (\bibinfo {year} {2020})}\BibitemShut {NoStop}%
\end{thebibliography}
\end{document}